\documentclass[11pt]{article}
\newcommand{\selfek}{
\begin{picture}(54,14)
\thicklines
\put(26,3.5){\circle{11}}
\multiput(25.8,-1.6)(0.5,0.5){12}{\circle*{1}}
\multiput(26.2,8.6)(-0.5,-0.5){12}{\circle*{1}}
\multiput(23.2,-1)(0.5,0.5){15}{\circle*{1}}
\multiput(28.8,8)(-0.5,-0.5){15}{\circle*{1}}
\put(20.5,3.5){\line(-1,0){17}}
\put(31.5,3.5){\line(1,0){19}}
\put(3.5,5.5){\footnotesize {\bf k}}
\put(42,5.5){\footnotesize -{\bf k}}
\multiput(12,6.5)(0.25,-0.5){13}{\circle*{1}}
\multiput(12,0.5)(0.25,0.5){13}{\circle*{1}}
\multiput(40,6.5)(-0.25,-0.5){13}{\circle*{1}}
\multiput(40,0.5)(-0.25,0.5){13}{\circle*{1}}
\end{picture} }
\newcommand{\selfekk}{
\begin{picture}(44,14)
\thicklines
\put(21,3.5){\circle{11}}
\multiput(20.8,-1.6)(0.5,0.5){12}{\circle*{1}}
\multiput(21.2,8.6)(-0.5,-0.5){12}{\circle*{1}}
\multiput(18.2,-1)(0.5,0.5){15}{\circle*{1}}
\multiput(23.8,8)(-0.5,-0.5){15}{\circle*{1}}
\put(15.5,3.5){\line(-1,0){12}}
\put(26.5,3.5){\line(1,0){14}}
\put(3.5,5.5){\footnotesize {\bf k}}
\put(32,5.5){\footnotesize -{\bf k}}
\end{picture} }
\newcommand{\selfeko}{
\begin{picture}(33,14)
\thicklines
\put(3.5,3.5){\line(1,0){26}}
\put(3.5,5.5){\footnotesize {\bf k}}
\put(21,5.5){\footnotesize -{\bf k}}
\end{picture} }
\newcommand{\selfekkk}{
\begin{picture}(65,14)
\thicklines
\multiput(21,3.5)(21,0){2}{\circle{11}}
\multiput(20.8,-1.6)(0.5,0.5){12}{\circle*{1}}
\multiput(21.2,8.6)(-0.5,-0.5){12}{\circle*{1}}
\multiput(18.2,-1)(0.5,0.5){15}{\circle*{1}}
\multiput(23.8,8)(-0.5,-0.5){15}{\circle*{1}}
\put(26.5,3.5){\line(1,0){10}}
\multiput(41.8,-1.6)(0.5,0.5){12}{\circle*{1}}
\multiput(42.2,8.6)(-0.5,-0.5){12}{\circle*{1}}
\multiput(39.2,-1)(0.5,0.5){15}{\circle*{1}}
\multiput(44.8,8)(-0.5,-0.5){15}{\circle*{1}}
\put(15.5,3.5){\line(-1,0){12}}
\put(47.5,3.5){\line(1,0){14}}
\put(3.5,5.5){\footnotesize {\bf k}}
\put(53,5.5){\footnotesize -{\bf k}}
\end{picture} }
\newcommand{\dshv}{
\begin{picture}(52,12)
\put(15,4){\line(-5,3){10}}
\put(15,4){\line(-5,-3){10}}
\multiput(15,4)(22,0){2}{\circle*{2}}
\multiput(15,4)(6,0){4}{\line(1,0){4}}
\put(37,4){\line(5,3){10}}
\put(37,4){\line(5,-3){10}}
\end{picture}  }
\newcommand{\vertice}{
\begin{picture}(30,12)
\put(15,4){\line(-5,3){10}}
\put(15,4){\line(-5,-3){10}}
\put(15,4){\circle*{2}}
\put(15,4){\line(5,3){10}}
\put(15,4){\line(5,-3){10}}
\end{picture}  }
\newcommand{\Gaussv}{
\begin{picture}(30,2)
\put(15,2){\line(-1,0){10}}
\put(15,2){\circle*{3}}
\put(15,2){\line(1,0){10}}
\end{picture}  }
\newcommand{\Gaussvx}{
\begin{picture}(30,12)
\put(15,2){\line(-1,0){10}}
\put(15,2){\circle*{3}}
\put(15,2){\line(1,0){10}}
\put(15,6){\circle{8}}
\end{picture}  }
\newcommand{\Gaussvxx}{
\begin{picture}(30,14)
\put(15,2){\line(-1,0){10}}
\put(15,2){\circle*{3}}
\put(15,2){\line(1,0){10}}
\put(15,6){\circle{8}}
\put(15,10){\circle*{3}}
\end{picture}  }
\newcommand{\Gaussvxxx}{
\begin{picture}(30,17)
\put(15,-1){\line(-1,0){10}}
\put(15,-1){\circle*{3}}
\put(15,-1){\line(1,0){10}}
\put(15,3){\circle{8}}
\put(15,7){\circle*{3}}
\put(15,11){\circle{8}}
\end{picture}  }
\newcommand{\Gaussvy}{
\begin{picture}(46,16)
\multiput(15,6)(16,0){2}{\circle*{2}}
\put(5,6){\line(1,0){36}}
\put(23,6){\circle{16}}
\end{picture}  }
\newcommand{\vvertice}{
\begin{picture}(50,12)
\linethickness{0.25pt}
\put(15,4){\line(-5,3){10}}
\put(15,4){\line(-5,-3){10}}
\multiput(15,4)(20,0){2}{\circle*{2}}
\bezier{100}(15,4)(25,12.27)(35,4)
\bezier{100}(15,4)(25,-4.27)(35,4)
\put(35,4){\line(5,3){10}}
\put(35,4){\line(5,-3){10}}
\end{picture}  }
\newcommand{\vverticeo}{
\begin{picture}(50,18)
\linethickness{0.25pt}
\put(15,4){\line(-5,3){10}}
\put(15,4){\line(-5,-3){10}}
\multiput(15,4)(20,0){2}{\circle*{2}}
\bezier{100}(15,4)(25,12.27)(35,4)
\bezier{100}(15,4)(25,-4.27)(35,4)
\bezier{100}(22,-2)(27,5)(32,12)
\put(35,4){\line(5,3){10}}
\put(35,4){\line(5,-3){10}}
\put(25,8){\circle*{3}}
\put(25,12){\circle{8}}
\end{picture}  }
\newcommand{\vverticeu}{
\begin{picture}(37,12)
\linethickness{0.25pt}
\multiput(12,4)(20,0){2}{\circle*{2}}
\bezier{100}(12,4)(22,12.27)(32,4)
\bezier{100}(12,4)(22,-4.27)(32,4)
\put(4.8,-2){\small {\bf 0}}
\end{picture}  }
\newcommand{\vverticeuq}{
\begin{picture}(37,12)
\linethickness{0.25pt}
\multiput(12,4)(20,0){2}{\circle*{2}}
\bezier{100}(12,4)(22,12.27)(32,4)
\bezier{100}(12,4)(22,-4.27)(32,4)
\put(4.8,-2){\small {\bf q}}
\end{picture}  }
\newcommand{\vverticeux}{
\begin{picture}(37,12)
\linethickness{1pt}
\multiput(12,4)(20,0){2}{\circle*{2}}
\bezier{7}(12,4)(22,12.27)(32,4)
\bezier{7}(12,4)(22,-4.27)(32,4)
\put(4.8,-2){\small {\bf 0}}
\end{picture}  }
\newcommand{\vverticeuxq}{
\begin{picture}(37,12)
\linethickness{1pt}
\multiput(12,4)(20,0){2}{\circle*{2}}
\bezier{7}(12,4)(22,12.27)(32,4)
\linethickness{0.25pt}
\bezier{100}(12,4)(22,-4.27)(32,4)
\put(4.8,-2){\small {\bf q}}
\end{picture}  }
\newcommand{\vverticeuxqq}{
\begin{picture}(37,12)
\linethickness{1pt}
\multiput(12,4)(20,0){2}{\circle*{2}}
\bezier{7}(12,4)(22,12.27)(32,4)
\bezier{7}(12,4)(22,-4.27)(32,4)
\put(4.8,-2){\small {\bf q}}
\end{picture}  }
\newcommand{\vverticeuo}{
\begin{picture}(37,18)
\linethickness{0.25pt}
\multiput(12,4)(20,0){2}{\circle*{2}}
\bezier{100}(12,4)(22,12.27)(32,4)
\bezier{100}(12,4)(22,-4.27)(32,4)
\put(4.8,-2){\small {\bf 0}}
\put(22,12){\circle{8}}
\put(22,8){\circle*{3}}
\end{picture}  }
\newcommand{\vverticeuu}{
\begin{picture}(37,12)
\linethickness{0.25pt}
\multiput(12,4)(20,0){2}{\circle*{2}}
\bezier{100}(12,4)(22,12.27)(32,4)
\bezier{100}(12,4)(22,-4.27)(32,4)
\put(4.8,-2){\small {\bf 0}}
\put(22,8){\circle*{4}}
\end{picture}  }
\newcommand{\vverticeoo}{
\begin{picture}(37,18)
\linethickness{1pt}
\multiput(12,4)(20,0){2}{\circle*{2}}
\bezier{7}(12,4)(22,12.27)(32,4)
\bezier{7}(12,4)(22,-4.27)(32,4)
\put(4.8,-2){\small {\bf 0}}
\put(22,12){\circle{8}}
\put(22,8){\circle*{3}}
\end{picture}  }

\newcommand{\vverticeuxx}{
\begin{picture}(37,12)
\multiput(12,4)(20,0){2}{\circle*{2}}
\linethickness{0.5pt}
\bezier{20}(12,4)(13.2,4.99)(14.57,5.73)
\bezier{20}(17.42,6.92)(18.91,7.37)(20.45,7.52)
\bezier{20}(23.55,7.52)(25.09,7.37)(26.58,6.92)
\bezier{20}(29.43,5.73)(30.8,4.99)(32,4)
\bezier{20}(12,4)(13.2,3.01)(14.57,2.27)
\bezier{20}(17.42,1.08)(18.91,0.63)(20.45,0.48)
\bezier{20}(23.55,0.48)(25.09,0.63)(26.58,1.08)
\bezier{20}(29.43,2.27)(30.8,3.01)(32,4)
\put(4.8,-2){\small {\bf 0}}
\end{picture}  }
\newcommand{\vverticexx}{
\begin{picture}(30,12)
\linethickness{1pt}
\multiput(5,4)(20,0){2}{\circle*{2}}
\bezier{7}(5,4)(15,12.27)(25,4)
\bezier{7}(5,4)(15,-4.27)(25,4)
\end{picture}  }
\newcommand{\vverticeuxk}{
\begin{picture}(37,12)
\linethickness{1pt}
\multiput(12,4)(20,0){2}{\circle*{2}}
\bezier{7}(12,4)(22,12.27)(32,4)
\bezier{7}(12,4)(22,-4.27)(32,4)
\put(4.8,-2){\small {\bf k}}
\end{picture}  }
\newcommand{\vverticeuyk}{
\begin{picture}(37,14)
\linethickness{1pt}
\multiput(12,4)(20,0){2}{\circle*{2}}
\bezier{7}(12,4)(22,12.27)(32,4)
\bezier{7}(12,4)(22,-4.27)(32,4)
\put(4.8,-2){\small {\bf k}}
\linethickness{0.25pt}
\bezier{100}(17,-2)(22,5)(27,12)
\end{picture}  }
\newcommand{\vverticeuyq}{
\begin{picture}(37,14)
\linethickness{1pt}
\multiput(12,4)(20,0){2}{\circle*{2}}
\bezier{7}(12,4)(22,12.27)(32,4)
\bezier{7}(12,4)(22,-4.27)(32,4)
\put(4.8,-2){\small {\bf q}}
\linethickness{0.25pt}
\bezier{100}(17,-2)(22,5)(27,12)
\end{picture}  }
\newcommand{\vverticer}{
\begin{picture}(30,12)
\put(13,4){\circle*{3}}
\put(19,4){\circle{12}}
\put(4.8,-2){\small {\bf 0}}
\end{picture}  }
\newcommand{\vverticerz}{
\begin{picture}(30,12)
\put(13,4){\circle*{3}}
\put(4.8,-2){\small {\bf 0}}
\linethickness{0.5pt}
\bezier{20}(13,4)(13,5.48)(13.69,6.79)
\bezier{20}(15.59,8.94)(16.81,9.78)(18.28,9.96)
\bezier{20}(21.13,9.61)(22.51,9.09)(23.49,7.98)
\bezier{20}(13,4)(13,2.52)(13.69,1.21)
\bezier{20}(15.59,-0.94)(16.81,-1.78)(18.28,-1.96)
\bezier{20}(21.13,-1.61)(22.51,-1.09)(23.49,0.02)
\bezier{20}(24.83,5.44)(25.4,4)(24.83,2.56)
\end{picture}  }
\newcommand{\vverticerr}{
\begin{picture}(33,14)
\multiput(12,4)(16,0){2}{\circle*{2}}
\put(12,4){\line(1,0){16}}
\put(20,4){\circle{16}}
\put(4.6,-2){\small {\bf 0}}
\end{picture}  }
\newcommand{\vverticery}{
\begin{picture}(33,14)
\multiput(12,4)(16,0){2}{\circle*{2}}
\multiput(14.5,4)(2.6667,0){5}{\circle*{1}}
\put(20,4){\circle{16}}
\put(4.6,-2){\small {\bf 0}}
\end{picture}  }
\newcommand{\vverticero}{
\begin{picture}(33,14)
\multiput(12,4)(16,0){2}{\circle*{2}}
\multiput(14.5,4)(2.6667,0){5}{\circle*{1}}
\put(20,4){\oval(16,16)[t]}
\put(4.6,-2){\small {\bf 0}}
\put(12.39,1.53){\circle*{1}}
\put(13.53,-0.7){\circle*{1}}
\put(15.3,-2.47){\circle*{1}}
\put(17.53,-3.61){\circle*{1}}
\put(20,-4){\circle*{1}}
\put(22.47,-3.61){\circle*{1}}
\put(24.7,-2.47){\circle*{1}}
\put(26.47,-0.7){\circle*{1}}
\put(27.61,1.53){\circle*{1}}
\bezier{50}(18,-6.4)(20,1)(22,7.4)
\end{picture}  }
\newcommand{\vverticerok}{
\begin{picture}(33,14)
\multiput(12,4)(16,0){2}{\circle*{2}}
\multiput(14.5,4)(2.6667,0){5}{\circle*{1}}
\put(20,4){\oval(16,16)[t]}
\put(4.6,-2){\small {\bf k}}
\put(12.39,1.53){\circle*{1}}
\put(13.53,-0.7){\circle*{1}}
\put(15.3,-2.47){\circle*{1}}
\put(17.53,-3.61){\circle*{1}}
\put(20,-4){\circle*{1}}
\put(22.47,-3.61){\circle*{1}}
\put(24.7,-2.47){\circle*{1}}
\put(26.47,-0.7){\circle*{1}}
\put(27.61,1.53){\circle*{1}}
\bezier{50}(18,-6.4)(20,1)(22,7.4)
\end{picture}  }

\newcommand{\vverticerrk}{
\begin{picture}(33,14)
\multiput(12,4)(16,0){2}{\circle*{2}}
\put(12,4){\line(1,0){16}}
\put(20,4){\circle{16}}
\put(4.6,-2){\small {\bf k}}
\end{picture}  }
\newcommand{\vverticerrksl}{
\begin{picture}(33,14)
\multiput(12,4)(16,0){2}{\circle*{2}}
\put(12,4){\line(1,0){16}}
\put(20,4){\circle{16}}
\put(4.6,-2){\small {\bf k}}
\bezier{100}(16,-6)(20,4)(24,14)
\end{picture}  }

\newcommand{\vverticeryk}{
\begin{picture}(33,14)
\multiput(12,4)(16,0){2}{\circle*{2}}
\multiput(14.5,4)(2.6667,0){5}{\circle*{1}}
\put(20,4){\circle{16}}
\put(4.6,-2){\small {\bf k}}
\end{picture}  }
\newcommand{\vverticeryksl}{
\begin{picture}(33,14)
\multiput(12,4)(16,0){2}{\circle*{2}}
\multiput(14.5,4)(2.6667,0){5}{\circle*{1}}
\put(20,4){\circle{16}}
\put(4.6,-2){\small {\bf k}}
\bezier{100}(16,-6)(20,4)(24,14)
\end{picture}  }
\newcommand{\vverticerxk}{
\begin{picture}(33,14)
\multiput(12,4)(16,0){2}{\circle*{2}}
\put(12,4){\line(1,0){16}}
\put(12.39,6.47){\circle*{1}}
\put(13.53,8.7){\circle*{1}}
\put(15.3,10.47){\circle*{1}}
\put(17.53,11.61){\circle*{1}}
\put(20,12){\circle*{1}}
\put(22.47,11.61){\circle*{1}}
\put(24.7,10.47){\circle*{1}}
\put(26.47,8.7){\circle*{1}}
\put(27.61,6.47){\circle*{1}}
\put(12.39,1.53){\circle*{1}}
\put(13.53,-0.7){\circle*{1}}
\put(15.3,-2.47){\circle*{1}}
\put(17.53,-3.61){\circle*{1}}
\put(20,-4){\circle*{1}}
\put(22.47,-3.61){\circle*{1}}
\put(24.7,-2.47){\circle*{1}}
\put(26.47,-0.7){\circle*{1}}
\put(27.61,1.53){\circle*{1}}
\put(4.6,-2){\small {\bf k}}
\end{picture}  }
\newcommand{\vverticerxksl}{
\begin{picture}(33,14)
\multiput(12,4)(16,0){2}{\circle*{2}}
\put(12,4){\line(1,0){16}}
\put(12.39,6.47){\circle*{1}}
\put(13.53,8.7){\circle*{1}}
\put(15.3,10.47){\circle*{1}}
\put(17.53,11.61){\circle*{1}}
\put(20,12){\circle*{1}}
\put(22.47,11.61){\circle*{1}}
\put(24.7,10.47){\circle*{1}}
\put(26.47,8.7){\circle*{1}}
\put(27.61,6.47){\circle*{1}}
\put(12.39,1.53){\circle*{1}}
\put(13.53,-0.7){\circle*{1}}
\put(15.3,-2.47){\circle*{1}}
\put(17.53,-3.61){\circle*{1}}
\put(20,-4){\circle*{1}}
\put(22.47,-3.61){\circle*{1}}
\put(24.7,-2.47){\circle*{1}}
\put(26.47,-0.7){\circle*{1}}
\put(27.61,1.53){\circle*{1}}
\put(4.6,-2){\small {\bf k}}
\bezier{100}(16,-6)(20,4)(24,14)
\end{picture}  }
\newcommand{\vverticetet}{
\begin{picture}(33,18)
\multiput(12,4)(16,0){2}{\circle*{2}}
\multiput(14.5,4)(2.6667,0){5}{\circle*{1}}
\put(12.39,6.47){\circle*{1}}
\put(13.53,8.7){\circle*{1}}
\put(15.3,10.47){\circle*{1}}
\put(17.53,11.61){\circle*{1}}
\put(20,12){\circle*{1}}
\put(22.47,11.61){\circle*{1}}
\put(24.7,10.47){\circle*{1}}
\put(26.47,8.7){\circle*{1}}
\put(27.61,6.47){\circle*{1}}
\put(12.39,1.53){\circle*{1}}
\put(13.53,-0.7){\circle*{1}}
\put(15.3,-2.47){\circle*{1}}
\put(17.53,-3.61){\circle*{1}}
\put(20,-4){\circle*{1}}
\put(22.47,-3.61){\circle*{1}}
\put(24.7,-2.47){\circle*{1}}
\put(26.47,-0.7){\circle*{1}}
\put(27.61,1.53){\circle*{1}}
\put(4.6,-2){\small {\bf k}}
\multiput(20.5,12)(0,-8){3}{\oval(4,4)[l]}
\put(20.5,16){\oval(4,4)[br]}
\multiput(19.5,8)(0,-8){2}{\oval(4,4)[r]}
\put(19.5,-8){\oval(4,4)[tr]}
\end{picture}  }
\newcommand{\vverticec}{
\begin{picture}(33,16)
\multiput(12,4)(16,0){2}{\circle*{2}}
\multiput(14.5,4)(2.6667,0){5}{\circle*{1}}
\put(12.39,6.47){\circle*{1}}
\put(13.53,8.7){\circle*{1}}
\put(15.3,10.47){\circle*{1}}
\put(17.53,11.61){\circle*{1}}
\put(20,12){\circle*{1}}
\put(22.47,11.61){\circle*{1}}
\put(24.7,10.47){\circle*{1}}
\put(26.47,8.7){\circle*{1}}
\put(27.61,6.47){\circle*{1}}
\put(12.39,1.53){\circle*{1}}
\put(13.53,-0.7){\circle*{1}}
\put(15.3,-2.47){\circle*{1}}
\put(17.53,-3.61){\circle*{1}}
\put(20,-4){\circle*{1}}
\put(22.47,-3.61){\circle*{1}}
\put(24.7,-2.47){\circle*{1}}
\put(26.47,-0.7){\circle*{1}}
\put(27.61,1.53){\circle*{1}}
\put(4.6,-2){\small {\bf k}}
\bezier{100}(16,-6)(20,4)(24,14)
\end{picture}  }
\newcommand{\vverticecc}{
\begin{picture}(33,16)
\multiput(12,4)(16,0){2}{\circle*{3}}
\linethickness{0.5pt}
\bezier{20}(12,4)(13.7,4)(15.4,4)
\bezier{20}(18.3,4)(20,4)(21.7,4)
\bezier{20}(24.6,4)(26.5,4)(28,4)
\bezier{20}(12,4)(12,5.826)(12.792,7.471)
\bezier{20}(15.012,10.255)(16.44,11.393)(18.22,11.799)
\bezier{20}(21.78,11.799)(23.56,11.393)(24.988,10.255)
\bezier{20}(27.208,7.471)(28,5.826)(28,4)
\bezier{20}(12,4)(12,2.174)(12.792,0.529)
\bezier{20}(15.012,-2.255)(16.44,-3.393)(18.22,-3.799)
\bezier{20}(21.78,-3.799)(23.56,-3.393)(24.988,-2.255)
\bezier{20}(27.208,0.529)(28,2.174)(28,4)
\put(4.6,-2){\small {\bf k}}
\bezier{100}(16,-6)(20,4)(24,14)
\end{picture}  }
\newcommand{\vverticex}{
\begin{picture}(50,12)
\linethickness{1pt}
\put(15,4){\line(-5,3){10}}
\put(15,4){\line(-5,-3){10}}
\multiput(15,4)(20,0){2}{\circle*{2}}
\bezier{7}(15,4)(25,12.27)(35,4)
\bezier{7}(15,4)(25,-4.27)(35,4)
\put(35,4){\line(5,3){10}}
\put(35,4){\line(5,-3){10}}
\end{picture}  }
\newcommand{\vverticey}{
\begin{picture}(50,14)
\linethickness{1pt}
\put(15,5){\line(-5,3){10}}
\put(15,5){\line(-5,-3){10}}
\multiput(15,5)(20,0){2}{\circle*{2}}
\bezier{7}(15,5)(25,13.27)(35,5)
\bezier{7}(15,5)(25,-3.27)(35,5)
\put(35,5){\line(5,3){10}}
\put(35,5){\line(5,-3){10}}
\linethickness{0.25pt}
\bezier{100}(20,-2)(25,5)(30,12)
\end{picture}  }

\newcommand{\sixwuu}{
\begin{picture}(60,17)
\put(15,5){\line(-5,3){10}}
\put(15,5){\line(-5,-3){10}}
\put(15,5){\line(-1,0){10}}
\put(25,5){\line(0,1){10}}
\put(14,5){\circle*{2}}
\multiput(17.5,5)(2.5,0){3}{\circle*{1}}
\multiput(25,5)(20,0){2}{\circle*{2}}
\put(45,5){\line(5,3){10}}
\put(45,5){\line(5,-3){10}}
\linethickness{1pt}
\bezier{7}(25,5)(35,13.27)(45,5)
\bezier{7}(25,5)(35,-3.27)(45,5)
\linethickness{0.25pt}
\bezier{100}(30,-2)(35,5)(40,12)
\end{picture}  }

\newcommand{\vverticeyy}{
\begin{picture}(50,14)
\linethickness{1pt}
\put(15,5){\line(-5,3){10}}
\put(15,5){\line(-5,-3){10}}
\multiput(15,5)(20,0){2}{\circle*{2}}
\bezier{7}(15,5)(25,13.27)(35,5)
\bezier{7}(15,5)(25,-3.27)(35,5)
\put(35,5){\line(5,3){10}}
\put(35,5){\line(5,-3){10}}
\linethickness{0.25pt}
\bezier{100}(22,-2)(27,5)(32,12)
\put(25,9){\circle*{4}}
\end{picture}  }
\newcommand{\vverticeyyx}{
\begin{picture}(50,14)
\linethickness{0.25pt}
\put(15,5){\line(-5,3){10}}
\put(15,5){\line(-5,-3){10}}
\multiput(15,5)(20,0){2}{\circle*{2}}
\bezier{100}(15,5)(25,13.27)(35,5)
\bezier{100}(15,5)(25,-3.27)(35,5)
\put(35,5){\line(5,3){10}}
\put(35,5){\line(5,-3){10}}
\bezier{100}(22,-2)(27,5)(32,12)
\put(25,9){\circle*{4}}
\end{picture}  }

\newcommand{\vverticez}{
\begin{picture}(50,14)
\put(15,5){\line(-5,3){10}}
\put(15,5){\line(-5,-3){10}}
\multiput(15,5)(20,0){2}{\circle*{2}}
\put(35,5){\line(5,3){10}}
\put(35,5){\line(5,-3){10}}
\linethickness{0.25pt}
\bezier{100}(15,5)(25,13.27)(35,5)
\bezier{100}(15,5)(25,-3.27)(35,5)
\bezier{100}(20,-2)(25,5)(30,12)
\end{picture}  }
\newcommand{\vverticew}{
\begin{picture}(50,14)
\put(15,5){\line(-5,3){10}}
\put(15,5){\line(-5,-3){10}}
\multiput(15,5)(20,0){2}{\circle*{2}}
\put(35,5){\line(5,3){10}}
\put(35,5){\line(5,-3){10}}
\linethickness{0.25pt}
\bezier{100}(20,-2)(25,5)(30,12)
\linethickness{0.5pt}
\bezier{20}(15,5)(16.2,5.99)(17.57,6.73)
\bezier{20}(20.42,7.92)(21.91,8.37)(23.45,8.52)
\bezier{20}(26.55,8.52)(28.09,8.37)(29.58,7.92)
\bezier{20}(32.43,6.73)(33.8,5.99)(35,5)
\bezier{20}(15,5)(16.2,4.01)(17.57,3.27)
\bezier{20}(20.42,2.08)(21.91,1.63)(23.45,1.48)
\bezier{20}(26.55,1.48)(28.09,1.63)(29.58,2.08)
\bezier{20}(32.43,3.27)(33.8,4.01)(35,5)
\end{picture}  }
\newcommand{\vverticezw}{
\begin{picture}(50,14)
\put(15,5){\line(-5,3){10}}
\put(15,5){\line(-5,-3){10}}
\multiput(15,5)(20,0){2}{\circle*{2}}
\put(35,5){\line(5,3){10}}
\put(35,5){\line(5,-3){10}}
\linethickness{0.25pt}
\bezier{100}(20,-2)(25,5)(30,12)
\bezier{100}(15,5)(25,-3.27)(35,5)
\linethickness{0.5pt}
\bezier{20}(15,5)(16.2,5.99)(17.57,6.73)
\bezier{20}(20.42,7.92)(21.91,8.37)(23.45,8.52)
\bezier{20}(26.55,8.52)(28.09,8.37)(29.58,7.92)
\bezier{20}(32.43,6.73)(33.8,5.99)(35,5)
\end{picture}  }
\newcommand{\vverticeww}{
\begin{picture}(50,14)
\put(15,5){\line(-5,3){10}}
\put(15,5){\line(-5,-3){10}}
\multiput(15,5)(20,0){2}{\circle*{2}}
\put(35,5){\line(5,3){10}}
\put(35,5){\line(5,-3){10}}
\linethickness{0.25pt}
\bezier{100}(15,5)(25,-3.27)(35,5)
\linethickness{0.5pt}
\bezier{20}(15,5)(16.2,5.99)(17.57,6.73)
\bezier{20}(20.42,7.92)(21.91,8.37)(23.45,8.52)
\bezier{20}(26.55,8.52)(28.09,8.37)(29.58,7.92)
\bezier{20}(32.43,6.73)(33.8,5.99)(35,5)
\end{picture}  }
\newcommand{\six}{
\begin{picture}(40,12)
\put(15,4){\line(-5,3){10}}
\put(15,4){\line(-5,-3){10}}
\multiput(14,4)(12,0){2}{\circle*{2}}
\put(5,4){\line(1,0){30}}
\put(25,4){\line(5,3){10}}
\put(25,4){\line(5,-3){10}}
\end{picture}  }
\newcommand{\sixx}{
\begin{picture}(45,12)
\put(15,4){\line(-5,3){10}}
\put(15,4){\line(-5,-3){10}}
\put(15,4){\line(-1,0){10}}
\multiput(14,4)(17,0){2}{\circle*{2}}
\multiput(17.5,4)(2.5,0){5}{\circle*{1}}
\put(30,4){\line(5,3){10}}
\put(30,4){\line(5,-3){10}}
\put(30,4){\line(1,0){10}}
\end{picture}  }
\newcommand{\sixxx}{
\begin{picture}(55,12)
\put(15,4){\line(-5,3){10}}
\put(15,4){\line(-5,-3){10}}
\put(15,4){\line(-1,0){10}}
\multiput(14,4)(27,0){2}{\circle*{2}}
\multiput(17.5,4)(2.5,0){9}{\circle*{1}}
\put(40,4){\line(5,3){10}}
\put(40,4){\line(5,-3){10}}
\put(40,4){\line(1,0){10}}
\put(27.5,4){\circle*{4}}
\end{picture}  }
\newcommand{\sixyyy}{
\begin{picture}(55,12)
\multiput(14,4)(27,0){2}{\circle*{2}}
\put(15,4){\line(-5,3){10}}
\put(15,4){\line(-5,-3){10}}
\put(5,4){\line(1,0){45}}
\put(40,4){\line(5,3){10}}
\put(40,4){\line(5,-3){10}}
\put(27.5,4){\circle*{4}}
\end{picture}  }
\newcommand{\sixo}{
\begin{picture}(55,14)
\multiput(14,4)(27,0){2}{\circle*{2}}
\put(15,4){\line(-5,3){10}}
\put(15,4){\line(-5,-3){10}}
\put(5,4){\line(1,0){45}}
\put(40,4){\line(5,3){10}}
\put(40,4){\line(5,-3){10}}
\put(27.5,4){\circle*{3}}
\put(27.5,8){\circle{8}}
\end{picture}  }
\newcommand{\sixz}{
\begin{picture}(45,12)
\put(15,4){\line(-5,3){10}}
\put(15,4){\line(-5,-3){10}}
\put(15,4){\line(-1,0){10}}
\multiput(14,4)(17,0){2}{\circle*{2}}
\multiput(14.5,4)(6,0){3}{\line(1,0){4}}
\put(30,4){\line(5,3){10}}
\put(30,4){\line(5,-3){10}}
\put(30,4){\line(1,0){10}}
\end{picture}  }
\newcommand{\sixy}{
\begin{picture}(77,16)
\put(15,4){\line(-5,3){10}}
\put(15,4){\line(-5,-3){10}}
\multiput(14,4)(17,0){2}{\circle*{2}}
\multiput(46,4)(17,0){2}{\circle*{2}}
\multiput(33.5,4)(2.5,0){5}{\circle*{1}}
\put(62,4){\line(5,3){10}}
\put(62,4){\line(5,-3){10}}
\multiput(15,4)(6,0){3}{\line(1,0){4}}
\multiput(46,4)(6,0){3}{\line(1,0){4}}
\put(31,4){\line(0,1){10}}
\put(46,4){\line(0,1){10}}
\end{picture}  }

\newcommand{\dshvv}{
\begin{picture}(74,12)
\put(15,4){\line(-5,3){10}}
\put(15,4){\line(-5,-3){10}}
\multiput(15,4)(16,0){2}{\circle*{2}}
\multiput(43,4)(16,0){2}{\circle*{2}}
\multiput(15,4)(6,0){3}{\line(1,0){4}}
\multiput(43,4)(6,0){3}{\line(1,0){4}}
\put(37,4){\circle{12}}
\put(59,4){\line(5,3){10}}
\put(59,4){\line(5,-3){10}}
\end{picture}  }
\newcommand{\dshvvx}{
\begin{picture}(74,14)
\put(15,4){\line(-5,3){10}}
\put(15,4){\line(-5,-3){10}}
\multiput(15,4)(16,0){2}{\circle*{2}}
\multiput(43,4)(16,0){2}{\circle*{2}}
\multiput(15,4)(6,0){3}{\line(1,0){4}}
\multiput(43,4)(6,0){3}{\line(1,0){4}}
\put(31.99,6.60){\circle*{1}}
\put(33.66,8.69){\circle*{1}}
\put(36.07,9.85){\circle*{1}}
\put(38.74,9.85){\circle*{1}}
\put(41.14,8.69){\circle*{1}}
\put(42.81,6.60){\circle*{1}}
\put(31.99,1.40){\circle*{1}}
\put(33.66,-0.69){\circle*{1}}
\put(36.07,-1.85){\circle*{1}}
\put(38.74,-1.85){\circle*{1}}
\put(41.14,-0.69){\circle*{1}}
\put(42.81,1.40){\circle*{1}}
\put(59,4){\line(5,3){10}}
\put(59,4){\line(5,-3){10}}
\linethickness{0.25pt}
\bezier{100}(31,-3.5)(37,4)(43,11.5)
\end{picture}  }
\newcommand{\dshvvv}{
\begin{picture}(64,18)
\put(15,6){\line(-5,3){10}}
\put(15,6){\line(-5,-3){10}}
\multiput(15,6)(16,0){2}{\circle*{2}}
\multiput(49,-2)(0,16){2}{\circle*{2}}
\multiput(15,6)(6,0){3}{\line(1,0){4}}
\multiput(49,-2)(0,6){3}{\line(0,1){4}}
\put(39,6){\oval(16,16)[l]}
\put(39,-2){\line(1,0){20}}
\put(39,14){\line(1,0){20}}
\end{picture}  }
\newcommand{\dshvvvx}{
\begin{picture}(64,20)
\put(15,4){\line(-5,3){10}}
\put(15,4){\line(-5,-3){10}}
\multiput(15,4)(16,0){2}{\circle*{2}}
\multiput(49,-4)(0,16){2}{\circle*{2}}
\multiput(15,4)(6,0){3}{\line(1,0){4}}
\multiput(49,-4)(0,6){3}{\line(0,1){4}}
\put(31.39,6.47){\circle*{1}}
\put(32.53,8.7){\circle*{1}}
\put(34.3,10.47){\circle*{1}}
\put(36.53,11.61){\circle*{1}}
\put(39,12){\circle*{1}}
\put(31.39,1.53){\circle*{1}}
\put(32.53,-0.7){\circle*{1}}
\put(34.3,-2.47){\circle*{1}}
\put(36.53,-3.61){\circle*{1}}
\put(39,-4){\circle*{1}}
\put(49,-4){\line(1,0){10}}
\put(49,12){\line(1,0){10}}
\multiput(41.5,12)(2.5,0){3}{\circle*{1}}
\multiput(41.5,-4)(2.5,0){3}{\circle*{1}}
\linethickness{0.25pt}
\bezier{100}(33.5,-6)(39.5,5)(45.5,16)
\end{picture}  }
\newcommand{\dshvvvv}{
\begin{picture}(50,16)
\multiput(5,-2)(0,16){2}{\line(1,0){40}}
\multiput(15,-2)(0,6){3}{\line(0,1){4}}
\multiput(35,-2)(0,6){3}{\line(0,1){4}}
\multiput(15,-2)(0,16){2}{\circle*{2}}
\multiput(35,-2)(0,16){2}{\circle*{2}}
\end{picture}  }
\newcommand{\dshvvvvx}{
\begin{picture}(50,18)
\multiput(5,-3)(0,16){2}{\line(1,0){10}}
\multiput(35,-3)(0,16){2}{\line(1,0){10}}
\multiput(17.5,-3)(2.5,0){7}{\circle*{1}}
\multiput(17.5,13)(2.5,0){7}{\circle*{1}}
\multiput(15,-3)(0,6){3}{\line(0,1){4}}
\multiput(35,-3)(0,6){3}{\line(0,1){4}}
\multiput(15,-3)(0,16){2}{\circle*{2}}
\multiput(35,-3)(0,16){2}{\circle*{2}}
\linethickness{0.25pt}
\bezier{100}(19,-6)(25,5)(31,16)
\end{picture}  }

\newcommand{\vvverticeyuu}{
\begin{picture}(70,14)
\linethickness{1pt}
\put(15,4){\line(-5,3){10}}
\put(15,4){\line(-5,-3){10}}
\multiput(15,4)(20,0){3}{\circle*{2}}
\bezier{7}(15,4)(25,16)(35,4)
\bezier{7}(15,4)(25,-8)(35,4)
\bezier{7}(35,4)(45,16)(55,4)
\bezier{7}(35,4)(45,-8)(55,4)
\put(55,4){\line(5,3){10}}
\put(55,4){\line(5,-3){10}}
\linethickness{0.25pt}
\bezier{100}(15,-2)(25,4)(35,10)
\bezier{100}(35,-2)(45,4)(55,10)
\end{picture}  }

\newcommand{\eight}{
\begin{picture}(50,12)
\put(5,4){\line(1,0){40}}
\multiput(15,4)(10,0){3}{\circle*{2}}
\multiput(15,-2)(10,0){3}{\line(0,1){12}}
\end{picture}  }
\newcommand{\eightx}{
\begin{picture}(50,14)
\multiput(17.5,4)(2.5,0){7}{\circle*{1}}
\multiput(15,4)(10,0){3}{\circle*{2}}
\multiput(15,-4)(10,0){3}{\line(0,1){16}}
\put(15,4){\line(-1,0){10}}
\put(35,4){\line(1,0){10}}
\end{picture}  }
\newcommand{\sixw}{
\begin{picture}(50,12)
\put(5,4){\line(1,0){40}}
\multiput(15,4)(20,0){2}{\circle*{2}}
\multiput(15,-2)(20,0){2}{\line(0,1){12}}
\put(25,4){\circle*{3}}
\end{picture}  }

\newcommand{\cycle}{
\begin{picture}(42,22)
\put(21,12){\circle*{2}}
\put(14.07,0){\circle*{2}}
\put(27.93,0){\circle*{2}}
\linethickness{0.25pt}
\bezier{100}(14.07,0)(9.07,0)(4.07,0)
\bezier{100}(27.93,0)(32.93,0)(37.93,0)
\bezier{100}(14.07,0)(11.57,-4.33)(9.07,-8.66)
\bezier{100}(27.93,0)(30.43,-4.33)(32.93,-8.66)
\bezier{100}(21,12)(18.5,16.33)(16,20.66)
\bezier{100}(21,12)(23.5,16.33)(26,20.66)
\linethickness{1pt}
\bezier{3}(14.07,0)(11.76,4)(14.07,8)
\bezier{3}(27.93,0)(30.24,4)(27.93,8)
\bezier{3}(14.07,8)(16.38,12)(21,12)
\bezier{3}(27.93,8)(25.62,12)(21,12)
\bezier{3}(14.07,0)(16.38,-4)(21,-4)
\bezier{3}(27.93,0)(25.62,-4)(21,-4)
\end{picture}  }
\newcommand{\fourbox}{
\begin{picture}(50,18)
\multiput(15,0)(20,0){2}{\line(0,1){12}}
\multiput(15,0)(0,12){2}{\line(1,0){20}}
\put(15,0){\line(-5,-2){10}}
\put(15,12){\line(-5,2){10}}
\put(35,0){\line(5,-2){10}}
\put(35,12){\line(5,2){10}}
\multiput(15,2)(0,2){5}{\line(1,0){20}}
\end{picture} }
\newcommand{\fourboxx}{
\begin{picture}(50,18)
\multiput(15,0)(20,0){2}{\line(0,1){12}}
\multiput(15,0)(0,12){2}{\line(1,0){20}}
\put(15,0){\line(-5,-2){10}}
\put(15,12){\line(-5,2){10}}
\put(35,0){\line(5,-2){10}}
\put(35,12){\line(5,2){10}}
\multiput(15,2)(0,2){5}{\line(1,0){20}}
\linethickness{0.25pt}
\bezier{100}(20,-4)(25,6)(30,16)
\end{picture} }
\newcommand{\fourboxz}{
\begin{picture}(47,18)
\multiput(12,0)(20,0){2}{\line(0,1){12}}
\multiput(12,0)(0,12){2}{\line(1,0){20}}
\put(5,-4){\small {\bf 0}}
\put(5,9){\small {\bf 0}}
\put(33.5,-4){\small {\bf 0}}
\put(33.5,9){\small {\bf 0}}
\multiput(12,2)(0,2){5}{\line(1,0){20}}
\end{picture} }
\newcommand{\vertexkq}{
\begin{picture}(39,22)
\put(12,4){\circle*{2}}
\multiput(28,-7)(0,22){2}{\circle*{2}}
\put(4.8,-2){\small {\bf k}}
\put(29.5,15){\small {\bf q}}
\linethickness{1pt}
\bezier{7}(12,4)(17.2,13.57)(28,15)
\bezier{7}(12,4)(17.2,-5.57)(28,-7)
\bezier{7}(28,-7)(22.4,4)(28,15)
\bezier{7}(28,-7)(33.6,4)(28,15)
\linethickness{0.25pt}
\bezier{100}(12,-2)(23,4)(34,10)
\bezier{100}(22,-1.5454)(28,1.7273)(34,5)
\end{picture}  }
\newcommand{\vertexzero}{
\begin{picture}(39,22)
\put(12,4){\circle*{2}}
\multiput(28,-7)(0,22){2}{\circle*{2}}
\put(4.8,-2){\small {\bf 0}}
\put(29.5,15){\small {\bf 0}}
\linethickness{1pt}
\bezier{7}(12,4)(17.2,13.57)(28,15)
\bezier{7}(12,4)(17.2,-5.57)(28,-7)
\bezier{7}(28,-7)(22.4,4)(28,15)
\bezier{7}(28,-7)(33.6,4)(28,15)
\end{picture}  }
\newcommand{\vertexzerosl}{
\begin{picture}(39,22)
\put(12,4){\circle*{2}}
\multiput(28,-7)(0,22){2}{\circle*{2}}
\put(4.8,-2){\small {\bf 0}}
\put(29.5,15){\small {\bf 0}}
\linethickness{1pt}
\bezier{7}(12,4)(17.2,13.57)(28,15)
\bezier{7}(12,4)(17.2,-5.57)(28,-7)
\bezier{7}(28,-7)(22.4,4)(28,15)
\bezier{7}(28,-7)(33.6,4)(28,15)
\linethickness{0.25pt}
\bezier{100}(22,0.75)(28,4)(34,7.25)
\end{picture}  }
\newcommand{\vertexone}{
\begin{picture}(39,22)
\put(12,4){\circle*{2}}
\multiput(28,-7)(0,22){2}{\circle*{2}}
\put(4.8,-2){\small {\bf 0}}
\put(29.5,15){\small {\bf 0}}
\linethickness{0.25pt}
\bezier{100}(12,4)(17.2,13.57)(28,15)
\bezier{100}(12,4)(17.2,-5.57)(28,-7)
\bezier{100}(28,-7)(22.4,4)(28,15)
\bezier{100}(28,-7)(33.6,4)(28,15)
\end{picture}  }
\newcommand{\vertextwo}{
\begin{picture}(39,22)
\put(12,4){\circle*{2}}
\multiput(28,-7)(0,22){2}{\circle*{2}}
\put(4.8,-2){\small {\bf 0}}
\put(29.5,15){\small {\bf 0}}
\linethickness{1pt}
\bezier{7}(28,-7)(22.4,4)(28,15)
\bezier{7}(28,-7)(33.6,4)(28,15)
\linethickness{0.25pt}
\bezier{100}(12,4)(17.2,13.57)(28,15)
\bezier{100}(12,4)(17.2,-5.57)(28,-7)
\bezier{100}(22,0.75)(28,4)(34,7.25)
\end{picture}  }
\newcommand{\vertexthree}{
\begin{picture}(39,22)
\put(12,4){\circle*{2}}
\multiput(28,-7)(0,22){2}{\circle*{2}}
\put(4.8,-2){\small {\bf 0}}
\put(29.5,15){\small {\bf 0}}
\linethickness{1pt}
\bezier{7}(28,-7)(22.4,4)(28,15)
\linethickness{0.25pt}
\bezier{100}(12,4)(17.2,13.57)(28,15)
\bezier{100}(12,4)(17.2,-5.57)(28,-7)
\bezier{100}(28,-7)(33.6,4)(28,15)
\end{picture}  }
\newcommand{\vertexfour}{
\begin{picture}(39,22)
\put(12,4){\circle*{2}}
\multiput(28,-7)(0,22){2}{\circle*{2}}
\put(4.8,-2){\small {\bf 0}}
\put(29.5,15){\small {\bf 0}}
\linethickness{1pt}
\bezier{7}(12,4)(17.2,13.57)(28,15)
\bezier{7}(12,4)(17.2,-5.57)(28,-7)
\bezier{7}(28,-7)(22.4,4)(28,15)
\linethickness{0.25pt}
\bezier{100}(28,-7)(33.6,4)(28,15)
\end{picture}  }
\newcommand{\vertexfive}{
\begin{picture}(39,22)
\put(12,4){\circle*{2}}
\multiput(28,-7)(0,22){2}{\circle*{2}}
\put(4.8,-2){\small {\bf 0}}
\put(29.5,15){\small {\bf 0}}
\linethickness{1pt}
\bezier{7}(12,4)(17.2,13.57)(28,15)
\bezier{7}(12,4)(17.2,-5.57)(28,-7)
\linethickness{0.25pt}
\bezier{100}(28,-7)(22.4,4)(28,15)
\bezier{100}(28,-7)(33.6,4)(28,15)
\end{picture}  }

\newcommand{\vertexuu}{
\begin{picture}(46,22)
\put(15,4){\circle*{2}}
\multiput(31,-7)(0,22){2}{\circle*{2}}
\put(15,4){\line(-5,3){10}}
\put(15,4){\line(-5,-3){10}}
\put(31,-7){\line(1,0){10}}
\put(31,15){\line(1,0){10}}
\linethickness{1pt}
\bezier{7}(15,4)(20.2,13.57)(31,15)
\bezier{7}(15,4)(20.2,-5.57)(31,-7)
\bezier{7}(31,-7)(25.4,4)(31,15)
\bezier{7}(31,-7)(36.6,4)(31,15)
\linethickness{0.25pt}
\bezier{100}(15,-2)(26,4)(37,10)
\bezier{100}(25,-1.5454)(31,1.7273)(37,5)
\end{picture}  }

\newcommand{\square}{
\begin{picture}(7,7)
\multiput(0,0)(7,0){2}{\line(0,1){7}}
\multiput(0,0)(0,7){2}{\line(1,0){7}}
\end{picture} }
\begin{document}

\title{ 
\textbf{Perturbative renormalization of the Ginzburg--Landau  model 
revisited}
}

\author{J. Kaupu\v{z}s
\thanks{E--mail: \texttt{kaupuzs@latnet.lv}} \\
Institute of Mathematics and Computer Science, University of Latvia\\
29 Rai\c{n}a Boulevard, LV--1459 Riga, Latvia}

\date{\today}

\maketitle

\begin{abstract}
The perturbative renormalization of the Ginzburg--Landau model is 
reconsidered based on the Feynman diagram technique.
We derive renormalization group (RG)
flow equations, exactly calculating all vertices appearing
in the perturbative renormalization of the $\varphi^4$ model up to the 
$\varepsilon^3$ order of the $\varepsilon$-expansion.
In this case, the 
$\varphi^2$, $\varphi^4$, $\varphi^6$, and $\varphi^8$ vertices appear. All these vertices are relevant.
We have tested the expected basic properties of the  RG flow, such as the semigroup property.
Under repeated RG transformation $R_s$,
appropriately represented RG flow on the critical surface converges to certain $s$--independent fixed point.
The Fourier--transformed two--point correlation function $G({\bf k})$ has been considered.
Although the $\varepsilon$-expansion of $X({\bf k})=1/G({\bf k})$ is well defined on 
the critical surface, we have revealed an inconsistency of the perturbative method
with the exact rescaling  of $X({\bf k})$, represented as an expansion in powers
of $k$ at $k \to 0$. We have discussed also some aspects of the perturbative
renormalization of the two--point correlation function slightly above the critical point.
Apart from the $\varepsilon$--expansion, we have tested and briefly discussed also a modified approach, 
where the $\varphi^4$ coupling constant $u$ is the expansion parameter at a fixed spatial 
dimensionality $d$.
\end{abstract}


\section{Introduction}
\label{intro}

The perturbative renormalization of the Ginzburg--Landau (or $\varphi^4$) 
model has a  long history (see~\cite{WF72,Ma,Justin,Kleinert}
and references therein). However, as mentioned in~\cite{Ma}, a complete
formulation of the renormalization group (RG) beyond the 
$\varepsilon^2$ order of the $\varepsilon$-expansion
(where $\varepsilon=4-d>0$, $d$ being the spatial dimensionality) met mathematical
difficulties, which could not be overcome. In fact, the $\varepsilon$-expansion
of the critical exponents beyond the lowest orders is based on an alternative
approach, which relies on the Callan-Symanzik equation~\cite{Cal,Sym1,Sym2}.
The latter one represents a scaling property of the $\varphi^4$ model~\cite{Cal}, 
and the method is based on a set of assumptions~\cite{Sym2}.
Since the perturbative RG theory is not rigorous, 
stringent tests of its validity and consistency make sense.
Our aim is to perform such tests.

We have tested the expected basic properties of the RG flow. 
Following the idea in~\cite{Delamotte}, we have checked the 
semigroup property $R_{s_1 s_2} \mu = R_{s_2} R_{s_1} \mu$, where $R_s$ is the 
RG operator with scale factor $s>1$ acting on the set of Hamiltonian parameters $\mu$.
We have tested the expected $s$--independence of the fixed point $\mu^*$,
as well as the scaling of the Fourier--transformed two--point correlation function
$G({\bf k})$ on the critical surface.

Another strategy of verification has been 
used in~\cite{BDH98}, considering a four--dimensional ($d=4$) model with Gaussian 
measure modified in such a way to simulate $d=4-\varepsilon$ dimensions 
with $\varepsilon$ small and positive. This method allows to control 
rigorously the remainder of the perturbation series.
The obtained results~\cite{BDH98} confirm  
the existence of the non--Gaussian fixed point at a distance 
$\mathcal{O}(\varepsilon)$ away from the Gaussian one 
as proposed earlier~\cite{WF72} by the $\varepsilon$--expansion. However, the expected 
decay of the two--point correlation function appears to 
be canonical (i.~e., Gaussian, see Introduction part 
in~\cite{BDH98}) in disagreement with that provided by the $\varepsilon$--expansion 
and observed in ordinary spin systems of dimensionality $d<4$ like, e.~g., three--dimensional and 
two--dimensional Ising models. In view of these observations, the classical 
(used in the $\varepsilon$--expansion) way of introduction of non-integer spatial 
dimensionality $d$ via an analytic continuation from $d$--dimensional hypercubes
appears to be more meaningful than that in~\cite{BDH98}.
In fact, there is no unique definition of the non-integer $d$. It can be 
introduced in a more physical way as a suitable ``fractal" dimension of an 
irregular lattice~\cite{TK,Ka}.

\section{Diagrammatic formulation of the renormalization}
\label{sec:diag}

As a starting point, we consider a 
$\varphi^4$ model with the Hamiltonian $H$ defined by
\begin{equation} \label{eq:Hx}
H/T= \int \frac{r}{2} \, \varphi^2({\bf x}) + \frac{c}{2} (\nabla \varphi({\bf x}))^2 
+ \frac{1}{8} \int\int \varphi^2 ({\bf x}_1) u({\bf x}_1-{\bf x}_2)
\varphi^2 ({\bf x}_2)  d{\bf x}_1 d{\bf x}_2 \;,
\end{equation}
where the order parameter $\varphi({\bf x})$ is an
$n$--component vector with components $\varphi_i({\bf x})$, depending on the
coordinate ${\bf x}$, and $T$ is the temperature.
The field $\varphi_j({\bf x})$ is given in Fourier representation by 
$\varphi_j({\bf x}) = V^{-1/2} \sum_{k<\Lambda}
\varphi_{j,{\bf k}} \, e^{i{\bf kx}}$, where $V=L^d$ is the volume
of the system, $d$ is the spatial dimensionality, and $\Lambda$ is the 
upper cut-off of the wave vectors. The Fourier-transformed Hamiltonian 
obeys the equation
\begin{eqnarray} \label{eq:H}
-\frac{H}{T} &=& -\frac{1}{2} \sum_{i,{\bf k}}  \left( r  +
c {\bf k}^2 \right) \mid \varphi_{i,{\bf k}} \mid^2  \\ 
&-& \frac{1}{8} V^{-1} \sum\limits_{i,j,{\bf k}_1,{\bf k}_2, {\bf k}_3}
\varphi_{i,{\bf k}_1} \varphi_{i,{\bf k}_2} u_{{\bf k}_1+{\bf k}_2}
\varphi_{j,{\bf k}_3} \varphi_{j,-{\bf k}_1 -{\bf k}_2 -{\bf k}_3} 
\nonumber \;.
\end{eqnarray}
In the Feynman diagram technique~\cite{Ma,K1}, the second term in~(\ref{eq:H}) is 
represented by a fourth order vertex \dshv (where field components 
with vanishing total wave vector are 
related to the solid lines and the remaining factor --- to the dashed 
line). In the following we will consider only a particular case 
$u({\bf x})=u \delta({\bf x})$ or $u_{\bf k}=u$. The fourth order vertex 
then can be depicted as \vertice by shrinking the dashed line to a 
point (node). We will mostly use this simplified 
notation. One has to remember, however, that two of the vertex lines 
are related to $i$--th component ($i=1, \ldots ,n$), and the other two lines --- to $j$--th 
component of a field vector (including the possibility 
$i=j$). It is shown explicitly in the representation 
\mbox{\dshv,} where these pairs of lines are separated.

 In the exact Wilson's RG equation the scale transformation, i.~e., the 
Kadanoff's transformation, is  performed by integrating over the Fourier
modes with wave vectors  obeying $\Lambda/s < k < \Lambda$. It is the first 
step of the full RG transformation. At this step the 
transformed Hamiltonian $H'$ is found from the equation
\begin{equation}
e^{-(H'/T)+AL^d} = \int e^{-H/T} \prod\limits_{i,\,\Lambda/s < k < \Lambda}
 d \varphi_{i,{\bf k}} \;,
\label{eq:WERGE}
\end{equation}
where $A$ is a constant. Note that $\varphi_{i,{\bf k}}$ is a complex 
number and $\varphi_{i,-{\bf k}}=\varphi^*_{i,{\bf k}}$ holds (since 
$\varphi_i({\bf x})$ is always real), so that 
the integration over $\varphi_{i,{\bf k}}$ in~(\ref{eq:WERGE}) means in 
fact the integration over real and imaginary parts of $\varphi_{i,{\bf k}}$ 
for each pair of conjugated wave vectors ${\bf k}$ and $-{\bf k}$. 
In practice Eq.~(\ref{eq:WERGE}) cannot be solved exactly. It is done 
perturbatively, as described in~\cite{Ma}. The 
perturbation terms can be found more easily by means of 
the Feynman diagrams.

In the perturbative approach, Hamiltonian is split in two 
parts $H = H_0 + H_1$, where $H_0$ is the Gaussian part and $H_1$ is 
the rest part considered as a small perturbation. The first 
and the second term on the right hand side of~(\ref{eq:H}) can be 
identified with $-H_0/T$ and $-H_1/T$, respectively. Since the term with 
$r$ also is considered as a small perturbation ($r \sim 
\varepsilon$ holds in the $\varepsilon$--expansion in $d=4-\varepsilon$ 
dimensions), it is   suitable to include it in $H_1$. In this case 
diagram expansions
are represented by vertices \dshv and  \mbox{\Gaussv,} where the latter
second--order vertex corresponds to the  term with $r$.  

It is convenient to normalize   Eq.~(\ref{eq:WERGE}) 
by $Z_s= \int \exp(-\widetilde H_0/T) \prod_{i,\,\Lambda/s < k < \Lambda} 
d \varphi_{i,{\bf k}}$, where $\widetilde H_0$ is the part of $H_0$
including only the terms with  $\Lambda/s < k < \Lambda$.
We have $\ln Z_s=A_s \, L^d$ at $L \to \infty$, where 
$A_s$ is independent of $L$, and the normalization yields
\begin{equation}
-(H'/T) + (A-A_s)L^d = -(H'_0/T) + \ln \langle \exp(-H_1/T) \rangle_0 \;, 
\label{eq:WERGE1}
\end{equation}
where $\langle \cdot \rangle_0$ means the Gaussian average over the 
field components with $\Lambda/s < k < \Lambda$, whereas $H'_0$ is the 
part of $H_0$ including only the components with $k< \Lambda/s$. 
Like free energy, $\ln \langle \exp(-H_1/T) \rangle_0$  
is represented perturbatively by the
sum over all connected Feynman diagrams made  of the vertices of $-H_1/T$
by coupling those solid lines, which are associated with wave vectors
obeying $\Lambda/s < k < \Lambda$,   according to the Wick's theorem.
A diagram can contain no coupled lines. There is also a set of diagrams 
with all lines coupled. The latter diagrams give a constant (independent of 
$\varphi_{i,{\bf k}}$) contribution which compensates the term $(A-A_s)L^d$ 
in~(\ref{eq:WERGE1}). The other contributions correspond to $-H'/T$.

Each term of $\ln \langle \exp(-H_1/T) \rangle_0$ comes from  a
diagram (or diagrams)  of certain topology and is given by a sum over wave vectors
which  fulfil certain constraints such that $k< \Lambda/s$ holds for 
uncoupled (external or outer) solid lines, associated with field components 
$\varphi_{i,{\bf k}}$, and $\Lambda/s < k < \Lambda$ holds  for
coupled solid lines. Besides, the sum of all wave vectors  coming into
any of the nodes is zero, and the same index $i$  is related
to the solid lines attached to one node of any vertex \mbox{\dshv}
or \mbox{\Gaussv.}
The Gaussian average
$G_0({\bf k}) = \langle \varphi_{i,{\bf k}} \varphi_{i,-{\bf k}} 
\rangle_0 = \langle \mid \varphi_{i,{\bf k}} \mid^2 \rangle_0$ 
is related to  a coupling line with wave vector ${\bf k}$. Here $G_0({\bf k})$ is the 
Fourier transform of the two--point correlation function in the Gaussian 
approximation. If only the term with ${\bf k}^2$ in~(\ref{eq:H}) is 
included in $H_0/T$, then we have $G_0({\bf k})=1/(ck^2)$. Including also the 
term with $r$, we have $G_0({\bf k}) = 1/(r+ck^2)$, but in this case 
we need finally to expand $G_0({\bf k})$ in terms of $r$. The second 
method yields the same results as the first one: such an expansion
generates the same terms, which are obtained in the first 
method by extending coupling lines in the diagrams originally 
constructed without vertices \Gaussv to include
all possible linear chains made of these vertices.
We will use the first, i.~e., more  diagrammatic  method.
Note that, when the  renormalization procedure
is repeated, a continuum of additional vertices appear in the expansion
of $-H_1/T$, including ones with explicit  wave--vector dependent factors
related to the solid lines, which then  also have to be taken into
account.

In fact, the first step of RG transformation implies the summation  over
wave vectors of the coupled lines in the diagrams, resulting in a
Hamiltonian  which depends on $\varphi_{i,{\bf k}}$ with $k<\Lambda/s$. 
The RG transformation has the second step~\cite{Ma}: changing of 
variables $\tilde {\bf k} = s{\bf k}$ and rescaling the field components 
$\varphi_{i,{\bf k}} \to s^{1-\eta/2} \varphi_{i, \tilde {\bf k}}$,
where $\eta$ is the critical exponent describing the $\sim k^{-2+\eta}$ 
singularity of the Fourier--transformed critical two--point correlation
function at $k \to 0$. The upper cut-off for the new wave vectors $\tilde
{\bf k}$ is the original one $\Lambda$, whereas the density of points in
the  $\tilde {\bf k}$--space corresponds to $s^d$  times decreased
volume. Therefore we make a  substitution $\tilde V = s^{-d} \, V$. In
the thermodynamic limit we can replace $\tilde V$ by $V$ consistently 
increasing the density  of points in the wave vector space. 
Finally, we set $\tilde {\bf k} \to {\bf k}$ and obtain a Hamiltonian in 
original notations.

\section{Renormalization up to the order of $\varepsilon^2$}
\label{sec:eps2}

\subsection{RG flow equations}
\label{subsec:RGf}

Here we consider RG flow equations including all terms
up to the order $\mathcal{O} \left( \varepsilon^2 \right)$, starting
with the renormalization of the coupling constant $u$. 
In  fact, this is the lowest
order of the theory, since it allows to find the  fixed point value
$u^*$ up to the order of $\varepsilon$. Our aim is to show that we can easily
recover the known results  using the diagrammatic approach described in
Sec.~\ref{sec:diag}. It allows also to write down unambiguously (i.~e., 
without intermediate approximations) the 
formulae for all perturbation terms.

Important statements here are that the
renormalized values of $r$ and  $u$ are quantities of order $\mathcal{O}(\varepsilon)$. 
Besides, $\eta= \mathcal{O} \left( \varepsilon^2 \right)$ holds within the
$\varepsilon$--expansion and, for any finite renormalization scale $s$,
the variation of $c$ is of order $\mathcal{O} \left( \varepsilon^2 \right)$~\cite{Ma}. 
Hence, performing the RG transformation $R_s$ with a finite $s$, the only diagrams of 
$\ln \langle \exp(-H_1/T) \rangle_0$ contributing to the renormalized 
coupling constant up to the order of $\varepsilon^2$ are \vertice  
and \vvertice (this statement can be verified in detail based
on a complete renormalization discussed further on). 
The first one provides the original $\varphi^4$ term in 
Eq.~(\ref{eq:H}), which is merely renormalized by factor 
$s^{\varepsilon-2\eta}$ at the second step of the RG transformation.  
The second diagram, which is constructed of two vertices 
\mbox{\vertice,} yields
\begin{eqnarray}
\vvertice &\to& s^{\varepsilon-2\eta} \left( \frac{u}{8c} \right)^2
V^{-1} \sum\limits_{i,j,{\bf k}_1,{\bf k}_2, {\bf k}_3}
\varphi_{i,{\bf k}_1} \varphi_{i,{\bf k}_2} 
\varphi_{j,{\bf k}_3} \varphi_{j,-{\bf k}_1 -{\bf k}_2 -{\bf k}_3} 
\times \nonumber \\
&&\hspace{3ex} \times [ (4n+16) \, Q( ({\bf k}_1+{\bf k}_2)/s,s) +16 \, Q( ({\bf k}_1+{\bf 
k}_3)/s,s)] \;. 
\label{eq:uu}
\end{eqnarray}
Here $Q$ is given by
\begin{equation}
Q({\bf k},s) = \frac{1}{(2 \pi)^d} \int\limits_{\Lambda/s < q < \Lambda}
q^{-2} \mid {\bf k}-{\bf q} \mid^{-2} \mathcal{F} ( \mid {\bf k}-{\bf q} 
\mid,s )  d^d q \;,
\label{eq:Q}
\end{equation}
where $\mathcal{F}(k,s)=1$ if $\Lambda/s < k < \Lambda$, and 
$\mathcal{F}(k,s)=0$ otherwise.
To obtain this result, we have deciphered the \vvertice diagram as 
a sum of three diagrams of different topologies made of vertices 
\mbox{\dshv,} i.~e., \mbox{\dshvv, \dshvvv, and \dshvvvv,}
providing the same topological picture \vvertice when shrinking the dashed lines 
to points. Note that any
loop made of solid lines of \dshv gives a factor $n$,
and one needs also to compute the combinatorial factors (see, e.~g., 
\cite{Ma,K1}). For the above three diagrams, the resulting factors are $4n$, $16$, and 
$16$, which enter the pre-factors of $Q$ in~(\ref{eq:uu}).

Quantity $Q({\bf k}/s,s)$ has a constant 
contribution 
\begin{equation}
Q({\bf 0},s) = K_d \, \Lambda^{-\varepsilon} 
(s^{\varepsilon}-1)/\varepsilon \;,
\label{eq:Q0}
\end{equation}
as well as a ${\bf k}$--dependent correction 
$\Delta(k,s)=Q({\bf k}/s,s)-Q({\bf 0},s)$ vanishing at $k=0$.
Here $K_d=S(d)/(2 \pi)^d$, where $S(d)=2 \pi^{d/2}/\Gamma(d/2)$ is 
the area of unit sphere in $d$ dimensions.
 
The term provided by the $Q({\bf 0},s)$ part of 
$Q({\bf k}/s,s)$ in~(\ref{eq:uu}) is identified with 
the $\varphi^4$ vertex, contributing to the renormalized coupling constant 
$u'$. It is consistent with the general form
\begin{equation}
V^{-1} \sum\limits_{i,j,{\bf k}_1,{\bf k}_2, {\bf k}_3}
\bar Q \left( {\bf k}_1,{\bf k}_2,{\bf k}_3,s,n,\varepsilon \right)
\varphi_{i,{\bf k}_1} \varphi_{i,{\bf k}_2} 
\varphi_{j,{\bf k}_3} \varphi_{j,-{\bf k}_1 -{\bf k}_2 -{\bf k}_3} \;, 
\label{eq:wf}
\end{equation}
of quartic ($\varphi^4$) terms, where
the contribution corresponding to the ordinary $\varphi^4$ vertex
is uniquely identified with one provided by the constant part
$\bar Q \left( {\bf 0,0,0},s,n,\varepsilon \right)$ of 
the weight function $\bar Q$. 

Taking into account also the  
contribution coming directly from \vertice vertex, the RG flow equation 
reads
\begin{equation}
u' = s^{\varepsilon-2\eta} \left[ u - u^2 \, \frac{K_d (n+8)}{2c^2 
\Lambda^{\varepsilon}}
\times \frac{s^{\varepsilon}-1}{\varepsilon} \right] 
+ \mathcal{O} \left( \varepsilon^3 \right) \;,
\label{eq:uflow0}
\end{equation}
where $u'$ is the renormalized and $u$ is the original coupling constant. 
The expansion in $\varepsilon$ yields the well known equation~\cite{Ma}
\begin{equation}
u' =  u + \varepsilon u \, \ln s  - u^2 \, B \, \ln s   
+ \mathcal{O} \left( \varepsilon^3 \right) \;,
\label{eq:uflow}
\end{equation}
where $B=K_4 (n+8)/(2c^2)$,
with the known fixed--point value at 
$u=u^*=B^{-1} \varepsilon + \mathcal{O} \left(\varepsilon^2 \right)$. 
 
 The correction term $\Delta(k,s)$ changes the renormalized Hamiltonian 
as follows:
\begin{eqnarray}
H/T &\to& (H/T) - s^{\varepsilon-2\eta} \left( \frac{u}{2c} \right)^2
V^{-1} \sum\limits_{i,j,{\bf k}_1,{\bf k}_2, {\bf k}_3}
\varphi_{i,{\bf k}_1} \varphi_{i,{\bf k}_2} 
\varphi_{j,{\bf k}_3} \varphi_{j,-{\bf k}_1 -{\bf k}_2 -{\bf k}_3} 
\times \nonumber \\
&&\hspace{16ex} \times [ ( 1 +n/4) \, \Delta( \mid {\bf k}_1+{\bf k}_2 
\mid ,s) +  \Delta( \mid {\bf k}_1+{\bf k}_3  \mid ,s) ] \;. 
\label{eq:correction}
\end{eqnarray}
It can be represented as a sum of two fourth--order vertices including 
pre-factors $\Delta( \mid {\bf k}_1+{\bf k}_2 \mid ,s)$ and
$\Delta( \mid {\bf k}_1+{\bf k}_3 \mid ,s)$, respectively, related to 
two of the vertex lines. 

In the following we will introduce a more suitable diagrammatic notation. 
In the one-component case $n=1$, the term appearing in~(\ref{eq:uu}) is represented as
\begin{equation}
s^{2 \varepsilon - 2 \eta} \, \frac{9}{16} u^2 \vverticex
\label{eq:ss}
\end{equation}
where \vverticex  is the Feynman diagram in which the Gaussian propagator $1/(ck^2)$ with 
$k \in [\Lambda,s \Lambda]$ is related to the dotted
coupling lines. In other words, the propagator is multiplied with the cut function
$\mathcal{F}(k/s,s)$. In general (also for other diagrams considered in our paper), 
the field components supplied with the factor
$V^{1-m}$ are related to $2m$ external lines, the rest of such factors being absorbed
in the ${\bf k}$--space integrals. An additional $s^{\varepsilon}$ factor in~(\ref{eq:ss}) comes
from the rescaling of wave vectors in such a way that the integration now takes place over
$k \in [\Lambda, \Lambda s]$ for the internal lines. This procedure is necessary in order to
represent the result as a Feynman diagram with vanishing sum of the wave vectors entering each node.

Subtracting from the ${\bf k}$--space integral its value calculated at
${\bf k=0}$ for the external lines, the result is represented as
\begin{equation}
s^{2 \varepsilon - 2 \eta} \, \frac{9}{16} u^2  \vverticey \;.
\end{equation}
Similar notation is applied throughout the paper, where the line crossing the diagram
indicates the subtraction of zero--${\bf k}$ contribution. 
The diagram representation of this kind has been proposed in~\cite{Hara_private}.
In the general $n$--component case, we denote by
\begin{equation}
\Sigma \left( \vverticey \right) = \frac{n}{9} \dshvvx + \frac{4}{9} \dshvvvx + \frac{4}{9} \dshvvvvx
\label{eq:nsig}
\end{equation} 
the sum of all such kind of diagrams, which reduce to \vverticey
when the dashed lines shrink to points. In~(\ref{eq:nsig})
the coefficients are normalized in such a way that their sum is $1$ at $n=1$.
Note that in the diagram technique with such vertices, one has
to count closed loops made of all kind of lines related to 
the propagator $1/(ck^2)$.
In this notation~(\ref{eq:correction}) becomes
\begin{equation}
H/T \to (H/T) - s^{2\varepsilon-2\eta} u^2 \, \frac{9}{16} \Sigma \left( \vverticey \right)  
\end{equation}
Analogous diagrammatic representation is possible for all vertices appearing in 
a repeated renormalization. 
In accordance with such notation, the ordinary $\varphi^4$ vertex
in~(\ref{eq:H}) will be represented as  $-(u/8) \dshv$.

For a complete renormalization up to the order of $\varepsilon^2$,
one has to take into account also other contributions of this order,
including the $\varphi^6$ vertex 
\begin{eqnarray}
\Sigma \left( \sixx \right) &\equiv& \sixy \nonumber \\
&=& c^{-1} V^{-2} \sum\limits_{i,j,l,{\bf k}_1,{\bf k}_2,{\bf k}_3,{\bf k}_4,{\bf k}_5}
\varphi_{i,{\bf k}_1} \varphi_{i,{\bf k}_2} 
\varphi_{j,{\bf k}_3} \varphi_{j,{\bf k}_4} \varphi_{l,{\bf k}_5}
\varphi_{l,-{\bf k}_1 -{\bf k}_2 -{\bf k}_3-{\bf k}_4-{\bf k}_5} 
\\
&& \hspace*{20ex} \times 
\mid {\bf k}_1+{\bf k}_2+{\bf k}_3 \mid^{-2}
\mathcal{F}(\mid {\bf k}_1+{\bf k}_2+{\bf k}_3 \mid/s,s)
\nonumber
\end{eqnarray}
  appearing due to the 
coupling of two vertices \mbox{\vertice,} as well as the quadratic term
of the general form 
$(1/2)\sum\limits_{i,{\bf k}} \tilde \theta({\bf k}) \mid \varphi_{i,{\bf k}} \mid^2$
provided by diagrams with two external lines. In the representation
$\tilde \theta({\bf k}) = r + c {\bf k}^2 + \theta({\bf k})$, it corresponds to 
the $\varphi^2$ vertex in~(\ref{eq:H}) with one additional term.   

In the rest part of this section we will consider
the structure of the renormalized Hamiltonian and the related RG flow equations, providing the 
proofs afterwards in Sec.~\ref{sec:proofRG2}.
 Thus, the renormalized Hamiltonian has the form
\begin{eqnarray}
\frac{H}{T} &=& \frac{1}{2} \sum\limits_{i,{\bf k}} 
\left( r + c {\bf k}^2 + \theta({\bf k}) \right) \mid \varphi_{i,{\bf k}} \mid^2
+ \frac{u}{8} \dshv \nonumber \\
&+& a_4 \; \Sigma \left( \vverticey \right) + a_6 \; \Sigma \left( \sixx \right)
+ \mathcal{O} \left( \varepsilon^3 \right) \;.
\label{eq:RHam}
\end{eqnarray}
Here the wave vectors of the dotted lines are in the interval $[\Lambda,\zeta \Lambda]$,
where the parameter $\zeta$ has the initial value $1$ and is transformed according to
\begin{equation}
\zeta' = s \, \zeta 
\label{eq:zeta}
\end{equation}
in a process of repeated renormalization. Here $\zeta$ is the previous and $\zeta'$ is the 
new (renormalized) value of this parameter after the current RG transformation (RGT) with the scale factor $s$.
The weight coefficients $a_4$ and $a_6$ in this order of the $\varepsilon$--expansion
obey simple RG flow equations
\begin{eqnarray}
a_4' &=& - \frac{9 u^2}{16} + \mathcal{O} \left(\varepsilon^3 \right) \label{eq:fl4} \\
a_6' &=& - \frac{u^2}{8} + \mathcal{O} \left(\varepsilon^3 \right)
\label{eq:fl6}
\end{eqnarray}
relating the new values of these parameters in each RGT 
to the previous value of $u$. The initial values of $a_4$ and $a_6$ are
not important, since the corresponding diagrams vanish at $\zeta=1$.
The RG flow equations for the Hamiltonian parameters $u$ and $r$ read
\begin{eqnarray}
u' &=& s^{\varepsilon} \left[ u - \frac{n+8}{2} u^2 \vverticeu  \right]  
+ \mathcal{O} \left( \varepsilon^3 \right) \label{eq:ud} \\
r' &=& s^2 \left[ \left( r + \frac{n+2}{2} u \vverticer \right) \left( 1 - \frac{n+2}{2} \, u  \vverticeu   \right)
\right. \nonumber \\
&&\left. \hspace*{-2ex}- \frac{n+2}{2} \left( u^2 \vverticerr - \frac{16}{3} a_4 \, \vverticero
- 24 a_6 \, \vverticery \right) \right] + \mathcal{O} \left(\varepsilon^3 \right) \,,
\label{eq:rd}
\end{eqnarray}
where \vverticeu is the diagram with amputated four external lines having zero wave vectors,
and the other diagrams are defined analogously. The wave vectors
of the internal solid lines are in the range $[\Lambda/s,\Lambda]$, whereas those of the dotted lines
--- in the range $[\Lambda,\Lambda \zeta]$.
Eq.~(\ref{eq:ud}) is the same as~(\ref{eq:uflow0}), only the irrelevant for this
order of the $\varepsilon$--expansion factor $s^{-2 \eta}$ is omitted, like also
the factor $s^{-\eta}$ in~(\ref{eq:rd}). Taking into account~(\ref{eq:fl4})--(\ref{eq:fl6})
and the fact that $u$ is renormalized only by  $\mathcal{O} \left(\varepsilon^2 \right)$ in one RGT, 
the diagram expression in the last line of~(\ref{eq:rd}) can be written as 
$- \frac{1}{2} (n+2) u^2 \left( \vverticerr + 3 \vverticero + 3 \, \vverticery \right)$.

The parameter $c$ enters all diagrams.
In the actual order of the $\varepsilon$--expansion its variation does not produce additional
terms in~(\ref{eq:ud}) and~(\ref{eq:rd}). However, one has to update its value according to
the equation
\begin{equation}
c' = s^{-\eta} \left( c - \frac{n+2}{2} \, \Pi \right) + \mathcal{O} \left(\varepsilon^3 \right) \;,
\label{eq:c}
\end{equation}
where 
\begin{equation}
\Pi = \lim\limits_{k \to 0} \frac{D(k)-D(0)}{k^2} 
\label{Pi}
\end{equation}
with
\begin{equation}
D(k) =  u^2 \left( \vverticerrk + 3 \vverticerxk  + 3 \vverticeryk \right)  \;.
\label{Pi1}
\end{equation}
It is true if the limit~(\ref{Pi}) exists.
Here the symbol ${\bf k}$ indicates the wave vector related to the amputated external line.
Finally, the RG flow equation for the function $\theta({\bf k})$ reads
\begin{equation}
\theta'({\bf k}) = s^2 \left( \theta({\bf k}/s) - \frac{n+2}{2} \Phi({\bf k}/s) \right) 
+ \mathcal{O} \left(\varepsilon^3 \right) \;,
\label{eq:thet}
\end{equation} 
where
\begin{equation}
\Phi({\bf k}) =  D(k)-D(0)-k^2 \Pi  \;.
\end{equation}
Using the diagram technique, the function $\theta({\bf k})$ can be represented as
\begin{equation}
 \theta({\bf k}) =  - \frac{n+2}{2} u^2 \vverticetet + \mathcal{O} \left(\varepsilon^3 \right) \;,
\label{eq:thetd}
\end{equation}
where the wavy line indicates that both the constant contribution and that proportional
to $k^2$ (with the proportionality coefficient determined at $k \to 0$) are subtracted.
Besides, the range of the wave vectors for the internal lines is $[\Lambda,\Lambda \zeta]$,
where $\zeta$ is determined after the actual RGT.

A question can arise about the existence of the limit~(\ref{Pi}).
In fact, it is a necessary condition for the stability
of the RG flow that a finite (or zero) limit in~(\ref{Pi}) exists. 
It can be shown directly that it exists for $s = 1+ds$ with small $ds$ and $\zeta = \infty$ 
(see Sec.~\ref{D1s}), as well as at small $ds$ and $\zeta = 1 + d\zeta$
with small $d \zeta$. Besides, the limit exists for each diagram separately in this case.
Consequently, the limit~(\ref{Pi}) exists at least
at the beginning of the renormalization process, performed in small steps.
In this paper we will test several important properties, in the case if
the limit~(\ref{Pi}) exists for any finite $s>1$. Besides, we will test the 
scenario, where this limit exists separately for the first diagram in~(\ref{Pi1}) and for
the sum of the remaining two diagrams. These, in fact, are conditions
at which a set of expected properties hold, as shown further on.

\subsection{Proof of the RG flow equations}
\label{sec:proofRG2}

In Sec.~\ref{subsec:RGf} the RG flow equations are given without proof.
Their proof consists of a verification that, at each RGT, the Hamiltonian 
keeps the form~(\ref{eq:RHam}) in accordance with the given update rules for all parameters. 
Consider first the $\varphi^6$ vertex at $n=1$, summing up all diagrams of this
topology, which appear in $H/T$ at a given RGT:
\begin{eqnarray}
&&s^{2 \varepsilon - 3 \eta} \left( -\frac{u^2}{8} \six + a_6 \sixz \right)
= -\frac{u^2}{8} \Big( \six  +  \sixz \Big) + \mathcal{O} \left(\varepsilon^3 \right) 
\nonumber \\
&&= -\frac{u^2}{8} \sixx  + \mathcal{O} \left(\varepsilon^3 \right)
= a'_6 \sixx  + \mathcal{O} \left(\varepsilon^3 \right) \;.
\label{eq:sumsix}
\end{eqnarray}
Here the wave vectors within $[\Lambda,\Lambda s]$, $[\Lambda s,\Lambda \zeta']$,
and $[\Lambda, \Lambda \zeta']$ correspond to the solid, dashed, and dotted coupling
lines, respectively, where $\zeta'=s \zeta$. (These dashed lines should not be confused with those 
of \mbox{\dshv.}) In~(\ref{eq:sumsix}), the vertex \six is produced
by coupling (and rescaling) two $\varphi^4$ vertices \vertice, whereas \sixz comes from the rescaling
of already existing before the given RGT $\varphi^6$ vertex. 
We have omitted the irrelevant for this order of the
$\varepsilon$--expansion rescaling factor resulting from the renormalization of parameter $c$. 
We have used the 
substitution $a_6 = - u^2/8 + \mathcal{O} \left(\varepsilon^3 \right)$, which
holds according to the assumption that our RG flow equation~(\ref{eq:fl6})
was satisfied and $u$ was changed only by $\mathcal{O} \left(\varepsilon^2 \right)$
in the previous RGT. It is true for the first RGT,
so that the relations~(\ref{eq:sumsix}) provide the proof of~(\ref{eq:fl6})
by induction. It is obviously valid for any finite number of RGT. 

The generalisation to the $n$--component case is trivial here,
since there is only one way how to couple two vertices \dshv to form the sixth-order
vertex.

Similarly, the summation of terms provided by the diagrams of 
\vvertice  topology, subtracting the contribution included in
the ordinary $\varphi^4$ vertex, gives us
\begin{eqnarray}
&&s^{2 \varepsilon - 2 \eta} \left( -\frac{9u^2}{16} \vverticez + a_4 \vverticew
+ 9 a_6 \vverticeww \right) \nonumber \\
&&= -\frac{9u^2}{16} \Big( \vverticez  +  \vverticew + 2 \vverticezw \Big) 
+ \mathcal{O} \left(\varepsilon^3 \right) 
\nonumber \\
&&= -\frac{9u^2}{16} \vverticey  + \mathcal{O} \left(\varepsilon^3 \right)
= a'_4 \vverticey  + \mathcal{O} \left(\varepsilon^3 \right) 
\label{eq:sumfour}
\end{eqnarray}
with the same $k$ intervals for different coupling lines as in~(\ref{eq:sumsix}).
It proves~(\ref{eq:fl4}) for a finite number $m$ of RG transformations at $n=1$. 
Here the three diagrams in the first line of~(\ref{eq:sumfour}) come from two \vertice vertices,
rescaling of the already existing vertex with the coefficient $a_4$, and from the $\varphi^6$
vertex via coupling its two lines.
Besides, $\vverticeww  \equiv  \vverticezw$ holds,
since the diagram \vverticeww vanishes if the external lines have zero wave vectors. 
This derivation is easily generalised to the $n$--component case represented by~(\ref{eq:RHam}):
the same relations, only multiplied with the corresponding weight factors in~(\ref{eq:nsig}),
hold for diagrams of each topology, when \vverticey  is deciphered as the
diagrams on the right hand side of~(\ref{eq:nsig}). 

The derivation of~(\ref{eq:ud}) has been already discussed in Sec.~\ref{subsec:RGf}.
Recall that the updated value of the coupling constant $u$ of the ordinary $\varphi^4$ vertex
is related to the constant part of the weight function in~(\ref{eq:wf}), which comes from
the rescaled already existing ordinary $\varphi^4$ vertex, as well as from
other diagrams with four external lines generated in the actual RGT. 
The coupling of two lines of the vertex \sixx  gives vanishing contribution
here, so that the coupling \vvertice is the only relevant one at the order of $\varepsilon^2$.  
It results in~(\ref{eq:ud}).

Similarly, the updated value of the parameter $r$ is identified with the constant part of the
weight function $\tilde \theta({\bf k})$ in the general representation of the quadratic term 
$(1/2)\sum\limits_{i,{\bf k}} \tilde \theta({\bf k}) \mid \varphi_{i,{\bf k}} \mid^2$,
and is generated by the diagrams with two external lines. It results in~(\ref{eq:rd}),
where the terms with $a_4$ and $a_6$ appear via coupling the lines of the vertices 
\vverticey and \sixx , respectively, whereas other diagrams in~(\ref{eq:rd})
are produced by the vertices \vertice and \mbox{\Gaussv.} 

 The RG flow equation~(\ref{eq:c}) for the parameter $c$ is similar to that for $r$
with the only difference that here we single out the contribution to
$\tilde \theta({\bf k})$, which is proportional to $k^2$. 
Subtracting both the constant contribution and that of $k^2$,
we arrive at the equation for $\theta({\bf k})$~(\ref{eq:thet}).
Its diagrammatic form~(\ref{eq:thetd}) is proven by summation of diagrams like, e.~g., 
in Eq.~(\ref{eq:sumfour}).

Finally, we check that all terms
up to the order of $\varepsilon^2$ are already included. 
In particular, the part of~(\ref{eq:RHam}) with
factor $\theta({\bf k})/2$ can be seen as the vertex \Gaussv supplied
with such a weight factor. However, its coupling to other vertices gives only a
contribution of order $\mathcal{O} \left(\varepsilon^3 \right)$. 
Recall that we do not consider the constant part of Hamiltonian, which is
independent of the field configuration and is represented by closed diagrams
without external lines.

\subsection{Estimation of 
${\bf k}$--space integrals}
\label{sec:estim}

It is important to know whether the ${\bf k}$--space integrals over the internal lines
of the diagrams, appearing in our representation of the  renormalized Hamiltonian, are
convergent when the RG transformation is repeated unlimitedly many times, i.~e., at $\zeta \to \infty$.

First, we note that the diagram \vverticeux in four dimensions, i.~e.,
\begin{equation}
 \vverticeux  = \frac{1}{(2 \pi)^4} \int\limits_{\Lambda <q< \Lambda \zeta} 
\frac{d^4 q}{c^2 q^4} = \frac{K_4}{c^2} \int\limits_{\Lambda}^{\Lambda \zeta} \frac{dq}{q} = 
\frac{K_4}{c^2} \ln \zeta
\end{equation}
diverges when the region $q \in [\Lambda,\Lambda \zeta]$ of wave vectors for the internal
lines is extended to $[\Lambda, \infty]$ at $\zeta \to \infty$.
Similar diagram, where the zero--${\bf k}$ contribution is subtracted, 
\begin{equation}
 J(k,\zeta) = \vverticeuyk  =  \vverticeuxk  -  \vverticeux 
\end{equation}
with ${\bf k}$ corresponding to the sum of wave vectors for the two amputated external lines,
is convergent at $d=4$, as well as in $d=4 -\varepsilon$ dimensions, where we have
\begin{eqnarray}
&& J(q,\zeta) = \frac{1}{c^2 (2 \pi)^d} \int\limits_{\Lambda <k< \Lambda \zeta} 
\left( \frac{\hat \mathcal{F}(\mid {\bf q +k}  \mid,\zeta)}{k^2 \mid {\bf q +k}  \mid^2} 
- \frac{1}{k^4} \right) d^d k \label{eq:Ik} \\
&=&  \frac{K_d}{c^2 \int\limits_{0}^{\pi} (\sin \theta )^{2-\varepsilon} d \theta}
\times \int\limits_{\Lambda}^{\Lambda \zeta} k^{1-\varepsilon} dk \int\limits_{0}^{\pi}
\left( \frac{\hat \mathcal{F} \left( \sqrt{q^2+2kq \cos \theta +k^2},\zeta \right) }
{q^2+2kq \cos \theta +k^2}  - \frac{1}{k^2} \right) (\sin \theta )^{2 -\varepsilon} d \theta
\nonumber
\end{eqnarray}
with $\hat \mathcal{F}(k,\zeta)=1$ if $\Lambda < k < \Lambda \zeta$, and 
$\hat \mathcal{F}(k,\zeta)=0$ otherwise. The last identity in~(\ref{eq:Ik}) is true for any integer
$d \ge 2$, according to the well known formula for integration in spherical coordinates, 
so that we can put here $\varepsilon$ as a continuous parameter in the usual sense
of the $\varepsilon$--expansion. 

In the following calculations of this subsection, we set $c=1$ and
$\Lambda=1$ for simplicity. Considering the large--$q$ asymptotic at $\zeta=\infty$,
we can set $\hat \mathcal{F}=1$ in~(\ref{eq:Ik}), which gives correct result
up to a term vanishing at $q \to \infty$. Further on, we use the decomposition
\begin{equation}
\frac{1}{k^2 \mid {\bf q + k} \mid^2} - \frac{1}{k^4} = f_0(k,q) + \Delta f({\bf k,q}) \;,
\label{eq:decomp}
\end{equation}
where 
\begin{equation}
f_0(k,q) = \left\{
\begin{array}{ccc}
k^{-2}q^{-2} - k^{-4} & , & k<q \\
0 & , & k>q
\end{array}
\right.
\label{eq:f0}
\end{equation}
An important property of $\Delta f({\bf k,q})$, defined by~(\ref{eq:decomp}) and~(\ref{eq:f0}), is
\begin{equation}
\int\limits_{k>1} \Delta f({\bf k,q}) d^4 k = 0 \qquad \mbox{at} \quad q \to \infty \;.
\label{eq:null}
\end{equation}
It is true because
\begin{equation}
\frac{1}{(2 \pi)^4} \int\limits_{k>1} 
\left( \frac{1}{k^2 \mid {\bf q + k} \mid^2} - \frac{1}{k^4} \right) d^4 k
= \frac{1}{(2 \pi)^4} \int\limits_{k>1} f_0(k,q) d^4 k = -\ln q + \frac{1}{2}
\end{equation}
holds at $q \to \infty$. The $-\ln q +1/2$ asymptotic is evident for the second integral.
For the first one, it is verified by integration over $k$ (via the substitution
$p=k+q \cos \theta$) and then -- over the angle $\theta$ 
in spherical coordinates. According to the decomposition~(\ref{eq:decomp}), we have
\begin{equation}
J(q,\infty) = K_d \left\{ \frac{q^{-\varepsilon}}{2-\varepsilon} - \frac{1-q^{-\varepsilon}}{\varepsilon} \right\}
+ \Delta J(q) \qquad \mbox{at} \quad q \to \infty \;,
\label{eq:Iqinf}
\end{equation}
where the term $K_d \left\{ \cdot \right\}$ is the contribution of $f_0(k,q)$, whereas $\Delta J(q)$ 
comes from $\Delta f({\bf k,q})$ and is equal to
\begin{equation}
\Delta J(q) = \frac{K_d \, \Gamma[(4-\varepsilon)/2]}{\sqrt{\pi} \, \Gamma[(3-\varepsilon)/2]}
\times \int\limits_1^{\infty} k^{-1-\varepsilon} dk \int\limits_0^{\pi}
\Delta \widetilde f(k/q,\theta) (\sin \theta)^{2-\varepsilon} d \theta \;,
\label{eq:diq}
\end{equation}
where
\begin{equation}
\Delta \widetilde f(x,\theta) = \frac{x^2}{1+2x \cos \theta + x^2} - 1 + 
\left( 1- x^2 \right) \Theta(1-x) \;.
\end{equation}
Here $\Theta(x)$ denotes the theta step function, and the known identity
\begin{equation}
\int\limits_0^{\pi} (\sin \theta)^x d \theta = \sqrt{\pi} \, \frac{\Gamma \left( \frac{x+1}{2} \right)}
{\Gamma \left( \frac{x+2}{2} \right)}
\end{equation}
has been used to obtain~(\ref{eq:diq}). According to~(\ref{eq:null}), $\Delta J(q)$ vanishes at $d=4$.
Using this property, from~(\ref{eq:diq}) we obtain 
\begin{equation}
\Delta J(q) = \frac{2}{\pi} K_d \, \mathcal{A} \, \left( -\varepsilon + \varepsilon^2 \ln q  \right)
+ \widetilde c \, \varepsilon^2 + \mathcal{O} \left( \varepsilon^3 \right) \;,
\label{eq:diqex}
\end{equation}
where
\begin{equation}
\mathcal{A} = \int\limits_0^{\infty} \frac{\ln x}{x} \, dx \int\limits_0^{\pi} 
\Delta \widetilde f(x,\theta) \sin^2 \theta \, d \theta +
\int\limits_0^{\infty} \frac{d x}{x} \int\limits_0^{\pi} 
\Delta \widetilde f(x,\theta) \sin^2 \theta \, \ln (\sin \theta ) d \theta
\end{equation}
and $\widetilde c$ is a constant at $q \to \infty$.
According to~(\ref{eq:Iqinf}) and~(\ref{eq:diqex}), we have
\begin{eqnarray}
J(q,\infty) &=& K_d \, \left[ - \ln q + \frac{1}{2} + \frac{\varepsilon}{2} \, \left( \ln^2 q - \ln q \right)
+ \left( \frac{1}{4} - \frac{2 \mathcal{A}}{\pi} \right) \left(\varepsilon - \varepsilon^2 \ln q  \right) \right. \nonumber \\
&+& \left. \frac{\varepsilon^2}{12}  \left(-2 \ln^3 q + 3 \ln^2 q \right) + \hat c \, \varepsilon^2 \right]
+ \mathcal{O} \left( \varepsilon^3 \right) \qquad \mbox{at} \quad q \to \infty \;,
\end{eqnarray}
where $\hat c$ is a constant. 

In a more general case, for $J(q,\zeta)$ we obtain
\begin{equation}
J(q,\zeta) = J(q,\infty) + \delta J(q,\zeta) \;,
\end{equation}
where $\delta J(q,\zeta)$ has the asymptotic form
\begin{equation}
\delta J(q,\zeta) = q^{-\varepsilon} \hat f(\zeta/q,\varepsilon) \qquad \mbox{at} \quad q, \zeta \to \infty \;,
\end{equation}
where
\begin{eqnarray}
\hat f(y,\varepsilon) &=& -\frac{K_d \, \Gamma[(4-\varepsilon)/2]}{\sqrt{\pi} \, \Gamma[(3-\varepsilon)/2]}
\left\{ \int\limits_0^y x^{-1-\varepsilon} dx \int\limits_0^{\pi} \frac{x^2 \, \Theta(1+2x \cos \theta + x^2 - y^2)}
{1+2x \cos \theta + x^2} \, (\sin \theta )^{2-\varepsilon} d \theta \right. \nonumber \\
&+& \left. \int\limits_y^{\infty} x^{-1-\varepsilon} dx \int\limits_0^{\pi} 
\left( \frac{x^2}
{1+2x \cos \theta + x^2} - 1 \right) \, (\sin \theta )^{2-\varepsilon} d \theta  \right\} \;.
\end{eqnarray}

Based on these results, we can evaluate the large--${\bf q}$  (i.~e., $C <q< \zeta$
for large enough $C$ at a given $k$) contribution to the diagram \vverticec in four dimensions as
\begin{equation}
\frac{1}{(2 \pi)^4} \int\limits_{C <q< \zeta}
[J(\mid {\bf k+q} \mid,\zeta)-J(q,\zeta)] \, q^{-2} d^4 q \;.
\end{equation}
It, in fact, diverges at $\zeta \to \infty$, and the divergent part comes
from the $-K_4 \, \ln q$ term in $J(q,\infty)$, providing the contribution
\begin{equation}
-\frac{K_4^2}{\pi} \int\limits_{C}^{\zeta}
q \, dq \int\limits_0^{\pi} \ln \left( 1+ \frac{2kq \cos \theta + k^2}{q^2} \right) 
\sin^2 \theta \, d \theta \;.
\label{eq:thetdiag}
\end{equation}
Using the expansion in powers of $k/q$, we see that~(\ref{eq:thetdiag}) logarithmically diverges at
$\zeta \to \infty$. However, if we subtract the term $\propto k^2$, the result is convergent.
It indicates that the diagram \vverticetet  in~(\ref{eq:thetd}) has finite value at $\zeta \to \infty$.

\vspace*{6pt}
 Similarly, we can treat the large--wave--vector contribution to the diagram \vertexkq further

\vspace*{5pt}
\noindent
considered in the renormalization up to the order of $\varepsilon^3$.
This graph contains the subtraction of the zero--vector contribution from the internal
block \mbox{\vverticexx,} as well as from the whole diagram. It is defined (at $c=\Lambda =1$) by
\begin{equation}
 \vertexkq = \frac{1}{(2 \pi)^4} \int\limits_{1 <q_1< \zeta}
\left\{ q_1^{-2} \mid {\bf q}_1 + {\bf k} \mid^{-2} 
\hat \mathcal{F} (\mid {\bf q}_1 + {\bf k} \mid, \zeta) \, 
J(\mid {\bf q}-{\bf q}_1 \mid,\zeta) - q_1^{-4} J(q_1,\zeta) \right\} d^4 q_1 
\label{eq:ddd}
\end{equation}
in four dimensions. Using the asymptotic estimates of $J(q,\infty)$ and $\delta J(q,\zeta)$, 
we find that the tail of large $q_1$ is convergent for any given ${\bf k}$ and ${\bf q}$.

The actual integrals~(\ref{eq:thetdiag}) and~(\ref{eq:ddd}) have been estimated in four dimensions ($d=4$). 
Those ones, which are convergent at $d=4$, are convergent also in $d= 4-\varepsilon$
dimensions for small (positive and also negative) $\varepsilon$ when considering the ${\bf k}$--space integrals
as continuous functions of $d$ in the usual sense of the $\varepsilon$--expansion.
It is because the contribution of large wave vectors changes only slightly.

\subsection{Testing the semigroup property}
\label{sec:semi}

Here we test the semigroup property
\begin{equation}
R_{s_1 s_2} \mu = R_{s_2} R_{s_1} \mu
\end{equation}
discussed already in Sec.~\ref{intro}.
In fact, we verify that the renormalized Hamiltonian after two subsequent
RGT with scale factors $s_1$ and $s_2$ is the same as after one RGT with
the scale factor $s_1 s_2$. It can be seen most easily for the vertices
\sixx  and \mbox{\vverticey,} as consistent with Eqs.~(\ref{eq:zeta}) to~(\ref{eq:fl6})
and the fact that $u$ is renormalized only by an amount of  
$\mathcal{O} \left(\varepsilon^2 \right)$.
The semigroup property for the coupling constant $u$ easily follows from~(\ref{eq:ud}).
After the first RGT we have
\begin{equation}
u' = s_1^{\varepsilon} \left[ u - \frac{n+8}{2} u^2 \vverticeux  \right]  
+ \mathcal{O} \left( \varepsilon^3 \right)  \;, 
\label{eq:u1}
\end{equation}
where the dotted line refers to the $k$ interval $[\Lambda/s_1,\Lambda]$.
After the second RGT the value of the coupling constant, denoted as $u''$,
becomes
\begin{equation}
u'' = s_2^{\varepsilon} \left[ u' - \frac{n+8}{2} u'^2 \, s_1^{-\varepsilon} \vverticeuxx  \right]  
+ \mathcal{O} \left( \varepsilon^3 \right) \;, 
\label{eq:u2}
\end{equation}
where the dashed line refers to the interval $[\Lambda/(s_1s_2),\Lambda/s_1]$.
Here the rescaling of the wave vectors from $k \in [\Lambda/s_2,\Lambda]$
to $k \in [\Lambda/(s_1s_2),\Lambda/s_1]$ has been performed for convenience.
Inserting~(\ref{eq:u1}) in~(\ref{eq:u2}), and taking into account that 
$u=\mathcal{O} \left(\varepsilon \right)$, we obtain
\begin{eqnarray}
u'' &=& (s_1s_2)^{\varepsilon} \left[ u - \frac{n+8}{2} u^2 \left\{ \vverticeux + \vverticeuxx \right\} \right]  
+ \mathcal{O} \left( \varepsilon^3 \right)  \nonumber \\ 
&=& (s_1s_2)^{\varepsilon} \left[ u - \frac{n+8}{2} u^2 \vverticeu  \right]  
+ \mathcal{O} \left( \varepsilon^3 \right) \;, 
\label{eq:u22}
\end{eqnarray}
where $k \in [\Lambda/(s_1s_2),\Lambda]$ corresponds to the solid coupling lines.
It is obvious that~(\ref{eq:u22}) is identical to the result of one RGT obtained
from~(\ref{eq:ud}) with $s=s_1s_2$, which proves the semigroup property.
Analogous diagram summation proves this property also in the next order
of the $\varepsilon$--expansion~\cite{Hara_private}.

 The semigroup property for the parameter $r$, when performing the RG transformations
of the initial Hamiltonian with $\zeta=1$, is proven straightforwardly by the same method.
Namely, the result of the second RGT (applying~(\ref{eq:rd})) is represented by the
diagrams containing the dotted and the dashed lines with 
$k \in [\Lambda/s_1,\Lambda]$ and $k \in [\Lambda/(s_1s_2),\Lambda/s_1]$, respectively.
These diagrams sum up to give ones with $k \in [\Lambda/(s_1s_2),\Lambda]$, which
correspond to those of the one--step renormalization with $s=s_1s_2$.
In distinction to the case of $u$, here more complicated terms appear,
including different lines in one diagram.
The situation is less trivial when starting with $\zeta>1$.
In this case all diagrams can be decomposed in such ones, which contain three types
of lines with $k \in [\Lambda,\Lambda \zeta]$,  
$k \in [\Lambda/s_1,\Lambda]$ and $k \in [\Lambda/(s_1s_2),\Lambda/s_1]$, respectively.
Then we prove the  semigroup property by checking that such diagram representation
for the two--step renormalization is identical (up to the considered order
of the $\varepsilon$--expansion) to that for the one--step renormalization.
The semigroup property for the parameters $c$ and $\theta({\bf k})$ is proven by this
method, as well.

\subsection{The fixed point}
\label{sec:fixp}

The RG flow described by the truncated perturbative equations of Sec.~\ref{subsec:RGf}
has certain fixed point as a steady state solution of the RG flow equations. We shall 
mark the fixed--point values by an asterisk, except the value of $c$,
since the fixed point exists for any given $c$, as it will be seen from the following 
analysis. The fixed--point value of the coupling constant $u$, i.~e.,
\begin{equation}
u^* = \frac{2c^2}{(n+8)K_4} \; \varepsilon + \mathcal{O} \left(\varepsilon^2 \right)
\label{eq:ufix}
\end{equation}
has been already mentioned in Sec.~\ref{subsec:RGf}. According to~(\ref{eq:fl4}) 
and~(\ref{eq:fl6}), we have $a_4^*=-(9{u^*}^2)/16 + \mathcal{O} \left(\varepsilon^3 \right)$
and  $a_6^*=-{u^*}^2/8 + \mathcal{O} \left(\varepsilon^3 \right)$. It is generally
expected that the fixed point is  independent of the
scale parameter $s$. In the exact renormalization, it is a consequence of the
semigroup property. In the perturbation theory, it is expected to hold
at each order of the expansion. The above considered fixed--point values obey
this requirement within the given accuracy. It is obviously true also for the $r^*$ value
in the lowest order of the expansion, i.~e.,
\begin{equation}
r^* = -\frac{c \Lambda^2}{2} \; \frac{n+2}{n+8} \; \varepsilon + \mathcal{O} \left(\varepsilon^2 \right) \;.
\label{eq:rfix}
\end{equation}

However, it is less obvious for the next--order correction to $r^*$, as well as for other 
parameters of the fixed--point Hamiltonian. Therefore, we have performed
some tests.
Using~(\ref{eq:ufix}), (\ref{eq:fl4}), and~(\ref{eq:fl6}), 
we find from~(\ref{eq:rd}) that $r^*$ is independent of $s$ 
within the error of $\mathcal{O} \left(\varepsilon^3 \right)$ if $u^*$ is $s$--independent
up to the $\varepsilon^2$ order and
\begin{equation}
f(s) = -\frac{3}{2} \frac{K_4^2 \Lambda^2}{c^3} \, \ln s + const \cdot \left(s^2-1 \right)
\label{eq:fff}
\end{equation}
holds, where
\begin{equation}
f(s) = s^2 \left( \vverticerr + 3 \vverticero + 3 \, \vverticery \right) \;.
\label{eq:fun}
\end{equation}
Here the solid lines have wave vectors within $k \in [\Lambda/s,\Lambda]$, whereas
the dotted lines -- within $k \in [\Lambda,\Lambda \zeta]$ at  $\zeta \to \infty$,
and all terms are evaluated at $d=4$.
On the other hand, we prove that the function~(\ref{eq:fun}) obeys the equation
\begin{equation}
s_2^2 f(s_1) + f(s_2) = f(s_1s_2) - 3 \vverticerz \times \vverticeux \times (s_1s_2)^2 \;,
\label{eq:equ}
\end{equation}
where the dashed lines refer to the interval $k \in [\Lambda/(s_1s_2),\Lambda/s_1]$
and the dotted lines --- to the interval $k \in [\Lambda/s_1,\Lambda]$.
The proof consists of a straightforward checking of the corresponding diagram
identity by the decomposition described at the end of Sec.~\ref{sec:semi}.
Note only that the wave vectors in the diagrams of $f(s_2)$ are rescaled by the factor $1/s_1$.
As a test of consistency, we verify that~(\ref{eq:fff}) is a solution of~(\ref{eq:equ}). 

As regards the parameter $c$, it appears as a common factor
in the fixed--point equation obtained by setting $c'=c$ and $u=u^*$ 
in~(\ref{eq:c}), taking into account that $u^* \propto c^2$, whereas the diagrams 
involved are proportional to $1/c^3$. Hence, if this equation is satisfied, then it holds
for any $c$. In fact, the equation $c'=c$ reduces to one for the exponent $\eta$:
\begin{equation}
\eta \ln s = -\frac{2(n+2)}{(n+8)^2 K_4^2} \, \widetilde{\Pi}(s) \; \varepsilon^2 
+ \mathcal{O} \left(\varepsilon^3 \right) \;,
\label{eq:eta}
\end{equation}
where 
\begin{equation}
\widetilde{\Pi}(s) = \lim\limits_{\zeta \to \infty} \lim\limits_{k \to 0}
\left\{k^{-2} \left. \left[ \vverticerrksl 
+3 \left( \vverticeryksl + \vverticerxksl \right)
\right] \right|_{c=1} \right\}
\label{eq:Pii}
\end{equation}
is calculated at $d=4$ with $q \in [\Lambda/s,\Lambda]$ corresponding to
the solid lines and $q \in [\Lambda,\Lambda \zeta]$ --- to the dotted lines
in the diagrams. This equation yields an universal critical exponent $\eta$
provided that the limit~(\ref{eq:Pii}) exists (recall the discussion at the end of 
Sec.~\ref{subsec:RGf}). In this case, the universality follows from the diagram identity
\begin{equation}
\widetilde{\Pi}(s_1) + \widetilde{\Pi}(s_2) = \widetilde{\Pi}(s_1s_2) \;,
\label{eq:identi}
\end{equation}
which implies that $\widetilde{\Pi}(s) \propto \ln s$ holds and, therefore, $\eta$
is independent of $s$. The identity~(\ref{eq:identi}) is proven by the same method 
as~(\ref{eq:equ}).
In principle, the critical exponent $\eta$ can be calculated from~(\ref{eq:eta})
at any given $s$. However, from a technical point of view, it is convenient to consider
the limit $s \to \infty$, since only the first diagram in~(\ref{eq:Pii})
provides the singular contribution $\sim \ln s$ in this case. Hence, we have
\begin{equation}
\eta = -\frac{2(n+2)}{(n+8)^2 K_4^2} \; \varepsilon^2  \times 
\lim\limits_{s \to \infty} \lim\limits_{k \to 0} \left\{
\frac{1}{k^2 \ln s} \left. \left( \vverticerrksl \right) \right|_{c=1} \right\}
+ \mathcal{O} \left(\varepsilon^3 \right) \;,
\label{eq:etax}
\end{equation}
if the limit in~(\ref{eq:etax}) exists.
In this case, rescaling the wave vectors by factor $s/\Lambda$, $\lim\limits_{s \to \infty} \lim\limits_{k \to 0} \{ \cdot \}$ 
in~(\ref{eq:etax}) reduces to the expansion coefficient at $k^2 \ln s$ in the asymptotic large--$s$ expansion of
the diagram \mbox{\vverticec,} in which $q \in [1,s]$ corresponds to the dotted
lines. This coefficient can be

\vspace*{5pt}
\noindent
 calculated from~(\ref{eq:thetdiag}) at $\zeta=s$, and is equal to $-K_4^2/4$.
It yields the well known result (see, e.~g.,~\cite{Ma})
\begin{equation}
\eta = \frac{1}{2} \frac{n+2}{(n+8)^2} \; \varepsilon^2 
+ \mathcal{O} \left(\varepsilon^3 \right) \;.
\label{eq:eta2}
\end{equation}

Closing this section, we note that the expected independence of the fixed--point 
function $\theta^*({\bf k})$ on the RGT scale parameter $s$
can be easily seen when using the diagrammatic representation~(\ref{eq:thetd}):
this diagram  converges to certain $s$--independent
value at $\zeta \to \infty$ (if the $\propto k^2$ contribution can be
indeed subtracted), as discussed in Sec.~\ref{sec:estim}.

\section{Renormalization up to the order of $\varepsilon^3$}
\label{sec:ep3}

\subsection{Vertices of the order 
$\mathcal{O} \left(\varepsilon^3 \right)$}
\label{sec:vertices}

According to our diagrammatic representation, vertices of the order
$\mathcal{O} \left(\varepsilon^3 \right)$ in the renormalized Hamiltonian are those
made by coupling three original vertices of~(\ref{eq:H}), as well as other diagrams of such topology.
Note that any diagram of the renormalized Hamiltonian containing a
sub-graph (which is not the whole diagram) of topology \mbox{\Gaussv, \Gaussvx,
 \Gaussvxx, \Gaussvxxx, \Gaussvy,} etc., i.~e., a sub-graph with two outgoing lines, 
can be non-vanishing only if the lines of 
this sub-graph are the internal ones of the whole diagram.
It is a consequence of two facts: 1) the sum of the wave vectors entering each
node in a Feynman diagram is vanishing; 2) by definition of the RGT, 
the internal and the external lines in the diagrams of the renormalized Hamiltonian always
have different values of $\mid {\bf k} \mid$.
Thus, it is easy to verify 
that non-vanishing connected diagrams made of two vertices
\Gaussv and one vertex \vertice can contain no more than two external lines. 
Moreover, the diagrams of 
order $\mathcal{O} \left(\varepsilon^3 \right)$ with at least four external lines can
have only the following topologies: \mbox{\eight, \sixw,} 
and those ones obtained by coupling the lines in these graphs.
In such a way, the renormalized Hamiltonian has the form
\begin{equation}
 H = H^{(2)} + H^{(4)} + H^{(6)} + H^{(8)} + \mathcal{O} \left(\varepsilon^4 \right) \;,
\label{eq:HHHH}
\end{equation}
where $H^{(2m)}$ includes all the $\varphi^{2m}$--type vertices representable by the diagrams
with $2m$ external lines. The corresponding algebraic representation is
\begin{equation}
 \frac{H^{(2m)}}{T} = V^{1-m} 
\sum\limits_{i_1, \ldots, i_m} \; \sum\limits_{{\bf k}_1,{\bf q}_1, \ldots, {\bf k}_m,{\bf q}_m}
Q_m \left( {\bf k}_1,{\bf q}_1, \ldots ,{\bf k}_m, {\bf q}_m \right)
\varphi_{i_1,{\bf k}_1} \varphi_{i_1,{\bf q}_1} \cdots \varphi_{i_m,{\bf k}_m} \varphi_{i_m,{\bf q}_m} \;. 
\label{eq:algebr}
\end{equation}
In such a form $Q_m \left( {\bf k}_1,{\bf q}_1, \ldots ,{\bf k}_m, {\bf q}_m \right)$ 
vanishes unless $\sum_i {\bf k}_i + {\bf q}_i = {\bf 0}$. The dependence of these weight factors on 
various parameters is not indicated here. 

As regards the quartic part $H^{(4)}$, we need to separate the contribution corresponding
the ordinary $\varphi^4$ vertex, which is provided by the constant part of the
weight function $Q_2$ in~(\ref{eq:algebr}) evaluated at ${\bf k}_i={\bf q}_i={\bf 0}$.
For this purpose we can represent each diagram of $H^{(4)}$ as
\begin{equation}
 \fourbox  = \fourboxx  + \fourboxz  \times \vertice \;.
\label{eq:decom}
\end{equation}
In this symbolic notation \fourbox
is any diagram with four external lines, whereas \fourboxx

\vspace*{2pt}
\noindent 
is the same diagram from which the contribution corresponding to the ordinary
$\varphi^4$ vertex is subtracted. The latter one is represented by the last term
in~(\ref{eq:decom}), where \fourboxz  is the 

\vspace*{2pt}
\noindent 
diagram with amputated external lines
having zero wave vectors. A particular example 
\begin{equation}
 \vverticex = \vverticey   + \vverticeux  \times \vertice
\end{equation}
refers to the diagrams already considered in Sec.~\ref{sec:eps2}.
A problem here is that not all diagrams of the kind \fourboxx are convergent when the integration region
for the internal lines $[\Lambda,\Lambda \zeta]$ is 

\vspace*{2pt}
\noindent
extended to infinity at $\zeta \to \infty$. To avoid possible unphysical divergence and instability
of the RG flow, a suitable representation of the 
renormalized Hamiltonian should be used, where all vertices are convergent at $\zeta \to \infty$.
It is reached by making appropriate zero--$\bf k$ subtractions not only from the entire $\varphi^4$
diagrams, but also from their internal parts, like, e.~g., \vertexuu 

\vspace*{2pt}
\noindent
and \mbox{\vvverticeyuu.} 
 In the latter case, it is not necessary to make an extra subtraction of the zero--${\bf k}$ contribution from the
whole diagram, since such a contribution is vanishing. Appropriate subtractions should be used
also for vertices of higher than $\varphi^4$ order to ensure their convergence at $\zeta \to \infty$.

\subsection{The renormalized Hamiltonian and RG flow equations}
\label{sec:RH3}

According to the analysis of Sec.~\ref{sec:ep3}, an appropriate representation of the renormalized Hamiltonian is
\begin{eqnarray}
 \frac{H}{T} &=& \frac{1}{2} \sum\limits_{i,{\bf k}} \left( r + c k^2 
+ \theta({\bf k}) \right) \mid \varphi_{i,{\bf k}} \mid^2
+ \frac{u}{8} \dshv + a_4^{(1)} \; \Sigma \left( \vverticey \right) \nonumber \\
&+& a_4^{(2)} \; \Sigma \left( \vvverticeyuu \right)  
+ a_4^{(3)} \; \Sigma \left( \vertexuu \right) + a_4^{(4)} \; \Sigma \left( \vverticeyy \right)
\nonumber \\
&+& a_6^{(1)} \; \Sigma \left( \sixx \right)
+ a_6^{(2)} \; \Sigma \left( \cycle \right)  
+ a_6^{(3)} \; \Sigma \left( \sixwuu \right) \nonumber \\
&+& a_6^{(4)} \; \Sigma \left( \sixxx \right)
+ a_8 \; \Sigma \left( \eightx \right) +  \mathcal{O} \left(\varepsilon^4 \right)
\label{eq:a8} \;,
\end{eqnarray}
where $\Sigma(\cdot)$ denotes the sum of all such diagrams
made of vertices \dshv and \Gaussv, 
which yield the given picture when the dashed lines shrink to points. Besides,
the combinatorial weight coefficients, including the $n$--dependent factors,
are normalized in such a way that their sum is $1$ at $n=1$.  The relations of the kind
\begin{eqnarray}
 \vverticeo  &=&  \vverticer  \times  \vverticeyyx \label{eq:rel1} \\
 \sixo  &=&  \vverticer  \times  \sixyyy  \label{eq:rel2}
\end{eqnarray}
have been applied to reduce the number of different kind of vertices in the representation
of the Hamiltonian. The wave vectors of dotted lines
are within $k \in [\Lambda,\Lambda \zeta]$ with $\zeta$ being updated as $\zeta' = s \, \zeta$ 
under the RG transformation $R_s$.

The other Hamiltonian parameters are updated as follows:
\begin{eqnarray}
u' &=& s^{\varepsilon - 2 \eta} \left[ u - \frac{n+8}{2} u^2 \vverticeu   
+ \frac{ n^2 + 6n + 20}{4} u^3 \left(\vverticeu \right)^2 \right. \nonumber \\
&+& (5n+22) u^3 \left( \vertexone + \vertextwo + 2 \vertexthree + 2 \vertexfour + \vertexfive \right) \nonumber \\
&+& \left. \frac{(n+2)(n+8)}{2} u^3 \left( \vverticeuo + \vverticeoo \right) 
+ (n+8) u^2 r \vverticeuu \right] + \mathcal{O} \left( \varepsilon^4 \right) \label{eq:ud1}  \\
\left( a_4^{(1)} \right)' &=& - \frac{9}{16} u^2 s^{2 \varepsilon}
\left( 1 - (n+8) u \vverticeu  \right) + \mathcal{O} \left( \varepsilon^4 \right) 
= - \frac{9}{16} u'^2 + \mathcal{O} \left( \varepsilon^4 \right)  \\
\left( a_4^{(2)} \right)' &=& \frac{27}{32} u^3 + \mathcal{O} \left( \varepsilon^4 \right)  \\
\left( a_4^{(3)} \right)' &=& \frac{27}{8} u^3 + \mathcal{O} \left( \varepsilon^4 \right)  \\
\left( a_4^{(4)} \right)' &=& \frac{9}{8} u^2 r' + \mathcal{O} \left( \varepsilon^4 \right)  \\
\left( a_6^{(1)} \right)' &=& - \frac{1}{8} u^2 s^{2 \varepsilon}
\left( 1 - (n+8) u \vverticeu  \right) + \mathcal{O} \left( \varepsilon^4 \right) 
= - \frac{1}{8} u'^2 + \mathcal{O} \left( \varepsilon^4 \right)  \\
\left( a_6^{(2)} \right)' &=& \frac{9}{16} u^3 + \mathcal{O} \left( \varepsilon^4 \right)  \\
\left( a_6^{(3)} \right)' &=& \frac{9}{8} u^3 + \mathcal{O} \left( \varepsilon^4 \right)  \\
\left( a_6^{(4)} \right)' &=& \frac{1}{8} u^2 r' + \mathcal{O} \left( \varepsilon^4 \right)  \\
a_8' &=& \frac{3}{16} u^3 + \mathcal{O} \left( \varepsilon^4 \right) \;.  
\end{eqnarray}
Here $k \in [\Lambda/s,\Lambda]$ holds for solid lines, and $r' = s^2 \left( r + \frac{n+2}{2} u \vverticer \right) 
+ \mathcal{O} \left( \varepsilon^2 \right)$ is the updated value of $r$ in the lowest order
of the $\varepsilon$--expansion. 
We have skipped the refined RG flow equations for $r$, $c$, and $\theta({\bf k})$, since those already considered
in Sec.~\ref{subsec:RGf} are sufficient for our further analysis.
These RG flow equations are proven by direct verification, as in Sec.~\ref{sec:proofRG2}.

The fixed--point values of the parameters $a_4^{(i)}$, $a_6^{(i)}$ and $a_8$
are trivially related to $u^*$ and $r^*$.
Namely, $\left( a_4^{(1)} \right)^* = -(9/16){u^*}^2 + o \left( \varepsilon^3 \right)$,
$\left( a_4^{(4)} \right)^* = (9/8){u^*}^2 r^* + o \left( \varepsilon^3 \right)$,
$\left( a_6^{(1)} \right)^* = -(1/8){u^*}^2 + o \left( \varepsilon^3 \right)$,
and similarly for other parameters. The value of $u^*$ is
\begin{equation}
u^* = \frac{2 c^2 \Lambda^{\varepsilon} \varepsilon}{(n+8)K_d} \times 
\left( 1 + a_1(n) \varepsilon + \mathcal{O} \left( \varepsilon^2 \right)  \right) \;.
\label{eq:uexp}
\end{equation}
The coefficient $a_1(n)$, calculated from~(\ref{eq:ud1}), (\ref{eq:rfix}), and~(\ref{eq:eta2}), is
\begin{equation}
a_1(n) = \frac{4(5n+22)}{K_4^2 (n+8)^2} \, Q - \frac{n+2}{(n+8)^2} \;.
\label{eq:a1}
\end{equation}
Here  $Q=\hat Q(s)/\ln s$, where $\hat Q$ is given by the diagram expression
\begin{equation}
\hat Q(s) = \left( \vertexone + \vertextwo + 2 \vertexthree + 2 \vertexfour + \vertexfive
- \frac{1}{2} \left(\vverticeu \right)^2 \right) 
\end{equation}
evaluated at $d=4$ and $c=1$.
The quantity $\hat Q(s)$ is proportional to $\ln s$, since $\hat Q(s_1s_2)= \hat Q(s_1) + \hat Q(s_2)$
holds at $\zeta \to \infty$ as a diagram identity, which can be verified by the method outlined in Sec.~\ref{sec:semi}.
Hence, $Q$ is $s$--independent and, therefore, $u^*$ is $s$--independent at least up to the
$\varepsilon^2$ order, as expected. It is convenient to consider the limit $s \to \infty$ to determine $Q$ from
\begin{eqnarray}
Q &=& \lim\limits_{s \to \infty} \left( \frac{\hat Q(s)}{\ln s} \right)
= \lim\limits_{s \to \infty} \left\{ \frac{1}{\ln s} \left[ \vertexone - \frac{1}{2} \left(\vverticeu \right)^2 \right] \right\}
\nonumber \\
&=& \lim\limits_{s \to \infty} \left\{ \frac{1}{\ln s} \left[ \vertexzero - \frac{1}{2} \left(\vverticeux \right)^2 \right] \right\}
= \lim\limits_{s \to \infty} \left\{ \frac{1}{\ln s} \left[ \vertexzerosl + \frac{1}{2} \left(\vverticeux \right)^2 \right] \right\}
\nonumber \\
&=& \lim\limits_{s \to \infty} \left\{ \frac{1}{\ln s} 
\left[ K_4^2 \int\limits_1^s \left(-\ln q + \frac{1}{2} \right) \frac{dq}{q}
+ \frac{1}{2} \left( K_4 \int\limits_1^s \frac{dq}{q} \right)^2 \right]  \right\}
= \frac{K_4^2}{2} \;.
\label{eq:QQQ}
\end{eqnarray}
The wave vectors are rescaled to $k \in [1,s]$ (i.~e., this interval corresponds to the dotted lines in
the diagrams of~(\ref{eq:QQQ})) and the zero--${\bf k}$
contribution of the internal diagram block is separated in the second line of~(\ref{eq:QQQ}).
The large--$k$ asymptotic estimate for \mbox{\vverticeuyk,} obtained in Sec.~\ref{sec:estim},
is used in the third line of~(\ref{eq:QQQ}). Inserting this result into~(\ref{eq:a1}), we obtain
\begin{equation}
a_1(n) = \frac{3(3n+14)}{(n+8)^2}  \;.
\end{equation}

Summarising the results of this section, one has to note that, at the $\varepsilon^3$ order, 
the renormalized Hamiltonian contains several extra vertices as compared to the initial
or bare Hamiltonian. All these terms, including the $\varphi^6$ vertices and the $\varphi^8$
vertex, are relevant in the sense that they do not vanish at the fixed point.
Moreover, they are always relevant in the same sense as the $\varphi^4$ vertex,
because of the relations between $a_6^{(i)}$, $a_8$ and $u$. In particular, 
if we consider the RG flow starting slightly away from the critical surface, 
then those terms are considered as relevant, which diverge
at large renormalization scales $\zeta \to \infty$. The shrinking terms are 
irrelevant in this usual sense. The $\varphi^4$ vertex is commonly considered as relevant
in $4-\varepsilon$ dimensions. Consequently, the $\varphi^6$ and $\varphi^8$ vertices 
are also relevant in the same (usual) sense.

\section{Expansion at a fixed spatial dimension $d$}
\label{sec:fixedd}

Apart from the $\varepsilon$--expansion, we have tested also an alternative approach,
where the coupling constant $u$ is considered as an expansion parameter
at a fixed spatial dimensionality, i.~e., fixed $\varepsilon$. In this sense
the discussed here approach is similar in spirit to that widely used for
calculations in three dimensions, $d=3$, as described in~\cite{Parisi,BGZ,GZ,BNGM,BNM}.
This known approach uses the Callan--Symanzik equation instead of the Wilson's equation~(\ref{eq:WERGE}).
However, if both are correct RG equations, then they should provide consistent results.
Here we use~(\ref{eq:WERGE}) to make the expansion in powers of $u$ at a fixed but small $\varepsilon$.

In this method, all the rescaling factors of the kind $s^{\varepsilon}$ have to be retained in 
their original form without the expansion in powers of $\varepsilon$. 
Following the approach of~\cite{Parisi,BGZ,GZ,BNGM,BNM}, the critical exponents should be
expanded in powers of the coupling constant, estimating their universal values at $u=u^*$. 
Therefore, the correction factors like $s^{-\eta}$
can be omitted in the first approximation, since one finds that $\eta = \mathcal{O} \left( u^2 \right)$.
Using this idea, and following the calculations in Sec.~\ref{subsec:RGf}, we determine the fixed--point
values of $u$ and $r$ in the lowest order of the theory, i.~e., $\tilde{u}^*$ and $\tilde{r}^*$, as
\begin{eqnarray}
 \tilde{u}^* &=& \frac{c^2 \Lambda^{\varepsilon}}{K_d \, s^{\varepsilon}} \;\; \frac{2}{n+8} \;\, \varepsilon \;, \\
\tilde{r}^* &=& - \frac{s^2-s^{\varepsilon}}{s^{2+\varepsilon} - s^{\varepsilon}} \;\;
\frac{c \Lambda^2}{2 -\varepsilon} \;\; \frac{n+2}{n+8} \;\; \varepsilon \;.
\end{eqnarray}
These values are obtained as the fixed--point solutions of~(\ref{eq:ud}) and~(\ref{eq:rd}),
retaining the relevant terms without their expansion in powers of $\varepsilon$.
In this case the terms up to the lowest--order diagrams, \vverticeu in~(\ref{eq:ud})
and \vverticer in~(\ref{eq:rd}), have been included.

As we see, the fixed--point values depend on the scale parameter $s$, thus pointing
to an internal inconsistency and incorrectness of this method. One has to note, however, that the
common method at fixed dimension~\cite{Parisi,BGZ,GZ,BNGM,BNM} uses the RG $\beta$-function.
A question arises whether it really
solves or only ``hides under carpet'' the problem.
The beta function describes
the renormalization of the coupling constant as a differential
 equation, when $s=1+ds$ at $ds \to 0$. In this sense,
 the beta-function approach is nothing but a particular case
 of our more general method, where $ds$ can be finite.
The actual consideration shows that there is something wrong
 at fixed $d$ and finite $ds$. Therefore, one can ask: why the method should work correctly
 at $ds \to 0$ if it does not work correctly at any finite $ds$?

\section{The two--point correlation function}
\label{sec:twopoint}

\subsection{The diagrammatic representation and scaling}

Consider now the Fourier--transformed two--point correlation function
$G({\bf k}) = \left\langle \mid \varphi_{i,{\bf k}} \mid^2 \right\rangle$. Its diagram expansion
contains all connected diagrams with two external lines, having wave vectors
${\bf k}$ and $-{\bf k}$. Thus we have
\begin{equation}
G({\bf k}) =  \selfeko  + \selfekk + \selfekkk + \ldots = 
\frac{G_0({\bf k})}{1-2 G_0({\bf k}) \widetilde \Sigma({\bf k})} \;,
\label{eq:Gexp}
\end{equation}
where \selfekk represents the sum of diagrams of this kind, which are
made of vertices of $-H/T$ and are irreducible in the sense that
they are not representable as a chain of such blocks. Here 
$\widetilde \Sigma({\bf k}) = \selfek$ is such irreducible diagram block
with amputated external lines. The quantity $-2 \widetilde \Sigma({\bf k})$
is known as self--energy~\cite{Ma}. The well known expansion~(\ref{eq:Gexp}) leads to the equation 
\begin{equation}
X({\bf k}) = ck^2 - 2 \widetilde \Sigma({\bf k}) \;,
\label{eq:XXX}
\end{equation}
where $X({\bf k}) = 1/G({\bf k})$ is the inverse of the two--point correlation
function, and $G_0({\bf k})$ is set to $1/(ck^2)$ within the $\varepsilon$--expansion.
In this case, the diagrams contain also insertions of the second--order vertex 
$(1/2)\sum\limits_{i,{\bf k}} (r+ \theta({\bf k})) \mid \varphi_{i,{\bf k}} \mid^2$.
The wave vectors within $q \in [0,\Lambda]$ are related to the solid coupling lines
between different vertices in these diagrams. The vertices themselves can contain 
internal dotted lines with $q \in [\Lambda,\Lambda \zeta]$, 
produced by the RG transformation, $\zeta$ being the renormalization scale
in the RG flow equations. 

As it is well known, the two--point correlation function rescales in accordance with 
\begin{equation}
X({\bf k};R_s H) = s^{2-\eta} X({\bf k}/s;H)
\label{eq:rescal} 
\end{equation}
after the RG transformation $R_s$, transforming the original Hamiltonian $H$ into the
renormalized one  $R_s H$. It is a general exact relation, trivially following from the
fact that the integration over the Fourier modes with $\Lambda/s < k < \Lambda$ in the first 
step of the RG transformation does not alter the correlation function for $k < \Lambda /s$,
whereas the $s$--dependent factors in~(\ref{eq:rescal}) compensate its rescaling
in the second step of the RG transformation (see Sec.~\ref{sec:diag}).

\subsection{The two--point correlation function at the fixed point}
\label{sec:corfix}

In the following, we will examine the two--point correlation function (or its inverse)
at the fixed point up to the $\varepsilon^2$ order.
According to~(\ref{eq:rescal}), $X({\bf k}) = a k^{2-\eta}$ holds within $k \le \Lambda$
for the fixed--point Hamiltonian $H^*$ (since $R_s H^* = H^*$) with some ${\bf k}$--independent 
constant $a$, which can depend on $n$, $\varepsilon$, and Hamiltonian parameters. It is consistent with~(\ref{eq:XXX}) if
\begin{equation}
a k^2 (1- \eta \ln k) = c k^2 - \frac{n+2}{2} {u^*}^2 \left\{ \vverticetet + \vverticerrksl  
+3 \left( \vverticeryksl  + \vverticerxksl  \right) \right\} 
 +  \mathcal{O} \left( \varepsilon^3 \right)
\label{eq:fixcorr}
\end{equation}
holds for $k \le \Lambda$, since $\widetilde \Sigma({\bf 0})$ vanishes at the fixed point.
Note that $q \in [0,\Lambda]$ refers to the solid lines and $q \in [\Lambda,\Lambda \zeta]$
at $\zeta \to \infty$ -- to the dotted lines in~(\ref{eq:fixcorr}).
The diagrams can be evaluated at $d=4$ in this case, and we can set $c=1$, as it 
gives merely a common factor. By definition of \mbox{\vverticetet,} we have
\begin{equation}
\vverticetet = \vverticec - k^2 R(\zeta) \;,
\label{eq:vertex}
\end{equation}
where
\begin{equation}
R(\zeta) = \lim\limits_{k \to 0} \left\{ \frac{1}{k^2} \vverticec \right\} 
= - \frac{K_4^2}{4} \ln \zeta + const \qquad \mbox{at} \quad d=4, \quad \zeta \to \infty \;,
\label{eq:RR}
\end{equation}
as consistent with the already considered estimation of such term (after rescaling
the wave vectors from $q \in [\Lambda,\Lambda \zeta]$ to $q \in [1,\zeta]$) at the end of Sec.~\ref{sec:fixp},
provided that the considered here limit (at $k \to 0$) exists. As in Sec.~\ref{sec:fixp}, this ensures
the expected properties.
Inserting~(\ref{eq:vertex}) into~(\ref{eq:fixcorr}), and summing up the diagrams,
we obtain (at $c=1$)
\begin{equation}
a k^2 (1- \eta \ln k) = k^2 - \frac{n+2}{2} {u^*}^2 f(k,\zeta)  
 +  \mathcal{O} \left( \varepsilon^3 \right) \qquad \mbox{at} \quad \zeta \to \infty \;,
\label{eq:fixcorr1}
\end{equation}
where
\begin{equation}
f(k,\zeta) = \vverticecc - k^2 R(\zeta) 
\end{equation}
with $q \in [0,\Lambda \zeta]$ corresponding to the dashed lines.
Using~(\ref{eq:RR}) and rescaling the wave vectors, we find that
\begin{equation}
f(k) = \lim\limits_{\zeta \to \infty} f(k,\zeta)
\end{equation}
obeys the equation
\begin{equation}
f(sk) = s^2 \left( f(k) + \frac{1}{4} K_4^2 \, k^2 \ln s \right)
\end{equation}
with the solution
\begin{equation}
f(k) = \mathcal{B} \, k^2 + \frac{1}{4} K_4^2 \, k^2 \ln k \;,
\label{eq:solu}
\end{equation}
where $\mathcal{B}$ is a constant.
According to this, Eq.~(\ref{eq:fixcorr1}) is indeed satisfied
with appropriate $a=a(n,\varepsilon)$ at $\eta$ given by~(\ref{eq:eta2}).
Thus, the $\varepsilon$--expansion of $X({\bf k})$ 
 at the fixed point is consistent with the 
scaling $X({\bf k}) = a k^{2-\eta}$ within $k \le \Lambda$  up to the order of $\varepsilon^2$, at least,
if the limit~(\ref{eq:RR}) exists.
Besides, such analysis provides one more method to determine the critical exponent $\eta$.

\subsection{The two--point correlation function on the critical surface}
\label{sec:crsurf}

The critical correlation function diverges at $k \to 0$, so that
$X({\bf 0})=0$ and $\widetilde \Sigma({\bf 0})=0$ must hold in this case, which means
that~(\ref{eq:XXX}) becomes
\begin{equation}
X({\bf k}) = c k^2 - 2 [\widetilde \Sigma({\bf k}) - \widetilde \Sigma({\bf 0})] \;.
\label{eq:XXXc}
\end{equation}
It is easy to verify that the condition
$\widetilde \Sigma({\bf 0})=0$ is indeed satisfied in the lowest order of the
$\varepsilon$--expansion, where the equation of the critical surface reads 
\begin{equation}
r = r_c(u) = -\frac{n+2}{2} u \vverticer + \mathcal{O} \left( \varepsilon^2 \right) 
\label{crsurf}
\end{equation}
with $k \in [0,\Lambda]$ for the solid lines. 
The condition $\widetilde \Sigma({\bf 0})=0$ should hold on the critical surface
also at higher orders of the $\varepsilon$--expansion. We have tested this expected property
up to the $\varepsilon^3$ order and have not revealed any contradiction.

According to~(\ref{eq:XXXc}), the equation for $X({\bf k})$ on the critical surface reads
\begin{equation}
X({\bf k}) = c k^2 - \frac{n+2}{2} u^2 \left\{ \vverticetet + \vverticerrksl  
+3 \left( \vverticeryksl  + \vverticerxksl  \right) \right\} 
 +  \mathcal{O} \left( \varepsilon^3 \right) \;.
\label{eq:XXXX}
\end{equation}
As already discussed in Sec.~\ref{subsec:RGf},
here we assume that the $\sim k^2$ contribution can be extracted from those kind of diagrams,
which appear in the RG flow equation for $c$, by taking the limit 
$\lim\limits_{k \to 0} \left\{ \frac{1}{k^2} \left( \cdot  \right) \right\}$.
Recall that the first diagram in~(\ref{eq:XXXX}) is obtained by such an extraction.

Based on~(\ref{eq:XXXX}), we will consider the rescaling of $X({\bf k})$
in the renormalization process, starting with the bare Hamiltonian at 
$(u-u^*)/u^* = \delta_0$ and $c=c_0$
and performing certain number $m$ of RG transformation $R_s$, in such a way that 
$(u-u^*)/u^*$ becomes equal to $\delta$, where
$\delta_0$ and $\delta$ are small constants. In this case, the fixed--point value $u^*$ of
the coupling constant is determined at the current value of the parameter $c$,
so that it is a slightly varying quantity due to the renormalization of $c$.
Such a choice is convenient, since we do not need to determine the fixed--point value
of $c$ after infinite number of RG transformations.
In the actually considered truncation of the $\varepsilon$--expansion, 
the renormalization of $u$ is given by~(\ref{eq:ud}), from which we evaluate 
$m=m(\delta_0,\delta,s,\varepsilon)$. In this case, $m \sim 1/\varepsilon$ is determined with
uncertainty of $\mathcal{O}(1)$, so that it can be rounded to appropriate integer value at any positive $s-1$ 
of order unity. The renormalization of $(u-u^*)/u^*$ can be performed at integer 
$m$ for $\varepsilon$--independent $\delta_0$ and $\delta$ up to any higher order
of the $\varepsilon$--expansion by adjusting $s$ as $s = s_0 + s_1 \varepsilon + s_2 \varepsilon^2 + \cdots \;$.
The entire renormalized Hamiltonian after $m \sim 1/\varepsilon$ RGT is well defined
within the $\varepsilon$--expansion, since we always can reach the required accuracy,
truncating the RG flow equations at high enough order. 

Our aim is to compare the perturbative rescaling of $X({\bf k})$ with the exact 
one~(\ref{eq:rescal}) at $k \to 0$, taking into account the expected asymptotic
form of $X({\bf k})$ on the critical surface. Namely, for the initial Hamiltonian we have
\begin{equation}
X({\bf k};H_0) = A(\delta_0,\varepsilon) \, k^{2-\eta} 
\left\{ 1 + \sum\limits_i a_i(\delta_0,\varepsilon) k^{\omega_i} 
+ \sum\limits_i \hat a_i(\delta_0,\varepsilon) k^{\hat \omega_i} \right\} \qquad \mbox{at} \quad k \to 0 \;,
\label{eq:Xsc0}
\end{equation}
where the argument $H_0$ means the initial Hamiltonian, $\omega_i$ are those correction--to--scaling
exponents, which tend to zero at $\varepsilon \to 0$, whereas $\hat \omega_i$ are those ones, which
tend to positive constants in this limit. The form~(\ref{eq:Xsc0}) is consistent with the usual
assumption that the correlation function can be expanded in powers 
of $k$ at $d<4$ for large renormalization scales $s \to \infty$. 
According to~(\ref{eq:rescal}), it is true for the bare Hamiltonian at $k \to 0$. 
Eq.~(\ref{eq:rescal}) then yields
\begin{equation}
X({\bf k};R_{s^m} H_0) = A(\delta_0,\varepsilon) \, k^{2-\eta} 
\left\{ 1 + \sum\limits_i a_i(\delta_0,\varepsilon) s^{-m \omega_i} k^{\omega_i} 
+ \sum\limits_i \hat a_i(\delta_0,\varepsilon) 
s^{-m \hat \omega_i} k^{\hat \omega_i} \right\} \quad \mbox{at} \quad k \to 0 \;,
\label{eq:Xsc1}
\end{equation}
where $R_{s^m} H_0$ denotes the renormalized Hamiltonian after $m$ RG transformations $R_s$.
The well known result of the $\varepsilon$--expansion states that the leading
correction--to--scaling exponent is 
\begin{equation}
\omega := \omega_1 = \varepsilon + \mathcal{O} \left( \varepsilon^2 \right) \;.
\label{eq:omegal}
\end{equation}

In the following, we will formulate as a theorem an important result,
stating certain consistency relations for the coefficients and exponents in~(\ref{eq:Xsc0}).

\vspace*{1ex}
\textbf{Theorem.} \hspace{0ex} 
\textit{
If the perturbative $\varepsilon$--expansion--based renormalization 
(repeating the RG transformation  $R_s$) of 
$X({\bf k}) = 1/G({\bf k})$ on the critical surface is consistent with~(\ref{eq:Xsc0})
and~(\ref{eq:Xsc1}) at any small $\varepsilon$--independent $\delta_0$ and $\delta$,
such that $0 < \delta/\delta_0 < 1$,
defining the initial ($\delta_0$) and the final ($\delta$) value of $(u-u^*)/u^*$
(with $u^*$ determined at the current $c$), then
\begin{eqnarray}
\omega_j &=& j \, \varepsilon + \mathcal{O} \left( \varepsilon^2 \right) \qquad : \qquad j \ge 1 
\label{eq:te1} \;, \\
A(\delta_0,\varepsilon) &=& c_0 + \mathcal{O}(\varepsilon) \;, \label{eq:tea} \\
a_j(\delta_0,\varepsilon) &=& b_j \, \varepsilon \left( \frac{\delta_0}{1+\delta_0}  \right)^j 
+ \mathcal{O} \left( \varepsilon^2 \right) \qquad : \qquad j \ge 1
\label{eq:te2}
\end{eqnarray}
hold, where $c_0$ is the initial value of $c$, and $b_j = -\eta_2 \, (j+1)/j$,
where $\eta_2$ is defined in $\eta = \eta_2 \, \varepsilon^2 + \mathcal{O}\left(\varepsilon^3 \right)$. 
} 

\vspace*{1ex}

 \textit{Proof.} Let us first determine the number $m=m(\delta_0,\delta,s,\varepsilon)$ of RGT 
required for transformation of $(u-u^*)/u^*$ from $\delta_0$ to $\delta$. From~(\ref{eq:ud}) we obtain
the updating rule for $\delta_{\ell} = (u_{\ell} - u^*_{\ell})/u^*_{\ell}$,
\begin{equation}
\frac{\delta_{\ell+1}}{\delta_{\ell}} = s^{-\varepsilon (1+\delta_{\ell}) + \mathcal{O} \left( \varepsilon^2 \right)} \;,
\label{eq:recud}
\end{equation}
where $u_{\ell}$ is the value of $u$ after $\ell$-th RGT, and $u^*_{\ell}= u^*(c_{\ell})$ is the
fixed--point value of $u$ determined at the current $c$ after $\ell$-th RGT, i.~e., at $c=c_{\ell}$.
Here we take into account that $c_{\ell+1} = c_{\ell} + \mathcal{O} \left( \varepsilon^2 \right)$
holds according to~(\ref{eq:c}) and therefore
$u^*_{\ell+1} = u^*_{\ell} + \mathcal{O} \left( \varepsilon^3 \right)$
follows from~(\ref{eq:uexp}). Denoting\linebreak
$x_{\ell} = \ln \mid \delta_{\ell} \mid$, Eq.~(\ref{eq:recud}) reduces to
\begin{equation}
x_{\ell+1} - x_{\ell} = - \varepsilon \, \left( 1 + \mathrm{sign}(\delta) 
\, e^{x_{\ell}} \right) \ln s + \mathcal{O} \left( \varepsilon^2 \right) \;.
\label{eq:xmm}
\end{equation}
The variation of $x_{\ell}$ becomes quasi--continuous in the limit $\varepsilon \to 0$, so that $x_{\ell}$
can be considered as a continuous function $x(\ell)$, and~(\ref{eq:xmm}) transforms into the differential equation
\begin{equation}
\frac{d x(\ell)}{d \ell} = - \varepsilon \left(1 + \mathrm{sign}(\delta) \, e^x \right) \ln s  
+ \mathcal{O} \left( \varepsilon^2 \right) \;.
\end{equation} 
Integrating this equation, we find the necessary number of RGT
\begin{equation}
m = \frac{1}{\varepsilon \ln s} \int\limits_{x_*}^{x_0} \frac{dx}{1+ \mathrm{sign}(\delta) \, e^x} + \mathcal{O}(1) 
= \frac{1}{\varepsilon \ln s} \ln \left[ \frac{\delta_0 (1+\delta)}{(1+\delta_0) \delta} \right] 
+ \mathcal{O}(1) \;,
\end{equation}
where $x_0= \ln \mid \delta_0 \mid$ is the initial and $x_* = \ln \mid \delta \mid$ is the final value of 
$\ln \mid \delta_{\ell} \mid$. It yields
\begin{equation}
s^{-m \varepsilon} = \frac{1+ \delta_0}{\delta_0} \times \frac{\delta}{1+\delta} + \mathcal{O}(\varepsilon) \;.
\label{eq:sm}
\end{equation}
Inserting~(\ref{eq:sm}) into~(\ref{eq:Xsc1}), we obtain
\begin{equation}
X({\bf k};R_{s^m} H_0) = A(\delta_0,\varepsilon) \, k^{2-\eta} 
\left\{ 1 + \sum\limits_i a_i(\delta_0,\varepsilon) 
\left[ \frac{\delta (1+\delta_0)}{(1+\delta) \delta_0} + \mathcal{O}(\varepsilon) \right]^{\omega_i/\varepsilon} 
k^{\omega_i}  + 
\mathcal{O} \left( e^{-\lambda /\varepsilon} k^{\hat \omega} \right) \right\} \;, 
\label{eq:Xex}
\end{equation}
where the terms with $\hat \omega_i$ are absorbed in the last remainder term, where $\lambda>0$
and $\hat \omega$ is the smallest exponent among $\hat \omega_i$.
This term is irrelevant in the $\varepsilon$--expansion.

From~(\ref{eq:XXXX}) we obtain 
\begin{equation}
X({\bf k};R_{s^m} H_0) = ck^2 - \frac{n+2}{2c^3} \left[u^*(1+ \delta) \right]^2
\left( \mathcal{B} \, k^2 + \frac{1}{4} K_4^2 \, k^2 \ln k  \right) + \mathcal{O} \left( \varepsilon^3 \right) \;.
\label{eq:Xper}
\end{equation}
The diagrams of~(\ref{eq:XXXX}) are evaluated in~(\ref{eq:Xper}) as in Sec.~\ref{sec:corfix} (cf.~Eq.~(\ref{eq:solu})),
since the difference between the cases $\zeta=\infty$ and $\zeta = s^{m(\delta_0,\delta,s,\varepsilon)}$ (with 
$q \in [\Lambda,\Lambda \zeta]$ for dotted lines) is irrelevant within
the $\varepsilon$--expansion.
Since $c$ is renormalized by $\mathcal{O} \left( \varepsilon^2 \right)$ in one RGT 
according to~(\ref{eq:c}), the renormalized $c$ value after $m \sim 1/\varepsilon$ RG transformations
in~(\ref{eq:Xper}) is $c=c_0 + \mathcal{O}( \varepsilon)$.

The consistency with~(\ref{eq:Xex})
(where $\omega_i \to 0$ at $\varepsilon \to 0$) implies that $X({\bf k};R_{s^m} H_0)$ can be expanded as
\begin{equation}
 X({\bf k};R_{s^m} H_0) = k^2 \left( \mathcal{B}_0(\delta_0,\delta,\varepsilon) 
+ \mathcal{B}_1(\delta_0,\delta,\varepsilon) \ln k + \mathcal{B}_2 (\delta_0,\delta,\varepsilon) (\ln k)^2 + \cdots \right)
\label{eq:BB}
\end{equation}
at $\varepsilon \to 0$. On the other hand,
the diagrammatic perturbative equation for $X({\bf k};R_{s^m} H_0)$ on the critical surface
can be represented in general (up to any order of the $\varepsilon$--expansion) as
\begin{equation}
 X({\bf k};R_{s^m} H_0)= c \, F({\bf k},\delta,\varepsilon) \;, 
\label{eq:Xrepr}
\end{equation}
where $c=c(\delta_0,\delta,\varepsilon)$, whereas $F({\bf k},\delta,\varepsilon)$
contains only integer powers of $\delta$ in the small--$\delta$--expansion.
It is because the critical value of $r$, as well as all Hamiltonian parameters
can be expanded in integer powers of $u=u^*(1+\delta)$, where $u^*=u^*(c)=c^2 u^*(1)$ holds 
with $\delta_0$-- and $\delta$--independent $u^*(1)$, and the diagrams associated with 
$u^{\ell}$ contain $2 \ell -1$ internal coupling lines giving the factor $c^{1-2 \ell}$. These are consequences
of the general structure of the RG flow equations. Hence, the small--$\delta$--expansion of the ratio 
$\mathcal{B}_i(\delta_0,\delta,\varepsilon)/\mathcal{B}_0(\delta_0,\delta,\varepsilon)$ in~(\ref{eq:BB})
contains only integer powers of $\delta$ for any $i \ge 1$. 
Assuming that $\omega_j = d_j \, \varepsilon + \mathcal{O} \left( \varepsilon^2 \right)$
holds with non-integer positive $d_j$ for some $j$, we arrive at a contradiction, since then
the $\varepsilon$--expansion in~(\ref{eq:Xex}) produces 
$\mathcal{B}_i(\delta_0,\delta,\varepsilon)/\mathcal{B}_0(\delta_0,\delta,\varepsilon)$
containing non-integer powers of $\delta$. The contradiction is obtained
also if we assume the existence of $\omega_j$ of a higher order than $\mathcal{O}(\varepsilon)$,
since  the $\varepsilon$--expansion of $\mathcal{B}_i(\delta_0,\delta,\varepsilon)/\mathcal{B}_0(\delta_0,\delta,\varepsilon)$
in~(\ref{eq:Xex}) produces powers of $\, \ln \delta$ in this case.
We can consider also a possibility that $\omega_j \sim \varepsilon^{\mu_j}$
with non-integer $\mu_j$ holds for some indices $j$. It does not lead to the consistency,
since the expansion of $F({\bf k},\delta,\varepsilon)$ contains only integer powers of $\varepsilon$,
coming from the $\varepsilon$--expansion of $u^*$ and diagrams evaluated at $c=1$.  
In such away, (\ref{eq:Xex}) can be consistent with the diagrammatic perturbative equation only 
if~(\ref{eq:omegal}) and~(\ref{eq:te1}) hold.

$X({\bf k};R_{s^m} H_0)=ck^2 + \mathcal{O} \left(\varepsilon^2 \right) = c_0 k^2 + \mathcal{O}(\varepsilon)$
holds according to~(\ref{eq:Xper}). Besides, the coefficient at $\ln k$ in~(\ref{eq:Xper})
is of order $\mathcal{O} \left( \varepsilon^2 \right)$, and  
it is independent of $\delta_0$ at this order. Eqs.~(\ref{eq:tea}) and~(\ref{eq:te2}) must hold
to satisfy these relations in~(\ref{eq:Xex}), performing the $\varepsilon$--expansion.
Finally, the equation 
\begin{equation}
- \eta + \varepsilon^2 \sum\limits_{j \ge 1} j \, b_j  \left( \frac{\delta}{1+\delta}  \right)^j 
= - \eta \, ( 1 + \delta )^2 + \mathcal{O} \left( \varepsilon^3 \right) \;,
\label{eq:te3}
\end{equation}
states the necessary consistency relation for the coefficient
at $\varepsilon^2 \ln k$ in~(\ref{eq:Xex}) and~(\ref{eq:Xper}), taking into account that 
$[(n+2)/(8c^3)] K_4^2 {u^*}^2 = c \, \eta + \mathcal{O} \left( \varepsilon^3 \right)$
holds according to~(\ref{eq:ufix}) and~(\ref{eq:eta2}).
From~(\ref{eq:te3}) we find $b_j = -\eta_2 \, (j+1)/j$, taking into
account that $\sum_{j=1}^{\infty} (j+1) x^j = 
\frac{\partial}{\partial x} \sum_{j=1}^{\infty} x^{j+1}= (2x-x^2)/(1-x)^2$ and setting here 
$x= \delta/(1+\delta)$.
\square

\subsection{A test of consistency for the correlation function}
\label{sec:furthertest}

\subsubsection{A diagrammatic consistency relation}

Based on the relations stated in the above theorem, here we perform some test 
for single RG transformation $R_s$ on the critical surface, starting with small 
$(u-u^*)/u^*=\delta_0$, defined as before.

We consider the asymptotic small--${\bf k}$ estimate of the inverse correlation function
defined as
\begin{equation}
X^{as}({\bf k};H_0) := A(\delta_0,\varepsilon) \, k^{2-\eta} 
\left\{ 1 + \sum\limits_i a_i(\delta_0,\varepsilon) k^{\omega_i}  \right\} 
\label{eq:Xsc00}
\end{equation}
obtained by neglecting the irrelevant (at $k \to 0$) terms with exponents 
$\hat \omega_i$ in~(\ref{eq:Xsc0}). Following~(\ref{eq:rescal}), it rescales exactly as
\begin{equation}
X^{as}({\bf k};R_{s} H_0) = A(\delta_0,\varepsilon) \, k^{2-\eta} 
\left\{ 1 + \sum\limits_i a_i(\delta_0,\varepsilon) s^{-\omega_i} k^{\omega_i} \right\} 
\label{eq:Xsc10}
\end{equation}
after the RG transformation $R_s$.
According to~(\ref{eq:te2}), we can write
\begin{equation}
a_i(\delta_0,\varepsilon) = b_i \, \varepsilon \left( \frac{\delta_0}{1+\delta_0}  \right)^i 
+ \mathcal{E}_i(\delta_0) \, \varepsilon^2 + \mathcal{O} \left( \varepsilon^3 \right) \qquad : \qquad i \ge 1 \;,
\end{equation}
where $\mathcal{E}_i(\delta_0)$ are some unknown coefficients.
Further on, we consider the expansion up to the $\varepsilon^2$ order for the ratio
$X^{as}({\bf k};R_{s} H_0) / X^{as}({\bf k};H_0)$, in which case these unknown terms cancel, i.~e.,
\begin{equation}
\frac{X^{as}({\bf k};R_s H_0)}{X^{as}({\bf k};H_0)} = 
1 - \varepsilon^2 \ln s \sum\limits_i i \, b_i \left( \frac{\delta_0}{1+ \delta_0} \right)^i
+ \mathcal{O} \left( \varepsilon^3 \right) = 1 + \eta \ln s \,
\left[ (1+ \delta_0)^2 -1 \right] + \mathcal{O} \left( \varepsilon^3 \right) \;.
\label{eq:att}
\end{equation}

In the following, we calculate the ratio $X^{as}({\bf k};R_{s} H_0) / X^{as}({\bf k};H_0)$
from the diagram equation~(\ref{eq:XXXX}) and compare the results.
In this case we need to include those terms, which
correspond to $X^{as}({\bf k};R_{s} H_0)$ and $X^{as}({\bf k};H_0)$ in
the $\varepsilon$--expansion. Thus, the
terms containing $k^2 (\ln k)^{\ell}$ with $\ell \ge 0$ must be included,
whereas those containing $k^{\mu} (\ln k)^{\ell}$ with $\mu >2$ have to be neglected.
Any quantity endowed with superscript ``\textit{as}" is calculated in this way. 
Furthermore, both $X^{as}({\bf k};R_{s} H_0)$ and $X^{as}({\bf k};H_0)$ are proportional
to the initial $c$ value $c_0$, so that we can calculate their ratio
at $c_0=1$ without loss of generality. According to the assumption
of the existence of limit 
$\lim\limits_{k \to 0} \left\{ \frac{1}{k^2} \left( \cdot  \right) \right\}$ for the diagrams entering
the RG flow equation for $c$, the term $\theta({\bf k})$, represented
by \mbox{\vverticetet,}  tends to zero faster than $\sim k^2$ at $k \to 0$. Therefore, 
it is neglected. Thus, the \linebreak
\vspace*{-8pt}

\noindent
diagram equation at $c_0=1$ yields
\begin{eqnarray}
X^{as}({\bf k};H_0) &=& k^2 - \frac{n+2}{2} {u^*}^2 (1+ \delta_0)^2 \left( \vverticerrksl \right)^{as} 
+ \mathcal{O} \left( \varepsilon^3 \right) \\
X^{as}({\bf k};R_{s} H_0) &=& c k^2 - \frac{n+2}{2} {u^*}^2 
(1+ \delta_0)^2 \left[ \left( \vverticerrksl \right)^{as} + \right. \nonumber \\
&&\left. 3 \left\{ \left( \vverticeryksl \right)^{as} + \left( \vverticerxksl \right)^{as} \right\}  \right] 
+ \mathcal{O} \left( \varepsilon^3 \right) \label{two} \;,
\end{eqnarray}
where $q \in [0,\Lambda]$ holds for the solid lines and $q \in [\Lambda,\Lambda s]$ -- for the dotted lines.
Here we have set $u=u^*(1+\delta_0)$ in both expressions, since $u^2$ after the RGT is varied only by
$\mathcal{O} \left( \varepsilon^3 \right)$. The quantity $\left( \vverticerrksl \right)^{as}$
cancels in  $X^{as}({\bf k};R_{s} H_0) / X^{as}({\bf k};H_0)$, calculated up to the $\varepsilon^2$ order.
The renormalized $c$ in~(\ref{two}), calculated from~(\ref{eq:c}), is
\begin{equation}
c = 1 - \eta \, \ln s - \frac{n+2}{2} {u^*}^2 (1+ \delta_0)^2 
\lim\limits_{k \to 0} \left. \left\{ \frac{1}{k^2} \left( \vverticerrksl \right) \right\} \right|_{d=4} 
+ \mathcal{O} \left( \varepsilon^3 \right)
\label{eq:ccc}
\end{equation}
with $q \in [\Lambda/s,\Lambda]$ corresponding to the solid lines in the diagram evaluated at $c=1$. If this diagram is
completed by $3 \left( \vverticeryksl + \vverticerxksl \right)$ with $q \in [\Lambda/s,\Lambda]$
for the solid lines and $q \in [\Lambda,\infty]$ for the dotted lines, then the obtained expression 
$\lim\limits_{k \to 0} \left. \left\{ \frac{1}{k^2} \left( \cdot  \right) \right\} \right|_{d=4}$
is exactly $\propto \ln s$, as it follows from the diagram identity~(\ref{eq:identi}) provided
that this limit, as well as the limit in~(\ref{eq:ccc}), exists. Recall that we test the scenario when this expected property holds. 
The proportionality coefficient is $-K_4^2/4$, in accordance with  the calculations at $s \to \infty$
in Sec.~\ref{sec:fixp}. Setting further $\Lambda = 1$ for simplicity, we can write
\begin{equation}
\lim\limits_{k \to 0} \left. \left\{ \frac{1}{k^2} \left( \vverticerrksl \right) \right\} \right|_{d=4}
= - \frac{K_4^2}{4} \ln s - 3 \lim\limits_{k \to 0} \frac{D(1/s,\infty;{\bf k})}{k^2} \;,
\label{limitsxx}
\end{equation}
where
\begin{equation}
D(\Lambda_1,\Lambda_2;{\bf k}) :=   
\vverticeryksl + \vverticerxksl 
\label{eq:DDD}
\end{equation}
with $q \in [\Lambda_1,1]$ for the solid lines and $q \in [1,\Lambda_2]$ for the dotted lines.

Summarizing all the derived here diagrammatic relations for   
$X^{as}({\bf k};R_{s} H_0)$ and $X^{as}({\bf k};H_0)$, and also 
$[(n+2)/(8c^3)] K_4^2 {u^*}^2 = c \, \eta + \mathcal{O} \left( \varepsilon^3 \right)$, we finally obtain
\begin{equation}
\frac{X^{as}({\bf k};R_s H_0)}{X^{as}({\bf k};H_0)} = 
 1 + \eta \left\{ 
\left[ (1+ \delta_0)^2 -1 \right]  \ln s 
- \frac{12}{K_4^2} (1 + \delta_0)^2 \, \frac{D(0,s;{\bf k}) 
- D(1/s,\infty;{\bf k})}{k^2}  \right\} + \mathcal{O} \left( \varepsilon^3 \right)
\label{eq:attp}
\end{equation}
at $k \to 0$.
It is evident that~(\ref{eq:att}) and~(\ref{eq:attp}) agree with each other if and only if
\begin{equation}
D^{as}(0,s;{\bf k}) = D^{as}(1/s,\infty;{\bf k})  \qquad \mbox{at} \quad d=4 \;,
\label{consistency}
\end{equation}
holds for $s>1$, at the condition that the limits in~(\ref{limitsxx}) exist, as expected.

\subsubsection{Scaling relation}
\label{sec:scaling}

Here we consider the exact scaling relation
\begin{equation}
 s^{2-2 \varepsilon} [D(0,\infty;{\bf k}/s) - D(1/s,\infty;{\bf k}/s)]
= [D(0,\infty;{\bf k}) - D(0,s;{\bf k})] \;.
\label{scaling}
\end{equation}
It is proven as follows.
Considering $D(0,\infty;{\bf k}/s) - D(1/s,\infty;{\bf k}/s)$, we decompose 
the $q \in [0,1]$ lines of the diagrams of $D(0,\infty;{\bf k}/s)$ in
two lines with $q \in [0,1/s]$ and $q \in [1/s,1]$. 
Similarly, for $D(0,\infty;{\bf k}) - D(0,s;{\bf k})$, we decompose 
the $q \in [1,\infty]$ lines of the diagrams of $D(0,\infty;{\bf k})$ in
$q \in [1,s]$ and $q \in [s,\infty]$.
Then, it is easy to see that  $D(0,\infty;{\bf k}/s) - D(1/s,\infty;{\bf k}/s)$
is represented by the sum of all such diagrams with three kind of lines, having
wave vectors within $[0,1/s]$, $[1/s,1]$ and $[1,\infty]$, respectively,
which contain at least one line within $[0,1/s]$ and at least one line
within $[1,\infty]$. Similarly, $D(0,\infty;{\bf k}) - D(0,s;{\bf k})$
is represented by the sum of all such diagrams with three kind of lines, having
wave vectors within $[0,1]$, $[1,s]$ and $[s,\infty]$, respectively,
which contain at least one line within $[0,1]$ and at least one line
within $[s,\infty]$. Rescaling the wave vectors in the diagrams of
$D(0,\infty;{\bf k}/s) - D(1/s,\infty;{\bf k}/s)$ by factor $s$ and
multiplying the result with $s^{2- 2 \varepsilon}$, we obtain just
the diagrams of $D(0,\infty;{\bf k}) - D(0,s;{\bf k})$. It proves~(\ref{scaling}).

According to~(\ref{scaling}), the condition~(\ref{consistency})
is satisfied if $D(0,\infty;{\bf k}) - D(1/s,\infty;{\bf k}) \propto k^2$
holds at $k \to 0$ in four dimensions. However, allowing a possibility that
$D(0,\infty;{\bf k})/k^2$ diverges at $k \to 0$, this condition might
be not satisfied.

\subsubsection{Evaluation of 
$D(1/s,\infty;{\bf k})$ in four dimensions}
\label{D1s}

In the following we consider the case $s=1+ds$, where $ds$ is small and positive,
and evaluate the diagrams of $D(1/s,\infty;{\bf k})$ at $d=4$. In particular,
the diagram \mbox{\vverticerxksl,} having wave vectors within $[1/s,1]$ for the solid line and $[1,\infty]$
for the dotted lines, can be calculated as follows:
\begin{equation}
 \vverticerxksl = \frac{2 K_4}{\pi} \int\limits_{1/s}^1 q dq \int\limits_0^{\pi}
[J(\mid {\bf q + k} \mid) - J(q)] \sin^2 \theta d \theta \;,
\end{equation}
where $\theta$ is the angle made by vectors ${\bf q}$ and ${\bf k}$ 
(i.~e., $\mid {\bf q + k} \mid^2 = q^2 +2 kq \cos \theta + k^2$) and
\begin{eqnarray}
 J(q) &=& \vverticeuyq = \frac{2 K_4}{\pi} \int\limits_1^{1+q} k dk \int\limits_0^{\hat{\theta}(q,k)}
\frac{\sin^2 \theta d \theta}{q^2 + 2kq \cos \theta + k^2} - K_4 \ln(1+q) \nonumber \\
&+& \frac{2 K_4}{\pi} \int\limits_{1+q}^{\infty} k dk \int\limits_0^{\pi}
\left( \frac{1}{q^2 + 2kq \cos \theta + k^2} - \frac{1}{k^2} \right) \sin^2 \theta d \theta 
\quad \mbox{for} \quad 0<q<2 \;,
\end{eqnarray}
where 
\begin{equation}
 \hat{\theta}(q,k) = \arccos \left( \frac{1-q^2-k^2}{2kq} \right) \;.
\end{equation}
Using the small--${\bf k}$ expansion
\begin{equation}
 J(\mid {\bf q + k} \mid) = J(q) + \frac{d J}{dq} \left( k \cos \theta + \frac{k^2}{2q} \sin^2 \theta \right)
+ \frac{1}{2} \frac{d^2 J}{dq^2} \, k^2 \cos^2 \theta + \mathcal{O} \left( k^3 \right) \;,
\end{equation}
we obtain
\begin{equation}
 \left( \vverticerxksl \right)^{as} = k^2 \times \frac{K_4}{8} \int\limits_{1/s}^1
\left( 3J'(q) + q J''(q) \right) dq \;. 
\label{d1}
\end{equation}
This method is certainly valid for small $ds$, since in this case $J(q)$ is an analytic function
of $q$ within the whole integration region, i.~e., around $q=1$.

The diagram $\vverticeryksl$ is calculated similarly:
\begin{equation}
 \left( \vverticeryksl \right)^{as} = k^2 \times \frac{K_4}{8} \int\limits_{1/s}^1
\left( 3 \tilde{J}'(q) + q \tilde{J}''(q) \right) dq \;, 
\label{d2}
\end{equation}
where
\begin{equation}
\tilde{J}(q) = \vverticeuxq = \frac{2K_4}{\pi} \int\limits_{1/s}^1 k dk \int\limits_0^{\hat{\theta}(q,k)}
\frac{\sin^2 \theta d \theta}{q^2 + 2kq \cos \theta + k^2} \;.
\end{equation}
Again, this method is valid for small $ds$, since $\tilde{J}(q)$ is an analytic function of $q$
around $q=1$.
Summing up~(\ref{d1}) and~(\ref{d2}) and expanding in a Taylor series of $ds$ at $ds \to 0$, we obtain
\begin{eqnarray}
 D^{as}(1/s,\infty;{\bf k}) &=& k^2 \times \frac{K_4}{8} \left\{ (3J'(1) + J''(1)) \, ds
+ \lim\limits_{ds \to 0} \left( \frac{3 \tilde{J}'(1) + \tilde{J}''(1)}{ds} \right)
\times (ds)^2 \right. \nonumber \\ 
&-& \left. \left( 3J'(1)+3J''(1)+ \frac{1}{2} J'''(1) \right) \, (ds)^2  
+ \mathcal{O} \left( (ds)^3 \right) \right\} \;.
\end{eqnarray}
We have evaluated $J(q)$, $\tilde{J}(q)$ and their derivatives at $q=1$:
$J(1) \simeq -0.214068 K_4$, $J'(1) \simeq -0.217996 K_4$, 
$J''(1) \simeq -0.012680 K_4$, $J'''(1) \simeq -0.0177 K_4$,
$\lim\limits_{ds \to 0} (\tilde{J}'(1)/ds) = \left( \frac{\sqrt{3}}{\pi} - \frac{2}{3} \right) K_4$
and $\lim\limits_{ds \to 0} (\tilde{J}''(1)/ds) \simeq -0.021539 K_4$.
Most of these values have been obtained calculating numerically the
corresponding integrals. These values satisfy certain expected relation, given in the next subsection,
and, thus, must be correct within the given numerical accuracy.

\subsubsection{Consistency relation for 
$D^{as}(0,s;{\bf k})$}

According to calculations in the previous subsection, $D(1/s,\infty;{\bf k})$ at small $ds$ is asymptotically
proportional to $k^2$ at $k \to 0$ with non-zero proportionality coefficient. Based on this
fact, and using diagram identities, here we derive a necessary consistency relation for $D^{as}(0,s;{\bf k})$ at 
which~(\ref{consistency}) can hold. As in previous section, here any diagram relation refers to the case $d=4$.

Let us introduce the notation
\begin{equation}
 Q(\Lambda;{\bf k}) := \vverticerrksl \;,
\label{def1}
\end{equation}
where the wave vectors of the coupling lines are within $[0,\Lambda]$.
In the following, we use the relation
\begin{equation}
 f(k) =  \vverticerrksl  
+3 \left( \vverticeryksl  + \vverticerxksl \right) 
\equiv Q(1;{\bf k}) + 3 D(0,\infty;{\bf k}) = \mathcal{B} \, k^2 + \frac{1}{4} K_4^2 \, k^2 \ln k 
\quad \mbox{at} \quad k \to 0 \;,
\label{eqqq}
\end{equation}
which follows from the equations of Sec.~\ref{sec:corfix}, taking into account
that \vverticetet decays faster than

\vspace*{3pt}
\noindent
$k^2$ at $k \to 0$ (according to the assumed expected properties
-- see Sec.~\ref{subsec:RGf}). Note that the wave vectors of the solid lines are within $[0,1]$ and 
those of the dotted lines are within $[1,\infty]$ in~(\ref{eqqq}). 

Since $D(1/s,\infty;{\bf k}) \propto k^2$ holds at $k \to 0$, 
we have $s^2 D(1/s,\infty;{\bf k}/s) = D(1/s,\infty;{\bf k})$ in this limit.
Assuming that~(\ref{consistency}) holds, we have $D(0,s;{\bf k}) = D(1/s,\infty;{\bf k})$ at $k \to 0$.
Inserting these relations into~(\ref{scaling}) at $\varepsilon=0$, we obtain
the asymptotic relation $s^2 D(0,\infty;{\bf k}/s) = D(0,\infty;{\bf k})$ for small $k$, 
which means that $D(0,\infty;{\bf k}) \propto k^2$ holds at $k \to 0$.
According to the latter, Eq.~(\ref{eqqq}) gives us
\begin{equation}
Q(1;{\bf k}) = \mathcal{B}_1 \, k^2 + \frac{1}{4} K_4^2 \, k^2 \ln k 
\quad \mbox{at} \quad k \to 0 \;,
\label{uuu}
\end{equation}
where $\mathcal{B}_1$ is some constant. Using~(\ref{uuu}) and the scaling relation
\begin{equation}
 Q(s;{\bf k}) = s^2 Q(1;{\bf k}/s) \;,
\end{equation}
we obtain
\begin{equation}
 Q(1+ds;{\bf k}) - Q(1;{\bf k}) =
- \frac{1}{4} K_4^2 \left( ds - \frac{1}{2} (ds)^2 + \mathcal{O} \left( (ds)^3 \right)
\right) \times k^2 \quad \mbox{at} \quad k \to 0 \;.
\label{qdiff}
\end{equation}

Consider now the diagram identity
\begin{equation}
 \vverticerrksl +3 \left( \vverticeryksl  + \vverticerxksl \right)  =
\vverticecc - \vverticec \;,
\label{identity}
\end{equation}
where $[0,1]$ corresponds to the solid lines, $[1,1+ds]$ -- to the
dotted lines and $[0,1+ds]$ -- to the dashed lines. The last diagram 
in~(\ref{identity}) can be evaluated by the method used in Sec.~\ref{D1s},
leading to the conclusion that it is $\propto k^2 (ds)^3$ at $k \to 0$
and small $ds$. Hence, according to the definitions~(\ref{def1}) and~(\ref{eq:DDD}),
the identity~(\ref{identity}) yields
\begin{equation}
 3 D(0,1+ds;{\bf k}) = Q(1+ds;{\bf k}) - Q(1;{\bf k}) + \mathcal{O} \left( k^2 (ds)^3 \right) \;.
\end{equation}
Inserting here~(\ref{qdiff}), we obtain
\begin{equation}
 D^{as}(0,1+ds;{\bf k}) =  \frac{1}{12} K_4^2 \left( -ds + \frac{1}{2} (ds)^2 + 
\mathcal{O} \left( (ds)^3 \right) \right) \times k^2 
\label{consistency1}
\end{equation}
at small $ds$. This relation has been derived assuming~(\ref{consistency}). Thus,
this is a necessary condition at which~(\ref{consistency}) can hold. 
According to~(\ref{consistency}) and ~(\ref{consistency1}), we must have also
\begin{equation}
 D^{as}(1/s,\infty;{\bf k}) =  \frac{1}{12} K_4^2 \left( -ds + \frac{1}{2} (ds)^2 + 
\mathcal{O} \left( (ds)^3 \right) \right) \times k^2  
\label{consistency2}
\end{equation}
for $s=1+ds$. The latter relation is really satisfied
in accordance with the numerical values listed at the end of Sec.~\ref{D1s}.

\subsubsection{Evaluation of 
$D(0,s;{\bf k})$ in four dimensions}

One of the diagrams of $D(0,s;{\bf k})$ at $d=4$ is
\begin{equation}
 \vverticeryksl = \frac{2 K_4}{\pi} \int\limits_1^s q dq \int\limits_0^{\pi}
[I(\mid {\bf q + k} \mid) - I(q)] \sin^2 \theta d \theta \;,
\end{equation}
where the wave vectors of solid and dotted lines are within
$[0,1]$ and $[1,s]$, respectively. Here
\begin{equation}
 I(q) = \vverticeuq = \frac{2K_4}{\pi} \int\limits_{q-1}^1
k dk \int\limits_{\hat{\theta}(q,k)}^{\pi} \frac{\sin^2 \theta d \theta}
{q^2 + 2kq \cos \theta + k^2} \quad \mbox{for} \quad 1<q<2 \;.
\label{Iq}
\end{equation}
A problem here is that $I(q)$, apparently, has a singularity at $q=1$. 
Considering the limit $\lim\limits_{\Delta \to 0} \lim\limits_{k \to 0}$,
the contribution of the region $1 + \Delta < q< s$ (for $1<s<2$)
can be evaluated by the small--${\bf k}$ expansion, as in Sec.~\ref{D1s},
whereas the contribution of the region $1 < q < 1 + \Delta$ has to be
considered separately. It yields
\begin{equation}
 \vverticeryksl = k^2 \times \frac{K_4}{8} \int\limits_1^s
\left( 3I'(q) + q I''(q) \right) dq + \mathcal{R}_1(k) \quad \mbox{at} \quad k \to 0 \;,
\label{eee}
\end{equation}
where $\mathcal{R}_1(k)$ is the contribution of $1  < q< 1+\Delta$ region.
It is important in our consideration if $\lim_{k \to 0} \mathcal{R}_1(k)/k^2 \ne 0$.
In any case, $\mathcal{R}_1(k)$ is independent of $s$, since $I(q)$ is $s$--independent
and also the borders of the region $1  < q< 1+\Delta$ do not contain $s$.
The ``regular'' contribution of the region $1 + \Delta < q< s$ can be represented
by the given integral, since $I'(q)$ is continuous and finite at $q=1$ and
$I''(q)$ is at least integrable. Moreover, it can be shown that 
${I''}_+(1) := \lim_{q \to 1} I''(q)$ (where ``$+$'' means tending to $1$ from above, $q>1$) 
has a finite value.

According to~(\ref{Iq}), for $1<q<2$, we have 
\begin{equation}
 I'(q) = \frac{K_4}{\pi} \left\{ -\int\limits_{q-1}^1 \sqrt{1- a^2(q,k)} \, b(q,k) dk 
- 4 \int\limits_{q-1}^1 k dk \int\limits_{\hat{\theta}(q,k)}^{\pi} 
\frac{(q+k \cos \theta) \sin^2\theta d\theta}{\left(q^2+2kq \cos \theta +k^2 \right)^2} \right\} \;,
\label{Iprim}
\end{equation}
\begin{eqnarray}
 && \hspace*{-6ex} I''(q) = \frac{K_4}{\pi} \left\{ -\frac{1}{2} \int\limits_{q-1}^1 
\frac{a(q,k) b^2(q,k)}{\sqrt{1- a^2(q,k)}} \, \frac{dk}{k} 
+ 2 \int\limits_{q-1}^1 \sqrt{1- a^2(q,k)} \left[\frac{1-k^2}{q^3} +  
b(q,k) \Big( q+k \, a(q,k) \Big) \right] dk \right. \nonumber \\
&-& \left. 4 \int\limits_{q-1}^1 k dk \int\limits_{\hat{\theta}(q,k)}^{\pi} 
\left(\frac{1}{\left(q^2+2kq \cos \theta +k^2 \right)^2} 
- \frac{4 (q+k \cos \theta )^2 }{\left(q^2+2kq \cos \theta +k^2 \right)^3} \right)
\sin^2 \theta d \theta \right\} \;,
\label{I2prim}
\end{eqnarray}
where $a(q,k)= \frac{1-q^2-k^2}{2kq}$ and $b(q,k)= \left(1 + \frac{1-k^2}{q^2} \right)$.
The integrand function is always finite for $I'(q)$, but it diverges at $q = 1$ and
$\theta \to \pi$ in the last integral of $I''(q)$. Therefore we have analysed the contribution
of the region $\theta_0 < \theta < \pi$ and $k_0 <k< 1$ to this integral, denoted as ${I''}_{\mathrm{sing}}(q)$,
where $\theta_0$ and $k_0$ are constants. These are chosen such that $\pi - \theta_0$ is small,
but $k_0$ can be any positive constant smaller than $1$ if $q \to 1$.
Since the remaining contribution to $I''(q)$ is certainly finite at $q \to 1$, 
it is clear that ${I''}_+(1)$ is finite if  ${I''}_{\mathrm{sing}}(q)$
is bounded at $q \to 1$. We have 
\begin{equation}
 {I''}_{\mathrm{sing}}(q) = \grave{I}''(q;k_0,1,\theta_0) \;,
\end{equation}
where
\begin{equation}
 \grave{I}(q;k_1,k_2,\theta_0) = \frac{2K_4}{\pi} \int\limits_{k_1}^{k_2}
k dk \int\limits_{\theta_0}^{\pi} \frac{\sin^2 \theta d \theta}
{q^2 + 2kq \cos \theta + k^2} \quad \;.
\label{Igq}
\end{equation}
The integration in~(\ref{Igq}) can be at least partly performed changing the integration order.
It yields
\begin{eqnarray}
 \grave{I}(q;k_1,k_2,\theta_0) &=&
\frac{2 K_4}{\pi} \int\limits_{\theta_0}^{\pi}
\left[ \frac{1}{2} \ln \left( \frac{(k_2+q \cos \theta)^2 + q^2 \sin^2 \theta}
{(k_1+q \cos \theta)^2 + q^2 \sin^2 \theta} \right) \right. \label{Igq1} \\ 
&-& \left. \cot \theta \left(\arctan \left(\frac{k_2+q \cos \theta}{q \sin \theta} \right) 
-\arctan \left(\frac{k_1+q \cos \theta}{q \sin \theta} \right)
\right) \right] \sin^2 \theta d \theta \nonumber \;.
\end{eqnarray}
From~(\ref{Igq1}) we find
\begin{eqnarray}
 \grave{I}''(q;k_1,k_2,\theta_0) &=& \frac{2K_4}{\pi} \int\limits_{\theta_0}^{\pi}
\left( \frac{k_2^2 - q^2 - 4k_2 \cos \theta (q+ k_2 \cos \theta)}
{\left(q^2+2k_2 q \cos \theta + k_2^2 \right)^2} \right. \nonumber \\ 
&-& \left. \frac{k_1^2 - q^2 - 4k_1 \cos \theta (q+ k_1 \cos \theta)}
{\left(q^2+2k_1 q \cos \theta + k_1^2 \right)^2}  \right) \sin^2 \theta d \theta \;.
\label{Igrave}
\end{eqnarray}
${I''}_{\mathrm{sing}}(q)$ is evaluated by setting here $k_1=k_0$ and $k_2=1$.
In this case, the second term gives constant contribution at $q \to 1$,
whereas the contribution of the first term for small $\hat{\Delta} = q-1$ and
small $\delta_0 = \pi - \theta_0$ can be evaluated as
\begin{equation}
 {I''}_{\mathrm{sing}}(q) \simeq 
\frac{2K_4}{\pi} \int\limits_0^{\delta_0}
\frac{2(\hat{\Delta} + \delta^2) \delta^2 d \delta}{\left(\hat{\Delta}^2 + \delta^2 \right)^2} \;.
\end{equation}
The used here approximation for the integrand function has vanishing
relative error at $\hat{\Delta} \to 0$ and $\delta = \pi - \theta \to 0$. Since the integrand function
is positively defined, this approximation gives vanishing relative
error for the integral at $\hat{\Delta} \to 0$ and $\delta \to 0$.
Taking into account that 
\begin{equation}
 \int\limits_0^{\theta_0} \frac{2 \hat{\Delta} \delta^2 d \delta}{\left(\hat{\Delta}^2 + \delta^2 \right)^2}
= \arctan \left( \frac{\theta_0}{\hat{\Delta}} \right) - \frac{\theta_0 \hat{\Delta}}{\hat{\Delta}^2 + \theta_0^2} \;, 
\end{equation}
and considering the limit $\hat{\Delta} \to 0$, we can conclude that ${I''}_+(1)$ is, indeed, finite.
Moreover, numerical analysis shows that $I''(q)$ is almost linear function of $q$ for small
and positive $q - 1$, indicating that ${I'''}_+(1)$ is also finite.

It can be shown that the contributions of the vicinities of singular
points (like $\mathcal{R}_1(k)$) are important at least for some of the considered here diagrams.
In particular, different estimates of $\left( \vverticerxksl \right)^{as}$ for $s=1+ds$ are inconsistent
with each other if these contributions are neglected. One of the possibilities is
\begin{eqnarray}
 \vverticerxksl &=&  
\frac{2 K_4}{\pi} \int\limits_1^{1+ds} q dq \int\limits_0^{\pi}
[\tilde{I}(\mid {\bf q + k} \mid) - \tilde{I}(q)] \sin^2 \theta d \theta \nonumber \\ 
&=& k^2 \times \frac{K_4}{8} \int\limits_1^{1+ds} 
\left(3 \tilde{I}'(q) + q \tilde{I}''(q) \right) dq + \mathcal{R}_2(k,s) \quad \mbox{at} \quad k \to 0 \;,
\label{viensx}
\end{eqnarray}
where 
\begin{equation}
\tilde{I}(q) = \vverticeuxq = \frac{2K_4}{\pi} \int\limits_1^{1+ds}
k dk \int\limits_{\hat{\theta}(q,k)}^{\pi} \frac{\sin^2 \theta d \theta}
{q^2 + 2kq \cos \theta + k^2} \quad \mbox{for} \quad 1<q<1+ds \;.
\label{viensy}
\end{equation}
The term $\mathcal{R}_2(k,s)$ is the contribution of vicinities of singular points
$q=1$ and $q=1+ds$.
Expressions for $\tilde{I}'(q)$ and $\tilde{I}''(q)$ are the same as those for
$I'(q)$ and $I''(q)$ (Eqs.~(\ref{Iprim}) and~(\ref{I2prim})) with the only
difference that the integration limits for $k$ now are from $1$ to $1+ds$. 

Another interesting possibility is to use the symmetry of the diagram \vverticerxksl
with respect to the dotted lines, i.~e., $\vverticerxksl = 2 \widetilde{\vverticerxksl \hspace*{8pt}}$,
where ``$\widetilde{\quad}$'' means the constraint that the wave 

\vspace*{2pt}
\noindent
vector of the upper dotted line
is larger in magnitude than that of the lower dotted line. It yields
\begin{eqnarray}
&& \vverticerxksl = \frac{4 K_4}{\pi} \int\limits_1^{1+ds} q dq \int\limits_0^{\pi}
[\bar{I}(\mid {\bf q + k} \mid,q) - \bar{I}(q,q)] \sin^2 \theta d \theta \label{2var} \\
&&= k^2 \times \lim\limits_{\Delta \to 0} \left[ \frac{K_4}{4} \int\limits_1^{1+ds} 
\left(3 \left. \frac{\partial \bar{I}(q,p)}{\partial q} \right|_{p=q-\Delta} + 
q \left. \frac{\partial^2 \bar{I}(q,p)}{\partial q^2} \right|_{p=q-\Delta} \right) dq \right] 
+ \mathcal{R}_2^*(k,s) \hspace*{7pt} \mbox{at} \hspace*{7pt} k \to 0 \nonumber \;,
\end{eqnarray}
where
\begin{equation}
\bar{I}(q,p) = \frac{2K_4}{\pi} \int\limits_1^p
k dk \int\limits_{\hat{\theta}(q,k)}^{\pi} \frac{\sin^2 \theta d \theta}{q^2 + 2kq \cos \theta + k^2} \;.
\label{2vary}
\end{equation}
The term  $\mathcal{R}_2^*(k,s)$ is the contribution of the region $0<q_1-q_2 < \Delta$, where ${\bf q}_1$ is 
the wave vector of the upper dotted line and ${\bf q}_2$ -- of the lower dotted line.
Expressions for $\partial \bar{I}(q,p)/\partial q$ and $\partial^2 \bar{I}(q,p)/\partial q^2$ are the same as those for
$I'(q)$ and $I''(q)$ (Eqs.~(\ref{Iprim}) and~(\ref{I2prim})) with the only
difference that the integration limits for $k$ now are from $1$ to $p$.

Using the same estimation techniques as for the diagram \mbox{\vverticeryksl,}
we find that 
\begin{eqnarray}
 \vverticerxksl &=& - k^2 \times \frac{K_4^2}{4} ds + \mathcal{R}_2(k,s) 
+ o(ds) \quad \mbox{at} \quad k \to 0 \;, \label{viens} \\
 \vverticerxksl &=& \hspace*{16ex} \mathcal{R}_2^*(k,s) 
+ o(ds)  \;, \label{divi}
\end{eqnarray}
where~(\ref{viens}) follows from~(\ref{viensx}) and (\ref{viensy}), 
whereas~(\ref{divi}) -- from~(\ref{2var}) and (\ref{2vary}).
The contribution of order $\mathcal{O}(ds)$ in~(\ref{viens})
comes from 
$$
k^2 \times \frac{K_4}{8} \int\limits_1^{1+ds} q \grave{I}''(q;1,1+ds,\theta_0) dq \;,
$$
where $\grave{I}''(q;1,1+ds,\theta_0)$ is the main contribution to $\tilde{I}''(q)$, given
by~(\ref{Igrave}), which surprisingly appears to be of order unity, i.~e., $-2 K_4 + \mathcal{O}(ds)$
for $1<q<1+ds$. As regards $\partial^2 \bar{I}(q,p)/\partial q^2$, the contribution
of similar origin is $\lim_{\Delta \to 0} \grave{I}''(q;1,q-\Delta,\theta_0)$, which is
not of order unity, since the main terms of~(\ref{Igrave}) cancel in this case.
It follows from~(\ref{viens}) and~(\ref{divi}) that at least one of the terms
$\mathcal{R}_2(k,s)$ and $\mathcal{R}_2^*(k,s)$ is important at $k \to 0$.  
The reason why such a term can appear to be important is that, for any given ${\bf k}$, 
the small--${\bf k}$ expansion might be convergent or valid only at large enough ($k$--dependent) 
distances from the singular point(s).

Further on, we will use~(\ref{2var})--(\ref{2vary}), since one can prove the following
statement for  $\mathcal{R}_2^*(k,s)$:

\vspace*{2ex}

\textbf{Lemma.} \hspace*{0.5ex} \textit{If, in four dimensions, 
$\mathcal{R}_2^*(k,s)$ is proportional to $k^2$ 
at $k \to 0$ (with non-zero proportionality coefficient), then it is also
proportional to $\ln s$, i.~e., $\mathcal{R}_2^*(k,s) \propto k^2 \ln s$ at $k \to 0$.}

\vspace*{1ex}

\textit{Proof.} \hspace*{0.5pt} Recall that $\mathcal{R}_2^*(k,s)$, evaluated in the limit
$\lim_{\Delta \to 0} \lim_{k \to 0}$, is given by $2 \vverticerxksl$, where the wave vector $1<q_1<s$ 
corresponds to the upper dotted line and $q_2$ -- to the lower dotted line with a
constraint $0<q_1-q_2<\Delta$. 
Let us denote by $\hat{\mathcal{R}}(k;\Delta,s_1,s_2,\Lambda_1,\Lambda_2)$
the term $2 \vverticerxksl$ with $s_1<q_1<s_2$, $0<q_1-q_2<\Delta$ and
wave vectors within $[\Lambda_1,\Lambda_2]$ for the solid line.
Then we have $\mathcal{R}_2^*(k,s) = \hat{\mathcal{R}}(k;\Delta,1,s,0,1) =
 \hat{\mathcal{R}}(k;\Delta,1,s,0,1/s) + \hat{\mathcal{R}}(k;\Delta,1,s,1/s,1)$.
The latter term can be well estimated by the small--${\bf k}$ expansion, used throughout here,
first calculating the block \mbox{\vverticeuxqq,} which is an analytic function
of $q$ within $q \in [1/s,1]$ for small $\Delta$. Besides, it vanishes at $\Delta \to 0$.
Therefore, we have $\hat{\mathcal{R}}(k;\Delta,1,s,1/s,1) = A(\Delta) \, k^2$ at $k \to 0$,
where $A(\Delta) \to 0$ at $\Delta \to 0$. Consequently, this term is negligible in
the considered limit at the conditions of the \textit{Lemma}. Rescaling
in this case yields $\hat{\mathcal{R}}(k;\Delta,1,s,0,1/s) = 
s^{-2} \hat{\mathcal{R}}(sk;s \Delta,s,s^2,0,1) = \hat{\mathcal{R}}(k;\Delta,s,s^2,0,1)$
at $k \to 0$.
Here $s \Delta$ is replaced by $\Delta$, since certain limit exists at $\Delta \to 0$
according to the conditions of the \textit{Lemma}.
Summarising these relations, we have
\begin{equation}
 \mathcal{R}_2^*(k,s) = \hat{\mathcal{R}}(k;\Delta,1,s,0,1)=\hat{\mathcal{R}}(k;\Delta,s,s^2,0,1)
\label{aaa}
\end{equation}
in the limit 
$\lim_{\Delta \to 0} \lim_{k \to 0}$. We can also use the diagram identities
\begin{equation}
 \hat{\mathcal{R}}(k;\Delta,1,s,0,1)+\hat{\mathcal{R}}(k;\Delta,s,s^2,0,1)
= \hat{\mathcal{R}}(k;\Delta,1,s^2,0,1) = \mathcal{R}_2^*(k,s^2) \;.
\label{bbb}
\end{equation}
Combining~(\ref{aaa}) and~(\ref{bbb}), we obtain
\begin{equation}
 \mathcal{R}_2^*(k,s^2) = 2 \, \mathcal{R}_2^*(k,s) \quad \mbox{for} \quad s>1
\end{equation}
and, consequently, $\mathcal{R}_2^*(k,s) \propto \ln s$. \square

Summarising the estimation of the diagrams of $D(0,s;{\bf k})$,
we conclude that the necessary consistency condition~(\ref{consistency1})
can be satisfied only if $\lim_{k \to 0} \mathcal{R}_1(k)/k^2 = 0$ holds,
since $\mathcal{R}_1(k)$ is a quantity of order $(ds)^0$ (independent of $ds$).
Furthermore, $\mathcal{R}_2^*(k,s) = C k^2 \ln s$ must hold to satisfy this
condition, where $C$ is a constant. Here we allow also a possibility $C=0$,
referring to the case where  $\lim_{k \to 0} \mathcal{R}_2^*(k)/k^2 = 0$.
If the latter limit is non-zero, then~(\ref{consistency1})
requires that  $\mathcal{R}_2^*(k) \propto k^2$ at $k \to 0$ holds, i.~e.,
the conditions of the \textit{Lemma} are satisfied and, therefore, 
$\mathcal{R}_2^*(k,s) = C k^2 \ln s$ holds with $C \ne 0$.
Taking into account these facts and the discussed here estimations of the
diagrams \vverticeryksl and \vverticerxksl (see Eqs.~(\ref{eee}) and~(\ref{2var})),
the condition~(\ref{consistency1})

\vspace*{2pt}
\noindent
 can be satisfied only if  
$D^{as}(0,s;{\bf k})$ has the form
\begin{eqnarray}
 D^{as}(0,s;{\bf k}) &=& k^2 \times \frac{K_4}{8} \left\{
(3I'(1) + {I''}_+(1)) \, ds + \left( 2 {I''}_+(1) + \frac{1}{2} {I'''}_+(1) \right)
\, (ds)^2 \right. \nonumber \\
&+& \left. C_1 \ln (1+ds) + C_2 (ds)^2  + \mathcal{O}\left( (ds)^3 \right) \right\} \;,
\label{ccc}
\end{eqnarray}
where $C_1 = 8C/K_4$ and
\begin{equation}
 C_2 = 2 \lim\limits_{ds \to 0} 
\left\{ \frac{1}{(ds)^2} \, \lim\limits_{\Delta \to 0} \int\limits_1^{1+ds} 
\left(\left. 3 \frac{\partial \bar{I}(q,p)}{\partial q} \right|_{p=q-\Delta} + 
q \left. \frac{\partial^2 \bar{I}(q,p)}{\partial q^2} \right|_{p=q-\Delta} \right) dq \right\} \;.
\end{equation}
Moreover, it is necessary that the expansion coefficients at $ds$ and $(ds)^2$ are consistent with
those in~(\ref{consistency1}). 

Let us first consider a possibility that $\lim_{k \to 0} \mathcal{R}_2^*(k,s)/k^2 =0$.
In this case $C_1=0$ and~(\ref{ccc}) is inconsistent with~(\ref{consistency1}) already
at the $\mathcal{O}(ds)$ order according to the numerical estimates
$I(1) \simeq 0.285932 K_4$, $I'(1) \simeq -1.217996K_4$, ${I''}_+(1) \simeq 4.9873 K_4$,
${I'''}_+(1) \simeq -22.02 K_4$. In this case, only $I'(1)$ and ${I''}_+(1)$ are important.
The above values satisfy certain relations
\begin{eqnarray}
 I(1) - J(1) &=& \frac{1}{2} K_4 \;, \\
 I'(1) - J'(1) &=& - K_4 \;, \\
 {I''}_+(1) - J''(1) &=& 5 K_4 \;, \\
 {I'''}_+(1) - J'''(1) &=& -22 K_4 \;, \\
 3 I'(1) + {I''}_+(1) &=& \frac{4}{3} K_4
\label{eq:relation}
\end{eqnarray}
within the given numerical accuracy, 
indicating that these values are not accidental.
Moreover, we have verified that the values of the derivatives, calculated 
directly from~(\ref{Iprim}) and~(\ref{I2prim}), are consistent with the ones
obtained by a numerical differentiation of $I(q)$ and $I'(q)$.
Therefore, we are confident about these values.

According to~(\ref{eq:relation}), the coefficient at $k^2 ds$ in~(\ref{ccc})
is $K_4^2/6$ if $C_1=0$, whereas it must be $-K_4^2/12$ 
if~(\ref{consistency1}) holds. The coefficient $C_1$ can be adjusted to reach
the consistency at the $\mathcal{O}(ds)$ order. In this case we set $C_1 = - 2K_4$,
and evaluate $C_2$ to check the consistency at the $\mathcal{O} \left( (ds)^2 \right)$
order. It is based on the observation that
\begin{eqnarray}
\lim\limits_{\Delta \to 0} \left. \frac{\partial \bar{I}(q,p)}{\partial q} \right|_{p=q-\Delta}
&=& B_1 \, (q-1) \quad \mbox{at} \quad q-1 \to 0 \;, \\
\lim\limits_{\Delta \to 0} \left. \frac{\partial^2 \bar{I}(q,p)}{\partial q^2} \right|_{p=q-\Delta}
&=& B_2 \, (q-1) \quad \mbox{at} \quad q-1 \to 0
\end{eqnarray}
 hold with some constants $B_1$ and $B_2$. Then we have
\begin{equation}
 C_2 = 3 B_1 + B_2 \;.
\end{equation}
Recall that $\partial \bar{I}(q,p)/\partial q$ and $\partial^2 \bar{I}(q,p)/\partial q^2$ are 
given by modified Eqs.~(\ref{Iprim}) and~(\ref{I2prim}), where the integration limits for $k$
are set from $1$ to $p$. The coefficient $B_1$ can be easily calculated analytically: 
$B_1= - \left( \frac{\sqrt{3}}{\pi} + \frac{1}{3} \right) K_4$.
For $B_2$, we use the decomposition
\begin{equation}
 B_2 = \widetilde{B}_2 + \hat{B}_2 \;,
\end{equation}
 where $\widetilde{B}_2$ comes from the integrals over $k$, whereas $\hat{B}_2$ --
from the double integral over $k$ and $\theta$. A simple calculation yields
 $\widetilde{B}_2 = \frac{2 K_4}{\sqrt{3} \, \pi}$. In the double integral for $\hat{B}_2$, 
we have $\hat{\theta}(q,k) = 2 \pi/3 + \mathcal{O}(ds)$, therefore this integration limit
can be set $2 \pi /3$ in the actual calculations up to the order of $(ds)^2$.
Then we perform the integration over $k$ to obtain the result in the form of~(\ref{Igrave}),
i.~e.,
\begin{eqnarray}
&& \lim\limits_{\Delta \to 0} \left[
\frac{2K_4}{\pi} \int\limits_{2 \pi/3}^{\pi}
\left( \frac{(q-\Delta)^2 - q^2 - 4(q-\Delta) \cos \theta (q+ (q-\Delta) \cos \theta)}
{\left(q^2+2(q-\Delta) q \cos \theta + (q-\Delta)^2 \right)^2} \right. \right. \nonumber \\ 
&&\left. - \left. \frac{1 - q^2 - 4 \cos \theta (q+  \cos \theta)}
{\left(q^2+2 q \cos \theta + 1 \right)^2}  \right) \sin^2 \theta d \theta \right]
= \hat{B}_2 \, (q-1) \quad \mbox{at} \quad q-1 \to 0 \;.
\label{haha}
\end{eqnarray}
Based on the numerical analysis of~(\ref{haha}), we have estimated
$\hat{B}_2 \simeq 5.654 K_4$. Note that
\begin{equation}
 3B_1 + \hat{B}_2 = 3K_4
\end{equation}
holds within the numerical error, indicating that $5.654$ is not an accidental value. 
Hence, $C_2= \left(3 + \frac{2}{\sqrt{3} \, \pi}  \right) K_4$
holds at least within the actual numerical accuracy.  
Summarising these estimates, the expansion coefficient at $(ds)^2$
in~(\ref{ccc}) within the given numerical accuracy is 
\begin{equation}
\frac{K_4}{8} \left\{ 2{I''}_+(1) + \frac{1}{2}{I'''}_+(1) - \frac{1}{2} C_1 + C_2 \right\} 
= \frac{5}{12} K_4^2 \;,
\end{equation}
if $C_1=-2K_4$. According to~(\ref{consistency1}),
the coefficient $1/24$ is expected instead of $5/12$ obtained here.
Again, it is very unlikely that such a simple rational coefficient
could be accidentally produced by an erroneous numerical calculation.
It shows that the necessary consistency condition~(\ref{consistency1}) and, consequently, 
also~(\ref{consistency}) is not satisfied.
According to the consideration in Sec.~\ref{sec:scaling}, it implies that
$D(0,\infty;{\bf k})/k^2$ diverges at $k \to 0$.

Recall that this result is obtained, assuming that finite (or zero) limits exist in~(\ref{Pi}), 
(\ref{eq:Pii}), (\ref{eq:etax}) and~(\ref{eq:RR}). This is the best scenario for the $\varepsilon$--expansion,
since divergent limits would imply the instability of the RG flow. Moreover,
the universality of the critical exponent $\eta$, determined from the fixed--point
equation $c'=c$ in Sec.~\ref{sec:fixp}, as well as the $\propto k^{-2+\eta}$ scaling of the
fixed--point correlation function considered in Sec.~\ref{sec:corfix}, is ensured by the existence of these limits.

The derived here contradiction implies that, even in this best case, the perturbative 
$\varepsilon$--expansion--based renormalization of $X({\bf k})$ on the critical surface is not fully
consistent with the one produced by the exact rescaling~(\ref{eq:rescal}) of the 
expected asymptotic expansion~(\ref{eq:Xsc0}) at small $k$. Note that the renormalization considered in
Sec.~\ref{sec:crsurf} is an infinite--scale ($m \to \infty$) renormalization at $\varepsilon \to 0$.
If the $\varepsilon$--expansion works correctly there,
then it turns out that an inconsistency appears on a finite renormalization scale
considered in this section. 
This contradiction shows that the $\varepsilon$--expansion fails to give correct results in the actual test, despite
of the fact that it is formally well defined and stable.

\subsection{The two--point correlation function above the critical point}
\label{sec:above}

Consider now the inverse of the two--point correlation function above the
critical point. It implies that $X({\bf k})$ has to be
positively defined for all ${\bf k}$ when the diagrams
are represented by ${\bf k}$--space integrals.
As throughout the paper, we consider the domain 
$r= \mathcal{O}(\varepsilon)$ and $u=\mathcal{O}(\varepsilon)$,
where the $\varepsilon$--expansion can be, in principle, valid. 
A problem here is such that the equation contains divergent diagrams.
Namely, we have
\begin{eqnarray}
X({\bf k}) &=& c k^2 - \frac{n+2}{2} u^2 \left\{ \vverticetet + \vverticerrk  
+3 \left( \vverticeryk  + \vverticerok  \right) \right\} \nonumber \\ 
 &+& \left( r + \frac{n+2}{2} u \vverticer \right) 
\left( 1 - \frac{n+2}{2} \, u  \vverticeu   \right) 
+  \mathcal{O} \left( \varepsilon^3 \right)  
\label{eq:Xabove}
\end{eqnarray}
where the diagram \vverticeu has to be calculated at $d=4$ at this order 
of the $\varepsilon$--expansion. It diverges as $\propto \int_0^{\Lambda} k^{-1}dk$
within the $\varepsilon$--expansion.
A standard idea, used in the perturbative renormalization, is to 
choose such values of the Hamilton parameters (coupling constants) at 
which the divergences cancel. It is also called renormalization.
However, in this simple example, the cancellation (at $u \ne 0$) occurs if and only if
$r + \frac{n+2}{2} u \vverticer = 0$ holds. This is exactly
the condition of the critical surface~(\ref{crsurf}).
Hence, the cancellation never occurs slightly above the critical point,
and such a renormalization method is not helpful here.
Another question is whether a correct result is obtained, if the divergent diagram
\vverticeu is simply discarded. The answer is negative. Indeed, it can be easily checked that
$X({\bf k})$ obtained in this way is inconsistent with the exact rescaling 
relation~(\ref{eq:rescal}). The latter finding is not surprising: in the diagrammatic perturbation theory
this relation, similarly as the semigroup property and the $s$--independence of the fixed point
of the RG transformation $R_s$, is a consequence of certain diagram rescaling and summation rules. 
These rules cannot work to give the expected result if some diagram is simply omitted.

In view of these facts, the applicability of such kind of perturbative renormalization,
where coupling constants are adjusted to cancel the divergent terms or, alternatively,
the divergent terms are simply discarded, is doubtful.

A serious idea is to perform the RG transformation until the
model becomes sufficiently off--critical. It has been applied
in the rigorous RG analysis in four dimensions~\cite{Hara_private,Hara}.
The only question, which might arise here, is how it 
can be done consistently within the
$\varepsilon$--expansion in $4-\varepsilon$ dimensions.

Certain method of calculation of the two--point correlation
function has been considered in~\cite{K1}, which also resolves the problem of divergent diagrams.
It, however, is not based on the $\varepsilon$--expansion, but on certain grouping of diagrams,
in such a way that the true correlation function instead of the Gaussian one is related to the coupling lines. 
It naturally ensures the convergence of the diagrams of the reorganized perturbation theory and
allows an analytic continuation of the obtained self consistent equations from the region $r>0$
to the vicinity of the critical point.

\section{Discussion}
\label{sec:discussion}

There are a lot of perturbative RG studies of the $\varphi^4$ model
made in the past (see~\cite{Ma,Justin,Kleinert,PV} for a review). 
They can be classified as approximative treatments
of the perturbation theory in view of our analysis, since a set of 
apparently irrelevant terms is neglected. Although a perturbative RG
approach to critical phenomena never could be considered as a rigorous method,
it is, nevertheless, possible to calculate exactly the perturbation terms
and to take them into account in a systematic way.
In particular, we find that there are many different terms to be included
in the renormalized Hamiltonian up to the order of $\varepsilon^3$, and 
these terms are relevant, as discussed at the end of Sec.~\ref{sec:ep3}.
To the contrary, one usually finds that only few terms, i.~e., those
included already in the initial Hamiltonian~(\ref{eq:H}) are relevant. 
A confusion about this might be caused by the fact that in the
usual (non-rigorous) treatments one looks for the behaviour of individual
terms and finds that most of them are shrinking in the renormalization procedure.
It, however, is not a rigorous and even not a valid argument, since
the $H^{(2m)}$ parts of~(\ref{eq:HHHH}) (as well as any part of $H^{(2m)}$ represented
as a sum over diagrams of certain topology) result from 
many such individual terms, summing up to yield relevant contributions.

The results of perturbative calculations up to the order of $\varepsilon^5$
are reported in literature. These, however, are based on the Callan--Symanzik
equation, but not on the Wilson's equation~(\ref{eq:WERGE}). The Callan--Symanzik
equation, in fact, represents a scaling property for the Hamiltonian
of the form~(\ref{eq:H}). The approach, thus, relies on the assumption that
all the relevant terms are included in the initial Hamiltonian,
only the existing here coupling constants being renormalized.
This assumption is not supported by our analysis: the perturbative renormalization
based on first principles (i.~e., avoiding such assumptions via using~(\ref{eq:WERGE})) produces
additional relevant terms. 

One believes that the $\varepsilon$--expansion, combined with the perturbative 
RG transformation, is able to describe correctly the small--${\bf k}$ asymptotic
of the correlation functions and the related critical phenomena.
The contradiction derived in Sec.~\ref{sec:furthertest} causes some doubts about it.
Due to the formal character of the $\varepsilon$--expansion, one might happen that
it describes the behaviour within any fixed range of non-zero wave vectors at $\varepsilon \to 0$ 
rather than the $k \to 0$ limit at a small, but fixed $\varepsilon$.
Indeed, the contradiction might be caused by the fact that the non--Gaussian part of the 
Hamiltonian cannot be considered as a small perturbation, like in the $\varepsilon$--expansion, 
if one considers just the small--${\bf k}$ contribution responsible for the critical fluctuations,
since the Fourier amplitudes of the relevant $\varphi_i({\bf x})$ configurations diverge
when $k \to 0$ at criticality.

\section{Conclusions}

\begin{enumerate}
\item
A diagrammatic formulation of the perturbative renormalization has been 
provided (Sec.~\ref{sec:diag}), which makes calculations of perturbation 
terms straightforward and transparent, avoiding any intermediate approximations.
\item 
The RG flow equations, including all terms up to the order of $\varepsilon^2$,
have been considered in Sec.~\ref{sec:eps2}. The tests of the expected properties, such as the
semigroup property and the existence of an independent of the scale parameter $s$ fixed point, 
have been performed. An expected scenario has been considered, where certain limits (Eqs.~(\ref{Pi}),
(\ref{eq:Pii}) and~(\ref{eq:etax})) exist. We have found that the tested here properties
hold in this case.
\item 
An alternative approach has been briefly considered in Sec.~\ref{sec:fixedd},
where $\varepsilon$ is fixed and the coupling constant $u$ is an expansion parameter.
It has been shown that the fixed point is not $s$--independent in this case.
The correctness of the known approach of the expansion at fixed dimension $d=3$
has been questioned in this respect.
\item 
At the $\varepsilon^3$ order, the renormalized Hamiltonian  contains several extra 
($\varphi^4$, $\varphi^6$, $\varphi^8$) vertices, as compared to the
bare Hamiltonian. All these terms are relevant, as discussed at the end of Sec.~\ref{sec:ep3}.
\item
The two--point correlation function has been considered in Sec.~\ref{sec:twopoint}
within the $\varepsilon$--expansion up to the order of $\varepsilon^2$.
It is consistent with the expected power--like behaviour at the fixed point (Sec.~\ref{sec:corfix})
if the limit~(\ref{eq:RR}) exists.
Nevertheless, a contradiction with the exact rescaling of the small--${\bf k}$ 
asymptotic expansion on the critical surface has been found (Sec.~\ref{sec:furthertest}).
\item
Some aspects of the perturbative renormalization of the two--point correlation function slightly
above the critical point have been briefly discussed in Sec.~\ref{sec:above}. 
\end{enumerate}
Concerning the points 4 and 5, several related issues have been discussed in Sec.~\ref{sec:discussion}.

\end{document}